\documentclass{ws-procs9x6}

\usepackage{amsfonts}
\usepackage{amssymb}
\usepackage{graphicx} 
\usepackage{color}

\definecolor{darkgreen}{rgb}{0,0.5,0}
\definecolor{purple}{rgb}{0.5,0,0.5}
\definecolor{nblue}{rgb}{0.0,0.0,0.50}
\definecolor{scarlet}{rgb}{1.0,0.2,0}

\usepackage[colorlinks=true, pdfstartview=FitV, linkcolor=purple, citecolor= purple, urlcolor=blue]{hyperref}

\newcommand{\sfrac}[2]{\mbox{\footnotesize $\displaystyle \frac{#1}{#2}$}} 
\newcommand{\lsim}{\mathrel{\rlap{\lower4pt\hbox{\hskip0pt$\sim$}} 
\raise1pt\hbox{$<$}}}           
\newcommand{\gsim}{\mathrel{\rlap{\lower4pt\hbox{\hskip0pt$\sim$}} 
\raise1pt\hbox{$>$}}}           

\renewcommand{\theequation}{\arabic{section}.\arabic{equation}}

\begin{document}

\title{Hadron Physics and \\ Dyson-Schwinger Equations}

\author{\footnotesize A.~H\"OLL,$\!$\footnotemark[2]\,\, %
C.\,D.~ROBERTS\footnotemark[1]\,\,\,\footnotemark[2]\,\, %
and S.\,V.~WRIGHT\footnotemark[1]
}

\address{%
\footnotemark[1]\ Physics Division, Argonne National Laboratory,\\ Argonne IL 60439, 
USA%
\\
\footnotemark[2]\ Institut f\"ur Physik, Universit\"at Rostock,\\ D-18051 Rostock, Germany
}

\maketitle

\abstracts{Detailed investigations of the structure of hadrons are essential for understanding how matter is constructed from the quarks and gluons of QCD, and amongst the questions posed to modern hadron physics, three stand out.  What is the rigorous, quantitative mechanism responsible for confinement?  What is the connection between confinement and dynamical chiral symmetry breaking?  And are these phenomena together sufficient to explain the origin of more than 98\% of the mass of the observable universe?  Such questions may only be answered using the full machinery of nonperturbative relativistic quantum field theory.  These lecture notes provide an introduction to the application of Dyson-Schwinger equations in this context, and a perspective on progress toward answering these key questions.}

\centerline{\rule{0.9\textwidth}{0.1ex}}

\section*{Table of Contents}
Section~\ref{intro} -- \textbf{Introduction} \dotfill\ \pageref{intro}

\hspace*{-\parindent}Section~\ref{DSEsection} -- \textbf{Dyson-Schwinger Equation Primer} \dotfill\ \pageref{DSEsection}

\hspace*{1.35\parindent}\ref{photonvacuum} - \emph{\textbf{Photon Vacuum Polarisation}} \dotfill\ \pageref{photonvacuum}

\hspace*{1.35\parindent}\ref{fermiongapequation} - \emph{\textbf{Fermion Gap Equation}} \dotfill\ \pageref{fermiongapequation}

\hspace*{-\parindent}Section~\ref{hadronphysics} -- \textbf{Hadron Physics} \dotfill\ \pageref{hadronphysics}

\hspace*{1.35\parindent}\ref{formulationofqcd} - \emph{\textbf{Aspects of QCD}} \dotfill\ \pageref{formulationofqcd}

\hspace*{1.35\parindent}\ref{emergentQCD} - \emph{\textbf{Emergent Phenomena}}\dotfill\ \pageref{emergentQCD}

\hspace*{-\parindent}Section~\ref{dsea} -- \textbf{Nonperturbative Tool in the Continuum} \dotfill\ \pageref{dsea}

\hspace*{1.35\parindent}\ref{dynamicalmass} - \emph{\textbf{Dynamical Mass Generation}}\dotfill\ \pageref{dynamicalmass}

\hspace*{1.35\parindent}\ref{confinement} - \emph{\textbf{Dynamical Mass and Confinement}}\dotfill\ \pageref{confinement}

\hspace*{-\parindent}Section~\ref{mesonproperties} -- \textbf{Meson Properties} \dotfill\ \pageref{mesonproperties}

\hspace*{1.35\parindent}\ref{coloured} - \emph{\textbf{Coloured Two- and Three-point  Functions}} \dotfill\ \pageref{coloured}

\hspace*{1.35\parindent}\ref{colourless} - \emph{\textbf{Colour-singlet Bound States}}
\dotfill\ \pageref{colourless}




\hspace*{-\parindent}Section~\ref{baryon} -- \textbf{Baryon Properties} \dotfill\ \pageref{baryon}

\hspace*{1.35\parindent}\ref{faddeev} - \emph{\textbf{Faddeev Equation}}
\dotfill\ \pageref{faddeev}


\hspace*{1.35\parindent}\ref{NDmasses} - \emph{\textbf{Nucleon and \mbox{\boldmath $\Delta$} Masses}}
\dotfill\ \pageref{NDmasses}



\hspace*{1.35\parindent}\ref{nucleonform} - \emph{\textbf{Nucleon Electromagnetic Form Factors}} \dotfill\ \pageref{nucleonform}





\hspace*{-\parindent}Section~\ref{epilogue} -- \textbf{Epilogue} \dotfill\ \pageref{epilogue}

\hspace*{-\parindent}\ref{Appendix1} -- \textbf{Euclidean Space} \dotfill \pageref{Appendix1}

\hspace*{2.5\parindent}\textbf{References} \dotfill \pageref{ibd1}

\centerline{\rule{0.9\textwidth}{0.1ex}}

\section{Introduction}
\label{intro}
A theoretical understanding of the phenomena of Hadron Physics requires the use of the full machinery of relativistic quantum field theory, which is based on the relativistic quantum mechanics of Dirac, and is currently the favoured way to reconcile quantum mechanics with special relativity.\footnote{In the following we assume that the reader is familiar with the notation and conventions of relativistic quantum mechanics.  For those for whom that is not the case we recommend  Ref.\,[\refcite{bd1}], in particular Chaps.~1-6.}

It is noteworthy that the unification of special relativity (viz., the requirement that the equations of physics be Poincar\'e covariant) and quantum mechanics took quite some time.  Indeed, questions still remain as to a practical implementation of an Hamiltonian formulation of the relativistic quantum mechanics of interacting systems.  The Poincar\'e group has ten generators: six associated with the Lorentz transformations (rotations and boosts); and four associated with translations.  Quantum mechanics describes the time evolution of a system with interactions and that evolution is generated by the Hamiltonian, or some generalisation thereof.  However, the Hamiltonian is one of the generators of the Poincar\'e group, and it is apparent from the Poincar\'e algebra that boosts do not commute with the Hamiltonian.  Hence the state vector calculated in one momentum frame will not be kinematically related to the state in another frame, a fact that makes a new calculation necessary in every momentum frame.  The discussion of scattering, which takes a state of momentum $p$ to another state with momentum $p^\prime$, is therefore problematic.\cite{kp91,co92} 

Moreover, relativistic quantum mechanics predicts the existence of antiparticles; i.e., the equations of relativistic quantum mechanics admit \emph{negative energy} solutions.  However, once one allows for negative energy, then particle number conservation is lost: 
\begin{equation}E_{\rm system} = E_{\rm system} + (E_{p_1} + E_{\bar{p}_1}) + \ldots\ \mbox{\emph{ad~infinitum}},
\end{equation}
where $E_{\bar k}= - E_{k}$.  This poses a fundamental problem for relativistic quantum mechanics: few particle systems can be studied, but the study of (infinitely) many bodies is difficult and no general theory currently exists.  

Relativistic quantum field theory provides a way forward.  In this framework the fundamental entities are fields, which can simultaneously represent infinitely many particles.  The neutral scalar field, $\phi(x)$, provides an example.  One may write
\begin{equation}
\phi(x) = \int\frac{d^3 k}{(2\pi)^3 }\frac{1}{2 \omega_k} \left[ a(k) {\rm e}^{-i k\cdot x} + a^\dagger(k) {\rm e}^{i k\cdot x} \right],
\end{equation}
where: $\omega_k=\sqrt{|\vec{k}|^2+m^2}$ is the relativistic dispersion relation for a massive particle; the four-vector $(k^\mu)= (\omega_k,\vec{k})$; $a(k)$ is an annihilation (creation) operator for a particle (antiparticle) with four-momentum $k$ ($-k$); and $a^\dagger(k)$ is a creation (annihilation) operator for a particle (antiparticle) with four-momentum $k$ ($-k$).  With this plane-wave expansion of the field one may proceed to develop a framework in which the nonconservation of particle number is not a problem.  That is crucial because key observable phenomena in hadron physics are essentially connected with the existence of \emph{virtual} particles.

Relativistic quantum field theory has its own problems, however.  For example, the question of whether a given quantum field theory is rigorously well defined is an \emph{unsolved} mathematical problem.  All relativistic quantum field theories admit analysis via perturbation theory, and perturbative renormalisation is a well-defined procedure that has long been used in Quantum Electrodynamics (QED) and Quantum Chromodynamics (QCD).  However, a rigorous definition of a theory means proving that the theory makes sense \emph{nonperturbatively}.  This is equivalent to proving that all the theory's renormalisation constants are nonperturbatively well-behaved.

An understanding of the properties of hadrons; viz., Hadron Physics, involves QCD.  This theory makes excellent sense perturbatively, as demonstrated in the Nobel Prize winning work on asymptotic freedom by Gross, Politzer and Wilczek.\cite{nobelQCD} However, QCD is not known to be a rigorously well-defined theory and hence it cannot yet truly be described as the theory of the strong interaction.

Nevertheless, the development of an understanding of observable phenomena cannot wait on mathematics.  Assumptions must be made and their consequences explored.  Practitioners therefore assume that QCD is (somehow) well-defined and follow where it may lead.  In experiment that means exploring and mapping the hadron physics landscape with well-understood probes, such as the electron at JLab; while in theory one employs established mathematical tools, and refines and invents others in order to use the Lagrangian of QCD to predict what should be observable real-world phenomena.

A primary aim of the world's current hadron physics programmes in experiment and theory is to determine whether there are any contradictions with what we can actually \emph{prove} in QCD.  Hitherto, there are none that are uncontroversial.\footnote{The pion's valence-quark distribution is one such contentious example.\protect\cite{cdrvalence,reimervalence}}  In this field the interplay between experiment and theory is the engine of discovery and progress, and the \emph{discovery potential} of both is high.  Much has been learnt in the last five years and one can safely expect that many surprises remain in Hadron Physics. 

QCD is a local gauge theory, and such theories are the keystone of contemporary hadron and high-energy physics.  They are difficult to quantise because one must deal with the gauge dependence, which is an extra non-dynamical degree of freedom.  The modern approach is to quantise these theories using the method of functional integrals, and Refs.\,[\refcite{iz80,pt84}] provide excellent descriptions.  The method of functional integration replaces canonical second-quantisation.  One may view this approach as originating in the path integral formulation of quantum mechanics.\cite{fh65}  NB. In general, mathematicians do not regard local gauge theory functional integrals as \emph{well-defined}.

In quantum field theory all physical amplitudes can be obtained from Green functions, which are expectation values of time-ordered products of fields measured with respect to the physical vacuum.\footnote{The physical or interacting vacuum is the analogue of the true ground state in quantum mechanics.} They describe all the characteristics of an interacting system.  The Green functions are obtained from generating functionals, the specification of which begins with the theory's action expressed in terms of the Poincar\'e invariant Lagrangian density.  

An analysis of the generating functional for interacting bosons proceeds almost classically.  The field variables and functional derivatives can be treated as ``c-numbers'', and a perturbative truncation of any Green function can be obtained in a straightforward manner.  A measure of clarity and rigour may be introduced by interpreting spacetime as a discrete lattice of points and recovering the continuum via a limiting procedure.  On these aspects, Appendix B of Ref.\,[\refcite{pt84}] is instructive.  Following this route it is plain that in perturbation theory the vacuum is trivial; viz., features such as dynamical symmetry breaking are impossible.

A complication is encountered in dealing with fermions; namely, fermionic fields do not have a classical analogue because classical physics does not contain anticommuting field variables.  In order to treat fermions using functional integrals one must employ Grassmann variables.  Reference [\refcite{fb66}] is the standard source for a rigorous discussion of Grassmann algebras, and Appendix B of Ref.\,[\protect\refcite{pt84}] is again instructive in this connection.

In order to illustrate some of the concepts described above we will work through an example: the case of a noninteracting Dirac quantum field.  The Lagrangian density for the free Dirac field is 
\begin{equation} 
\label{freeDiracL} L_0^\psi(x)= 
\bar\psi(x) \left( i \partial 
\!\!\!\slash\; -m \right) \psi(x)\,. 
\end{equation} 
Consider therefore the functional integral
\begin{eqnarray}
\nonumber
W[\bar\Xi,\Xi] &= &\int [D\bar\psi(x)] [D\psi(x)]\; 
{\rm e}^{\displaystyle i \!\int d^4 x\,  \bar\psi(x) \left( i \partial \!\!\!\slash\; -m + i \eta^+ \right)\psi(x)}\\
&& \times\; {\rm e}^{\displaystyle i \! \int d^4 x  \left(\bar\psi(x) \Xi(x) + \bar\Xi(x)\psi(x)\right)},
\label{Wfermion}
\end{eqnarray}
where $\eta\to 0^+$ as the last step in any calculation.\footnote{$\eta$ is a convergence factor, which is necessary to define the integral.  It subsequently appears in propagators and thereby implements Feynman boundary conditions, as discussed, e.g., in Ref.\,[\protect\refcite{pt84}], App.\,B.}  This is the generating functional for complete $n$-point Green functions in the quantum field theory.  Here ``complete'' means that the $n$-point Green function, $G(x_1, x_2, \ldots , x_n)$, will include contributions from products of lower-order Green
functions ($m_i,m_j, {\rm etc.}<n$); i.e., disconnected diagrams.  In Eq.\,(\ref{Wfermion}), $\bar\psi(x)$, $\psi(x)$ are identified with the generators of $G$, a Grassmann algebra with involution: the latter means that an inner-product of sorts is defined. There is a minor additional complication here -- the spinor degree-of-freedom is implicit; i.e., to be explicit, one should write 
\begin{equation}
\prod_{r=1}^4 [D\bar\psi_r(x)] \prod_{s=1}^4 [\psi_s(x)].
\end{equation}
However, that only adds a finite matrix degree-of-freedom to the problem, which may easily be handled.  In Eq.~(\ref{Wfermion}) we have introduced anticommuting sources: $\bar\Xi(x)$, $\Xi(x)$, which are also elements in $G$.

To evaluate the free-field functional integral, which is Gaussian, one writes
\begin{equation} 
O(x,y) = (i \partial \!\!\!\slash\; -m + i \eta^+) \delta^4(x-y) 
\end{equation} 
and observes that the solution of 
\begin{equation}
\int d^4w \, O(x,w) \, P(w,y) = \mbox{\boldmath $I$}_{\rm D} \,\delta^4(x-y)\,;
\end{equation}
i.e., the inverse of the operator $O(x,y)$, is precisely the free-fermion propagator: 
\begin{equation} 
P(x,y) = S_0(x-y) = 
\int\frac{d^4 p}{(2\pi)^4} \, 
{\rm e}^{-i p \cdot(x-y)} \, S_0(p)\,,
\end{equation} 
with
\begin{equation}
S_0(p)= \frac{p\!\!\slash\, + m}{p^2-m^2+i \eta^+}\,. 
\end{equation}
NB.\ This can be verified by substitution, using $\{\gamma_\mu ,\gamma_\nu\} = 2 g_{\mu\nu} \mbox{\boldmath $I$}_{\rm D}$.  It is true in general that in the absence of external sources, $n$-point functions are translationally invariant.

One can now rewrite Eq.~(\ref{Wfermion}) in the form 
\begin{eqnarray}
\nonumber 
\rule{-5em}{0ex}W[\bar\Xi,\Xi] & = & \int [D\bar\psi(x)] [D\psi(x)]\, {\rm e}^{ \displaystyle i \!\int d^4 x d^4y\, \bar\psi^\prime(x) \,
O(x,y)\, \psi^\prime(y)}\\
&& \times\, {\rm e}^{ \displaystyle -i \!\int d^4 x d^4y\,  \bar\Xi(x) \, S_0(x-y) \, \Xi(y)},
\end{eqnarray}
wherein
\begin{equation}
\begin{array}{rrcl} 
&\bar\psi^\prime(x)  & :=&   \bar\psi(x) + \int d^4 w \, \bar\Xi(w)\,
S_0(w-x)\,,\\
& \psi^\prime(x)  & :=&  \psi(x) + \int d^4 w \,  S_0(x-w) \,\Xi(w)\,. 
\end{array}
\end{equation} 
The new fields $\bar\psi^\prime(x)$ and $\psi^\prime(x)$ are still in $G$, and are related to the original variables by a unitary transformation.  Thus the change of variables introduces only a unit Jacobian and hence
\begin{eqnarray} 
\nonumber \lefteqn{ W[\bar\Xi,\Xi] =
{\rm e}^{\displaystyle- i \!\int d^4 x d^4y\, \bar\Xi(x)\,S_0(x-y)\, \Xi(y)}} \\
 && \times
\int [{D}\bar\psi^\prime(x)] [{ D}\psi^\prime(x)]\; 
{\rm e}^{\displaystyle i \!\int d^4 x d^4y\,  \bar\psi^\prime(x) O(x,y) \psi^\prime(y) }  .
\end{eqnarray} 
The expression in the second line of this equation is a standard functional integral:
\begin{equation}
{\rm e}^{\displaystyle i \!\int d^4 x d^4y\,  \bar\psi^\prime(x) O(x,y) \psi^\prime(y) } = {\rm Det}[O(x-y)]\,,
\end{equation}
where ``${\rm Det}$'' is a generalisation of the concept of a matrix determinant.  One therefore arrives at 
\begin{eqnarray} 
{W[\bar\Xi,\Xi]} & = & { \frac{1}{ N_0^\psi}\; {\rm e}^{\displaystyle - i \int d^4 x
d^4y\, \bar\Xi(x)\,S_0(x-y)\, \Xi(y)}}, \label{FFGF}
\end{eqnarray} 
where 
\begin{equation}
{ N}_0^\psi := {\rm Det} [i S_0(x-y) ].
\end{equation}
Clearly, with the definition in Eq.\,(\ref{Wfermion}), $\displaystyle { N}_0^\psi \, \left. W[\bar\Xi,\Xi] \right|_{\bar\Xi=0=\Xi} = 1$.  This is not a convenient normalisation and it is therefore customary to redefine $W[\bar\Xi,\Xi]$ so that ${ N}_0^\psi$ is included in the measure ``$ [{D}\bar\psi^\prime(x)] [{ D}\psi^\prime(x)]$'' and 
\begin{equation}
\left. W[\bar\Xi,\Xi] \right|_{\bar\Xi=0=\Xi} = 1.
\end{equation}

The two-point Green function for the free-fermion quantum field theory is now easily obtained: 
\begin{eqnarray} 
\nonumber \lefteqn{ \left. \frac{\delta^2 W[\bar\Xi,\Xi]}{i 
\delta\bar\Xi(x) \,(-i)\delta\Xi(y)}\right|_{\bar\Xi=0=\Xi} 
=:   \frac{\langle 0 | T\{\hat\psi(x) \hat{\bar\psi}(y)\} | 0\rangle } 
{\langle 0 | 0 \rangle } }\\ 
  & &  \!\!\!\!\!\!\!\! =  \int [{ D}\bar\psi(x)] [{ D}\psi(x)]\, \psi(x) 
\bar\psi(y)\, {\rm e}^{\displaystyle i \int d^4 x\, \bar\psi(x) \left( i
\partial \!\!\!\slash\; -m + i \eta^+ \right) \psi(x) }.
\end{eqnarray}
The functional differentiation in Eq.\,(\ref{FFGF}) is straightforward and yields 
\begin{equation}
\left. \frac{\delta^2 W[\bar\Xi,\Xi]}{i 
\delta\bar\Xi(x) \,(-i)\delta\Xi(y)}\right|_{\bar\Xi=0=\Xi} 
= i \, S_0(x-y) \,; \label{propfreeF}
\end{equation} 
i.e., the inverse of the Dirac operator.  

It is useful to have systematic procedure for the \textit{a priori} elimination of disconnected parts from $n$-point Green functions because the recalculation of $m < n$-point Green functions at every stage is inefficient.  The generating functional for ``connected'' $n$-point Green functions, $Z[\bar\Xi,\Xi]$, is defined via: 
\begin{equation} 
{ W[\bar\Xi,\Xi] =: \exp\left\{ i Z[\bar\Xi,\Xi] \right\}} \,.
\end{equation} 
It follows immediately from Eq.\,(\ref{FFGF}) that 
\begin{equation}
\label{S0Z}
Z[\bar\Xi,\Xi] = - \int d^4 x \,d^4y\; \bar\Xi(x)\,S_0(x-y)\, \Xi(y)\,.
\end{equation}
This equation states that for a noninteracting field, there is one, and only one, connected Green function; namely, the free particle propagator, which is the simplest possible Green function.  NB.\ It is a property of theories based on a Grassmann algebra with involution that one-point Green functions for fermions are identically zero in the absence of external sources.

At this point it is useful to illustrate what is meant by the functional determinant introduced above; i.e., ${\rm Det}[O]$, where $O$ is an integral operator.  This is relevant owing to the importance of this fermion determinant and its absence, e.g., in numerical simulations of lattice-regularised QCD.  Consider a translationally invariant operator
\begin{equation} 
O(x,y) = O(x-y) = \int\!\frac{d^4 p}{(2\pi)^4} \, O(p) \,{\rm e}^{-i p\cdot(x-y)}. 
\end{equation} 
Then, for any function $f$ that may be expressed as a power series on some domain:
\begin{equation}
\label{fsum}
f(x) = \sum_{i=0}^{\infty} f_i x^i\,,
\end{equation}
we have 
\begin{eqnarray} 
\nonumber f[O(x-y)] & = & \int\!\frac{d^4 p}{(2\pi)^4}\, \left\{ f_0 + f_1\, 
O(p) + f_2\, O(p)^2 + [\ldots] \right\}\,{\rm e}^{-i p\cdot (x-y)}\\ 
& = & \label{foperator2} \int\!\frac{d^4 p}{(2\pi)^4}\, f(O(p))\,{\rm e}^{-i 
p\cdot(x-y)}\,. 
\end{eqnarray} 

This expression may be applied directly to ${ N}_0^\psi = {\rm Det} [i S_0(x-y) ]$.  To that end we proceed by noting  
\begin{eqnarray} 
S_0(p)& = & m \,\Delta_0(p^2) \left[ 1 + \frac{p\!\!\slash}{m} \right]\,, 
\; \Delta_0(p^2) = \frac{1}{p^2-m^2 + i \eta^+}\,,\\
\Rightarrow  S_0(x-y)& =& \int d^4 w\, m \,\Delta_0(x-w) \, { F}(w-y)\,, 
\end{eqnarray} 
with $\Delta_0(x-y)$ the obvious Fourier transform of $\Delta_0(p^2)$, and 
\begin{equation} 
{ F}(x-y) = \int\!\frac{d^4 p}{(2\pi)^4}\, \left[ 1 + 
\frac{p\!\!\slash}{m} \right] \,{\rm e}^{-i p\cdot(x-y)}\,. 
\end{equation} 

We now remark that for a bilocal operator $P(x,y)$
\begin{equation}
{\rm Tr} P := \int d^4x\, {\rm tr} P(x,x)\,,
\end{equation}
where ``tr'' indicates a trace over whatever matrix structure is present.  Thus
Eq.\,(\ref{fsum}) entails that ${\rm TrLn}[ O ]= {\rm Ln Det}[ O]$, in addition to the more obvious result ${\rm Ln} [AB] = {\rm Ln} [A] + {\rm Ln} [B]$.  Hence
\begin{equation} 
{\rm Tr} {\rm Ln} \,i S_0(x-y) = {\rm Tr} \left\{ {\rm Ln} \,i\, m 
\Delta_0(x-y) 
+ {\rm Ln} \left[ \delta^4(x-y) + { F}(x-y)\right] \right\}\,. 
\end{equation} 

Applying Eq.\,(\ref{foperator2}) to the second term above one obtains
\begin{eqnarray} 
\nonumber{\rm Tr} {\rm Ln} \left[ \delta^4(x-y) + {
F}(x-y)\right]  & = &  
\int d^4 x\, \int\!\frac{d^4 p}{(2\pi)^4}\,{\rm tr} \ln\left[1 + { 
F}(p)\right]\\ 
& = &\int d^4 x\, \int\!\frac{d^4 p}{(2\pi)^4}\,2\,\ln\left[1 - 
\frac{p^2}{m^2}\right], \label{trlnfermion} 
\end{eqnarray} 
while the first term gives\footnote{In both cases the multiplicative factor $\int d^4 x$ simply measures the (infinite) volume of spacetime.  The factor poses no problems in a properly regularised theory.}
\begin{eqnarray} 
{\rm Tr} {\rm Ln} \, i m \Delta_0(x-y) & = & \int d^4 x\, \int\!\frac{d^4 
p}{(2\pi)^4}\, 2\,\ln\left[ i\, m \Delta_0(p^2)\right]^2\,.
\end{eqnarray} 
Combining this result with Eq.\,(\ref{trlnfermion}) yields
\begin{equation}
{\rm Ln} \,{ N}_0^\psi =  {\rm Tr} {\rm Ln} \,i S_0(x-y) = \int d^4 
x\,\int\!\frac{d^4 p}{(2\pi)^4}\, 2\,\ln \Delta_0(p^2)\,,
\label{fermionDet}
\end{equation}
where the factor ``2'' reflects the spin-degeneracy of the free-fermion's 
eigenvalues.  NB.\ Upon the inclusion of a ``colour'' degree-of-freedom, as in QCD, this would become ``2 $N_c$,'' where $N_c$ is the number of colours.  

The interpretation of Eq.\,(\ref{fermionDet}) is straightforward.  As is the case for finite dimensional matrices; viz., 
\begin{equation}
\label{matrixdet}
\ln \det M = \ln \prod_i \lambda_i^M = \sum_i \ln \lambda_i^M\,,
\end{equation}
the logarithm of the determinant of an operator is simply the logarithm of the product of the operator's eigenvalues, which is equivalent to the sum of the logarithms of these eigenvalues.  In our particular instance, there is a continuum of eigenvalues for the inverse of the free Dirac operator.  For each value of three-momentum $\vec{p}$, we have two spins and a positive and negative energy solution.  The product of these four eigenvalues is described by the function $\Delta_0(p^2)^2$.  Finally, the integral over momentum in Eq.\,(\ref{fermionDet}) is the analogue of the sum in Eq.\,(\ref{matrixdet}).  This picture generalises to the case of more complicated Dirac operators.

Thus far we have made little mention of gauge-boson fields; namely, photons, gluons, etc.  A generating functional for gauge field $n$-point functions can be constructed.  The primary difficulty in this instance is the problem of gauge fixing, which is not yet fully resolved.  The so-called Faddeev-Popov determinant is one part of the solution.  That determinant can be expressed through the introduction of dynamical ghost fields.  We will not write more on the issue herein.\footnote{A pedagogical introduction is provided in Appendix~B of Ref.\,[\protect\refcite{pt84}] and a contemporary perspective may be traced from Ref.\,[\protect\refcite{agwGribov}] and references therein.}  The omission is not crucial for our development.  At this point, the basic qualitative ideas of the functional integral formulation of relativistic quantum field theory have been presented.

\section{Dyson-Schwinger Equation Primer}
\label{DSEsection}
\setcounter{equation}{0}
It has long been known that from the field equations of quantum field theory one can derive a system of coupled integral equations interrelating all of a theory's Green functions \cite{dyson49,schwinger51}.  This collection of a countable infinity of equations is called the complex of Dyson-Schwinger equations (DSEs).  It is an intrinsically nonperturbative structure, which is vitally important in proving the renormalisability of quantum field theories.  Moreover, at its simplest level the complex provides a generating tool for perturbation theory.  In the context of quantum electrodynamics (QED) we will illustrate a nonperturbative derivation of two DSEs.  The derivation of others follows the same pattern.

\subsection{Photon Vacuum Polarisation}
\label{photonvacuum}
The vacuum polarisation is an essentially quantum field theoretical effect and an important part of the Lamb shift.  It may be derived from the action for QED with $N_f$ flavours of electromagnetically active fermions:
\begin{eqnarray} 
\nonumber
S[A_\mu,\psi,\bar\psi] & = & 
\int\, d^4 x\;
\left[\,\sum_{f=1}^{N_f}\, \bar\psi^f(x)  \left( i\not\!\partial - m_0^f + e_0^f
\not\!\!A \right)\psi^f(x) \right.\\
&& \left. - \frac{1}{4} F_{\mu\nu}(x) F^{\mu\nu}(x) -\frac{1}{2\xi_0}\,
 \partial^\mu  A_\mu(x)   \,\partial^\nu A_\nu(x) \right]. \label{Sqed} 
\end{eqnarray}  
The action is manifestly Poincar\'e covariant.  $\bar\psi^f(x)$, $\psi^f(x)$ are elements of a Grassmann algebra with involution that describe the fermion degrees of freedom, with $m_0^f$ representing the fermions' bare masses.  $A_\mu(x)$ describes the gauge boson [photon] field, with
\begin{equation}
F_{\mu\nu} = \partial_\mu A_\nu - \partial_\nu A_\mu
\end{equation}
and $\xi_0$ the bare gauge fixing parameter.  Interactions in the theory are generated by a simple coupling term in the first line of Eq.\,(\ref{Sqed}): $\bar\psi^f e_0^f \not\!\!A \psi^f$, which is linear in all the field variables and has a coupling constant, $e_0^f$, that represents the fermion charges.  (NB.\ Throughout we use $c=1=\hbar$, in which case $e_0^f$ has mass-dimension zero.  To describe an electron the physical charge $e_f < 0$.)  
 
One can write the complete generating functional for this theory:
\begin{eqnarray} 
\nonumber  W[J_\mu,\Xi,\bar\Xi\,] & =&  \int\![{ D}A_\mu] \, 
[{ D}\psi] [{D}\bar\psi] \,\exp \left\{i\int\! d^4 x\, S[A_\mu,\psi,\bar\psi\,] \right.\\ %
& &  \left. + \, J^\mu(x) A_\mu(x) + \bar\Xi^f(x) \psi^f(x) + \bar\psi^f(x) 
\Xi^f(x) \rule{0em}{3.0ex}\right\}, \label{WGFqed} 
\end{eqnarray} 
where $J_\mu$ is an external source for the electromagnetic field, and $\Xi^f$, $\bar\Xi^f$ are external sources for the fermion fields that are, of course, elements in the Grassmann algebra.  As we noted in Sec.\,\ref{intro}, it is advantageous to work with the generating functional of connected Green functions; i.e., $Z[J_\mu,\Xi,\bar\Xi\,]$ defined via 
\begin{equation} 
\label{WZdef} W[J_\mu,\Xi,\bar\Xi\,] =: \exp\left\{ i  Z[J_\mu,\Xi,\bar\Xi]\right\}\,,
\end{equation} 
and that is how we proceed.

The derivation of a DSE now follows simply from an observation that the integral of a total derivative vanishes, given appropriate boundary conditions; e.g., 
\begin{eqnarray} 
\nonumber 0 & =&  \int\![{ D}A_\mu] \, [{ D}\psi] [{ D}\bar\psi] \; \frac{\delta}{\delta A_\mu(x)}\;
\exp \displaystyle \bigg( i S[A_\mu,\psi,\bar\psi\,]  \\
&&  + i \int\,d^4 x\;
\left[ \bar\psi^f\Xi^f + \bar\Xi^f\psi^f + A_\mu J^\mu \right] 
\bigg) \\
%
%
\nonumber & = &\int\![{ D}A_\mu] \, [{ D}\psi] [{
    D}\bar\psi]\,\left\{ \frac{\delta S}{\delta A_\mu(x)} + J_\mu(x) \right\}
    \\   
& & \times 
        \exp\left\{i\left( S[A_\mu,\psi,\bar\psi\,] + 
    \int\,d^4 x\;\left[ \bar\psi^f\Xi^f + \bar\Xi^f\psi^f + A_\mu J^\mu \right] 
                \right)\right\} \nonumber\\ 
& = & \left\{ \frac{\delta S}{\delta A_\mu(x)} 
        \left[\frac{\delta}{i\delta J }, 
        \frac{\delta}{i\delta\bar\Xi}, -\frac{\delta}{i\delta\Xi}\right] 
        + J_\mu(x) \right\} W[J_\mu,\Xi,\bar\Xi] ~, 
\label{ELeqn} 
\end{eqnarray} 
where the last line has meaning as a functional differential operator acting
on the generating functional.

One now needs to differentiate Eq.~(\ref{Sqed}):
\begin{equation} 
\frac{\delta S}{\delta A_\mu(x)} =\left[ \partial_\rho \partial^\rho g_{\mu\nu} 
- \left( 1- \frac{1}{\xi_0}\right) \partial_\mu \partial_\nu\right] 
A^\nu(x) + \sum_f e_0^f \, \bar\psi^f (x)\gamma_\mu \psi^f(x) \,,
\end{equation} 
in which case Eq.\,(\ref{ELeqn}) becomes
\begin{eqnarray} 
\nonumber \lefteqn{ -J_\mu(x) = \left[ \partial_\rho \partial^\rho g_{\mu\nu}
- \left( 1-  \frac{1}{\xi_0}\right) \partial_\mu \partial_\nu\right]
\frac{\delta  Z}{\delta J_\nu(x)} }\\ 
\nonumber &&  +  \sum_f e_0^f\,\left(  - \frac{\delta Z}{\delta\Xi^f(x)} 
        \gamma_\mu \frac{\delta Z}{\delta \bar\Xi^f(x)} 
+ \frac{\delta}{\delta\Xi^f(x)} 
        \left[\gamma_\mu \frac{\delta \, i Z}{\delta \bar\Xi^f(x)} 
        \right]\right) ,\\ 
&& \label{FldEqn} 
\end{eqnarray} 
where we have eliminated a common factor of $W[J_\mu,\Xi,\bar\Xi]$.  Equation~(\ref{FldEqn}) represents a compact form of the nonperturbative
equivalent of Maxwell's equations.

A valuable step now is the introduction of a generating functional for one-particle-irreducible (1PI) Green functions, $\Gamma[A_\mu,\psi,\bar\psi\,]$.  It is obtained from $Z[J_\mu,\Xi,\bar\Xi\,]$ via a Legendre transformation; namely, 
\begin{equation} 
Z[J_\mu,\Xi,\bar\Xi\,] = {\Gamma[A_\mu,\psi,\bar\psi\,]} + \int\!d^{4} x\left[ 
\bar\psi^f\Xi^f + \bar\Xi^f\psi^f + A_\mu J^\mu \right]. \label{Legendre_transf} 
\end{equation} 
A 1PI $n$-point function or ``proper vertex'' does not contain any contribution that becomes disconnected when a single connected $m$-point Green function is removed; e.g., via functional differentiation.  This means that no diagram which represents or contributes to a given proper vertex separates into two disconnected diagrams if only one connected propagator is cut.  (A detailed explanation can be found in Ref.\,[\refcite{iz80}], pp.\,289-294.) 

Consider Eq.\,(\ref{WGFqed}) and observe that
\begin{equation} 
\label{deltaZ} \frac{\delta Z}{\delta J^\mu(x)} = A_\mu(x)\,,\; 
\frac{\delta Z}{\delta \bar\Xi(x)} = \psi(x)\,,\; 
\frac{\delta Z}{\delta \Xi(x)} = -\bar\psi(x)\,, 
\end{equation} 
where here the external sources are \textbf{nonzero}.  It follows that $\Gamma$ in 
Eq.~(\ref{Legendre_transf}) must satisfy 
\begin{equation} 
\label{deltaGamma} \frac{\delta \Gamma}{\delta A^\mu(x)} = - J_\mu(x)\,,\; 
\frac{\delta \Gamma}{\delta \bar\psi^f(x)} = - \Xi^f(x)\,,\; 
\frac{\delta \Gamma}{\delta \psi^f(x)} = \bar\Xi^f(x)\,. 
\end{equation} 
Here one must bear in mind that since the sources are not zero then, e.g., 
\begin{equation} 
A_\rho(x)= A_\rho(x;[J_\mu,\Xi,\bar\Xi\,]) \;\Rightarrow\; \frac{\delta
A_\rho(x)}{\delta J^\mu(y)} \neq 0\,,
\end{equation} 
with analogous statements for the Grassmannian functional derivatives.  NB.\ It follows from the general properties of Grassmann variables that if one sets $\bar\psi = 0 = \psi$ after differentiating $\Gamma$ then the result is zero \textit{unless} there are equal numbers of $\bar\psi$ and $\psi$ derivatives.  

Consider now the operator and matrix product ($r$, $s$, $t$ are spinor labels)
\begin{equation} 
\label{SGamma} 
\left. - \int\! d^4z \, { \frac{\delta^2 Z}{\delta \Xi_r^f(x) 
      \bar\Xi_t^h(z)} }
       \, \frac{\delta^2\Gamma}{\delta \psi_t^h(z) \bar\psi_s^g(y)} 
        \right|_{\begin{array}{c} \Xi=\bar\Xi=0 \\ 
                                \psi=\overline{\psi}=0 
                \end{array}}\,. 
\end{equation} 
Using Eqs.~(\ref{deltaZ}), (\ref{deltaGamma}), this simplifies as follows: 
\begin{eqnarray} 
\nonumber & = & \left. \int\! d^4z \, 
\frac{\delta \psi^h_t(z)}{\delta \Xi_r^f(x)}\, \frac{\delta\Xi^g_s(y)}{\delta 
\psi_t^h(z)} 
        \right|_{\begin{array}{c} \Xi=\bar\Xi=0 \\ 
                                \psi=\overline{\psi}=0 
                \end{array}} \\
& = &  \left. \frac{\delta\Xi^g_s(y)}{\delta \Xi_r^f(x)} 
        \right|_{\begin{array}{c} \psi=\overline{\psi}=0 
                \end{array}}
 =\delta_{rs}\, \delta^{fg}\, \delta^4(x-y)\,. \label{SGammaA}
\end{eqnarray} 
This result is useful in Maxwell's equation, Eq.~(\ref{FldEqn}), whereupon substitution and setting $\bar\Xi=0=\Xi$ subsequently yields
\begin{eqnarray}
\nonumber \left. \frac{\delta\Gamma}{\delta A^\mu(x)}\right|_{\psi=\overline{\psi}=0} & = &
\left[ \partial_\rho \partial^\rho g_{\mu\nu} - \left( 1- 
\frac{1}{\xi_0}\right) 
\partial_\mu \partial_\nu\right] A^\nu(x)\\
&& 
- \, i \,\sum_f e_0^f {\rm tr}\left[ 
\gamma_\mu S^f(x,x;[A_\mu]) \right], \label{FEppta} 
\end{eqnarray} 
wherein we have made the identification (no summation on $f$) 
\begin{equation} 
S^f(x,y;[A_\mu]) =  - \, { \frac{\delta^2 Z}{\delta \Xi^f(y) \bar\Xi^f(x)} }
= \frac{\delta^2 Z}{\delta \bar\Xi^f(x) \Xi^f(y)},
\label{SfA}
\end{equation} 
which is an obvious consequence of Eq.\,(\ref{S0Z}) for a noninteracting theory and the natural definition of the connected fermion two-point function in an interacting theory.  Clearly, the vacuum fermion propagator or connected fermion two-point function is
\begin{equation} 
\label{SAeq0} S^f(x,y):=S^f(x,y;[A_\mu=0])\,. 
\end{equation} 
Such vacuum Green functions are the keystones of quantum field theory.

As a direct consequence of Eqs.\,(\ref{SGamma}), (\ref{SGammaA}), it is apparent that the inverse of the Green function in Eq.\,(\ref{SfA}) is
\begin{equation} 
\label{SfinverseA} S^f(x,y;[A])^{-1} =  \left. \frac{\delta^2 
\Gamma}{\delta\psi^f(x) \delta\bar\psi^f(y)} \right|_{\psi=\overline{\psi}=0} 
\;.
\end{equation} 
This illustrates a general property: functional derivatives of the generating functional for 1PI Green functions are always related to the associated $n$-point function's inverse.  

To continue, we differentiate Eq.~(\ref{FEppta}) with respect to $A_\nu(y)$ and set $J_\mu(x)=0$ to obtain
\begin{eqnarray} 
\nonumber \lefteqn{
\left. \frac{\delta^2 \Gamma}{\delta A^\mu(x) \delta A^\nu(y)}  
\right|_{\begin{array}{c} A_\mu=0 \\ \psi=\overline{\psi}=0
\end{array}}}\\
\nonumber 
& = & \left[ \partial_\rho \partial^\rho g_{\mu\nu} - \left( 1- 
\frac{1}{\xi_0}\right) \partial_\mu \partial_\nu\right] \, \delta^4(x-y) \\
 & &
 -  i \sum_f e_0^f {\rm tr}\left[ 
\gamma_\mu \frac{\delta }{\delta A_\nu(y)} \, \left( \left. \frac{\delta^2 
\Gamma}{\delta\psi^f(x) \delta\bar\psi^f(x)} 
\right|_{\psi=\overline{\psi}=0}\right)^{-1}\right]\,.  \label{deltaFEppta} 
\end{eqnarray} 
The left-hand-side (lhs) of this equation is easily understood:  Eq.~(\ref{SfinverseA}) expresses the inverse of the fermion propagator and here, in analogy, we have the inverse of the gauge-boson propagator 
\begin{equation} 
\label{Dinverse} (D^{-1})_{\mu\nu}(x,y) := \left.\frac{\delta^2 \Gamma}{\delta 
A^\mu(x) \delta A^\nu(y)} \right|_{\begin{array}{c} A_\mu=0 \\ 
\psi=\overline{\psi}=0 
\end{array}}. 
\end{equation} 

The right-hand-side (rhs), however, requires simplification before interpretation.  First observe that
\begin{eqnarray} 
\nonumber \lefteqn{ - \frac{\delta }{\delta A_\nu(y)} \, \left( \left. 
\frac{\delta^2 \Gamma}{\delta\psi^f(x) \delta\bar\psi^f(x)} 
\right|_{\psi=\overline{\psi}=0}\right)^{-1} }\\ 
\nonumber & & 
= \int\!d^4 u d^4 w\, \left( 
\left.\frac{\delta^2 \Gamma}{\delta\psi^f(x) \delta\bar\psi^f(w)} 
\right|_{\psi=\overline{\psi}=0} \right)^{-1}\, \\
&& \times  \frac{\delta }{\delta A_\nu(y)} 
\frac{\delta^2 \Gamma}{\delta\psi^f(u) \delta\bar\psi^f(w)}\, 
\left( \left.\frac{\delta^2 \Gamma}{\delta\psi^f(w) \delta\bar\psi^f(x)} 
\right|_{\psi=\overline{\psi}=0} \right)^{-1} \! .
\label{dAAinv}
\end{eqnarray} 
This is merely an analogue of a result pertaining to finite dimensional matrices: 
\begin{eqnarray} 
\nonumber \frac{d}{dx} \left[ A(x) A(x)^{-1} = \mbox{\boldmath $I$} \right] & = 
0  = &  \frac{d A(x)}{dx} \, A(x)^{-1} + A(x)\,\frac{d A(x)^{-1}}{dx} 
 \\ 
\nonumber
& \Rightarrow & \frac{d A(x)^{-1}}{dx} = - A(x)^{-1} \,\frac{d 
A(x)}{dx}\,A(x)^{-1}\,. \\
\end{eqnarray} 

Equation (\ref{dAAinv}) contains a new 1PI three-point function; namely, 
\begin{equation} 
\label{Gammadef} e_0^f \Gamma_\mu^f(x,y;z) := \frac{\delta }{\delta A^\mu(z)} 
\frac{\delta^2 \Gamma}{\delta\psi^f(x) \delta\bar\psi^f(y)}\,,
\end{equation} 
which is the proper fermion-gauge-boson vertex.  At leading order in perturbation theory 
\begin{equation} 
\Gamma_\nu^f(x,y;z) = \gamma_\nu\,\delta^4(x-z) \, \delta^4(y-z)\,, 
\end{equation} 
a result that can be obtained via explicit calculation of the functional derivatives in Eq.~(\ref{Gammadef}).  

It is now evident that the second derivative of the generating functional for one-particle-irreducible Green functions, $\Gamma[A_\mu,\psi,\bar\psi]$, gives the inverse-fermion and -photon propagators, and the third derivative gives the proper photon-fermion vertex.  It is a general rule that all derivatives of $\Gamma[A_\mu,\psi,\bar\psi]$, higher than two, produce a proper vertex, where the number and type of derivatives gives the number and type of proper Green functions that the vertex can connect.

At this point it is useful to introduce the gauge-boson \textit{vacuum polarisation}:
\begin{equation} 
\Pi_{\mu\nu}(x,y) =  i \sum_f (e_0^f)^2  \int\,d^{4}z_1\,d^{4}z_2\, 
 {\rm tr}\left[ \gamma_\mu S^f(x,z_1)\Gamma^f_\nu(z_1,z_2;y) 
        S^f(z_2,x)\right].
\label{PPTeq} 
\end{equation}
It is the photon's \emph{self-energy} and describes the modification of the gauge-boson's propagation characteristics owing to the presence of virtual particle-antiparticle pairs in quantum field theory.  It is an essential element of quantum electrodynamics and, e.g., plays an important part in the description of a physical process such as $\rho^0\to e^+ e^- $.  With the aid of Eq.\,(\ref{PPTeq}), Eq.\,(\ref{deltaFEppta}) can be written in a compact form: 
\begin{equation} 
(D^{-1})_{\mu\nu}(x,y) = \left[ \partial_\rho \partial^\rho g_{\mu\nu} - \left( 
1- \frac{1}{\xi_0}\right) 
\partial_\mu \partial_\nu\right] \delta^4(x-y) + \Pi_{\mu\nu}(x,y)\,. 
\label{fullDinverse} 
\end{equation} 

\begin{figure}[t]
\centerline{%
\includegraphics[width=0.70\textwidth]{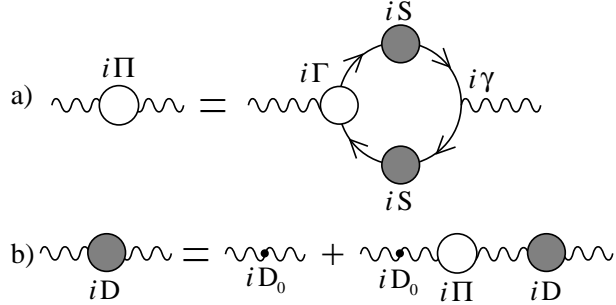}
}
\caption{\label{photonDSE} Dyson-Schwinger equation for the photon propagator.  Diagram a) expresses the photon self-energy in terms of the connected fermion two-point functions (dressed-propagators -- lines with filled circle) and the one-particle-irreducible electron-photon vertex (proper three-point function -- junction with the open circle).  (``$i \gamma$'' denotes the bare quark-photon vertex.)  Diagram b) expresses the iteration of the proper self-energy \emph{driving term} to obtain the connected photon propagator.  (Adapted from Ref.\,[\protect\refcite{cdragw}].)}
\end{figure}

The two-point Green function (propagator) for a noninteracting gauge boson field is plainly given by Eq.\,(\ref{fullDinverse}) in the absence of the photon's self-energy; viz., $\Pi\equiv 0$, and thus in momentum space, making use of the translational invariance, 
\begin{equation} 
D_0^{\mu\nu}(q)=\frac{-g^{\mu\nu}+(q^\mu q^\nu/[q^2+i\eta^+])}{q^2+i\eta^+} 
- \xi_0\frac{q^\mu q^\nu}{(q^2+i\eta^+)^2}.
\label{photonfree} 
\end{equation} 
It follows that Eq.~(\ref{fullDinverse}) can be written $iD =iD_0 +  iD_0 \,i\Pi\,i D$, and thus we have our first DSE, which is represented diagrammatically in Fig.\,\ref{photonDSE}.

In the presence of interactions; i.e., for $\Pi_{\mu\nu}\neq 0$ in Eq.~(\ref{fullDinverse}), one finds 
\begin{equation} 
D^{\mu\nu}(q)=\frac{-g^{\mu\nu}+(q^\mu q^\nu/[q^2+i\eta^+])}{q^2+i\eta^+} 
\frac{1}{1+\Pi(q^2)} - \xi_0\frac{q^\mu q^\nu}{(q^2+i\eta^+)^2}.
\label{photon_propagator} 
\end{equation} 
In obtaining this result we used the ``Ward-Takahashi identity'' for the photon vacuum polarisation; namely, 
\begin{eqnarray}
&&  q_\mu\, \Pi^{\mu\nu}(q) = 0  = \Pi^{\mu\nu}(q) \, q_\nu\\
\label{PiWTI} &\Rightarrow &\Pi^{\mu\nu}(q) = \left(-g^{\mu\nu}q^2+q^\mu q^\nu\right) \, \Pi(q^2)\,. 
\end{eqnarray} 
The quantity $\Pi(q^2)$ may be described as the polarisation scalar.  It is independent of the gauge parameter, $\xi_0$, in QED.  On the subject of the gauge parameter, $\xi_0=1$ is called the ``Feynman gauge.'' It is useful in perturbative calculations because it simplifies the $\Pi(q^2)=0$ gauge boson propagator enormously.  In nonperturbative applications, however, $\xi_0=0$, the ``Landau gauge,'' is most useful because it ensures the gauge boson propagator is itself transverse.  NB.\ Landau gauge is a fixed point of the renormalisation group.

Ward-Takahashi identities (WTIs) are relations satisfied by combinations of Green functions.  They are an essential consequence of a theory's local 
gauge invariance; i.e., local current conservation, and play a crucial role.  The WTIs can be proved directly from the generating functional and have physical implications.  For example, Eq.~(\ref{PiWTI}) ensures that the photon remains massless in the presence of charged fermions.\footnote{This is analysed, e.g., in Ref.\,[\protect\refcite{bpr91}], which also discusses aspects of gauge covariance, a modern understanding of which may be traced from Refs.\,[\protect\refcite{br93,dong}] to Refs.\,[\protect\refcite{raya,raya1,raya2}].}  A discussion of WTIs can be found in Ref.\,[\refcite{iz80}], pp.\,407-411, and Ref.\,[\refcite{bd2}], pp.\,299-303.  In their generalisation to non-Abelian gauge field theories, WTIs are described as Slavnov-Taylor identities and in this guise they are discussed in Ref.\,[\protect\refcite{pt84}], Chap.\,2.

As we have observed, in the absence of external sources, Eq.\,(\ref{PPTeq}) can easily be represented in momentum space because then the two- and three-point functions appearing therein must be translationally invariant and hence can simply be expressed in terms of Fourier amplitudes; viz.,
\begin{equation} 
\label{PiMom} i\Pi_{\mu\nu}(q)= - \,\sum_f (e_0^f)^2\int \frac{d^4 
\ell}{(2\pi)^4} {\rm 
tr}[(i\gamma_\mu)(iS^f(\ell))(i\Gamma_\nu^f(\ell,\ell+q))(iS^f(\ell+q))] \,. 
\end{equation} 
The ability to express 1PI functions via a single integral makes momentum space representations the most widely used in continuum calculations. 

With Eq.\,(\ref{PiMom}) we have reached our first goal: the DSE for the photon vacuum polarisation.  In QED this quantity is directly related to the running coupling
constant, which is a connection that makes its importance obvious.  In QCD there are other contributions but the polarisation scalar is nevertheless a key component in the evaluation of the strong running coupling.
 
\subsection{Fermion Gap Equation}
\label{fermiongapequation}
Equation~(\ref{FldEqn}) is a nonperturbative generalisation of Maxwell's equation in quantum field theory.  Its derivation provides the pattern by which one can obtain an equivalent generalisation of Dirac's equation:
\begin{eqnarray} 
\nonumber 0 & = &\int\![{ D}A_\mu] \, [{ D}\psi] [{ D}\bar\psi]
\,\frac{\delta}{\delta \bar\psi^f(x)}  \; 
\exp \bigg( i S[A_\mu,\psi,\bar\psi\,] \\
&& + i \int\,d^4 x\;  \left[ \bar\psi^g\Xi^g + \bar\Xi^g\psi^g + A_\mu J^\mu \right] \bigg) \nonumber\\ 
\nonumber & = &\int\![{ D}A_\mu] \, [{ D}\psi] [{
    D}\bar\psi]\,\left\{ \frac{\delta S}{\delta\bar\psi^f(x)} + \Xi^f(x)
    \right\}      \\ 
& & \times 
    \exp\left\{i\left( S[A_\mu,\psi,\bar\psi] + 
    \int\,d^4 x\;\left[ \bar\psi^g\Xi^g + \bar\Xi^g\psi^g + A_\mu J^\mu \right] 
                \right)\right\} \nonumber\\ 
& = & \left\{ \frac{\delta S}{\delta\bar\psi^f(x)} 
        \left[\frac{\delta}{i\delta J }, 
        \frac{\delta}{i\delta\bar\Xi}, -\frac{\delta}{i\delta\Xi}\right] 
        + \eta^f(x) \right\} W[J_\mu,\Xi,\bar\Xi\,]
\end{eqnarray}
and hence
\begin{equation}
0 = \left[ \Xi^f(x) + \left( i \not\!\partial - m_0^f + e_0^f \gamma^\mu 
\,\frac{\delta}{i \delta J^\mu(x)}\right) \frac{\delta}{i 
\delta\bar\Xi^f(x)}\right] W[J_\mu,\Xi,\bar\Xi],
\label{ELeqnpsi} 
\end{equation} 
which is the nonperturbative functional equivalent of Dirac's equation that we sought.

The next step is to apply a functional derivative with respect to $\Xi^f$: 
$\delta/\delta \Xi^f(y)$, which yields 
\begin{equation} 
0 = \delta^4(x-y) W[J_\mu] - \left( i \not\!\partial - m_0^f + e_0^f \gamma^\mu 
\,\frac{\delta}{i \delta J^\mu(x)}\right) W[J_\mu] \, S^f(x,y;[A_\mu])\,,
\end{equation} 
after setting $\Xi^f = 0 = \bar\Xi^f$, where $W[J_\mu]:= W[J_\mu,0,0]$ and
$S(x,y;[A_\mu])$ is defined in Eq.~(\ref{SfA}).  Now, using Eqs.~(\ref{WZdef}), (\ref{deltaGamma}), this can be rewritten
\begin{equation} 
\label{gapeqa} \delta^4(x-y) - \left( i \not\!\partial - m_0^f +e_0^f \not\!\! A (x;[J]) + 
e_0^f \gamma^\mu \,\frac{\delta}{i \delta J^\mu(x)}\right)\, S^f(x,y;[A_\mu])
= 0\,,
\end{equation} 
which defines the nonperturbative connected two-point fermion Green function.

As the electromagnetic four-potential vanishes in the absence of an external source; i.e., $A_\mu(x;[J=0]) = 0$, it remains only to exhibit the content of the remaining functional differentiation in Eq.~(\ref{gapeqa}), which can be accomplished using Eq.~(\ref{dAAinv}): 
\begin{equation} 
\nonumber \frac{\delta}{i \delta J_\mu(x)}\, S^f(x,y;[A_\mu]) =
\int\!d^4z \frac{\delta A^\nu(z)}{i \delta J_\mu(x)} 
\frac{\delta}{\delta A^\nu(z)} \left( \left. \frac{\delta^2 
\Gamma}{\delta\psi^f(x) \delta\bar\psi^f(y)} 
\right|_{\psi=\overline{\psi}=0}\right)^{-1}
\end{equation}
\begin{eqnarray} 
\nonumber & &  \!\!\!\! = - e_0^f \int d^4 z \,d^4 u \, d^4 w\; \frac{\delta A^\nu(z)}{i \delta J_\mu(x)}\, 
S^f(x,u)\,\Gamma_\nu(u,w;z)\,S^f(w,y) \\ 
 & &\!\!\!\! = - e_0^f \int d^4 z \,d^4 u \, d^4 w\; i 
D^{\mu\nu}(x-z)\,S^f(x,u)\,\Gamma_\nu(u,w;z)\,S^f(w,y).
\end{eqnarray} 
In the last line we set $J=0$ and used Eq.~(\ref{Dinverse}).  It is now evident that in the absence of external sources Eq.~(\ref{gapeqa}) is equivalent to
\begin{eqnarray} 
\nonumber \lefteqn{\delta^4(x-y) = \left( i \not\!\partial - m_0^f \right) \,S^f(x,y)}\\
&&\!\!\!\!\!\!\!\! -i\, (e_0^f)^2 \int d^4 z \,d^4 u \, d^4 w\,
D^{\mu\nu}(x,z)\,\gamma_\mu  \, S^f(x,u)\, \Gamma_\nu(u,w;z)\,
S^f(w,y). \label{gapeqb}  
\end{eqnarray}

In Eq.~(\ref{PPTeq}) the proper photon vacuum polarisation was introduced to re-express the DSE for the gauge boson propagator.  One can analogously define a proper fermion self-energy: 
\begin{equation} 
\label{Sigmaf} \Sigma^f(x,z) = i (e_0^f)^2 \int d^4 u \,d^4 w \; 
D^{\mu\nu}(x,z)\,\gamma_\mu \, S^f(x,u)\, \Gamma_\nu(u,w;z)\,, 
\end{equation} 
in which case Eq.~(\ref{gapeqb}) assumes the form 
\begin{equation} 
\label{gapeqc} \int d^4z\,\left[ \left( i \not\!\partial_{x} - m_0^f \right) 
\delta^4(x-z) - \Sigma^f(x,z) \right] \,S^f(z,y) = \delta^4(x-y)\,. 
\end{equation} 
Once more using the property that Green functions are translationally invariant in 
the absence of external sources, Eq.\,(\ref{Sigmaf}) becomes
\begin{equation} 
\label{GapEqQED} -i \Sigma^f(p) =  (e_0^f)^2 \, \int\! \frac{d^4 
\ell}{(2\pi)^4}\, 
[iD^{\mu\nu}(p-\ell)]\, [i\gamma_\mu]\, [i S^f(\ell)]\, [i\Gamma_\nu^f(\ell,p)] 
\,. 
\end{equation} 
It now follows from Eq.~(\ref{gapeqc}) that in momentum space the connected fermion two-point function is 
\begin{equation} 
\label{SfpD}
S^f(p) = \frac{1}{\not\!\!p  - m_0^f - \Sigma^f(p) + i \eta^+}\,. 
\end{equation} 

\begin{figure}[t]
\centerline{\includegraphics[width=0.70\textwidth]{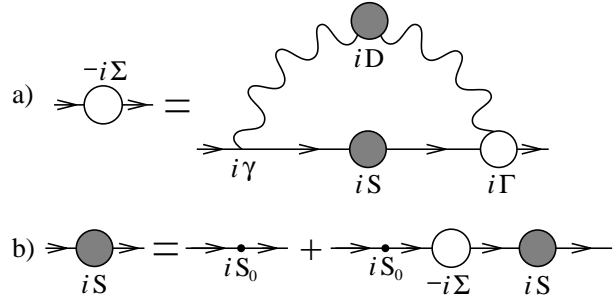}}
\caption{\label{figqdse}  Dyson-Schwinger equation for the fermion propagator.  Diagram a) expresses a charged fermion's self-energy in terms of that fermion's connected  two-point function (Eq.\,(\protect\ref{SfpD}), dressed-propagator -- straight line with filled circle), the photon's connected two-point function (Eq.\,(\protect\ref{photon_propagator}), dressed-propagator -- wavy line with filled circle) and the one-particle-irreducible electron-photon vertex (proper three-point function -- junction with the open circle).  (``$i \gamma$'' denotes the bare quark-photon vertex.)  Diagram b) expresses the iteration of the proper self-energy \emph{driving term} to obtain the connected fermion propagator.  (Adapted from Ref.\,[\protect\refcite{cdragw}].)}
\end{figure}

Equation~(\ref{GapEqQED}), depicted in Fig.\,\ref{figqdse}, is the \textbf{exact} \textit{Gap Equation}.  It describes the manner in which the propagation characteristics of a charged fermion moving through the ground state of QED (the QED vacuum) are altered by the repeated emission and reabsorption of virtual photons.  It is evident that Eqs.\,(\ref{PiMom}) and (\ref{GapEqQED}) are coupled via Eqs.\,(\ref{photon_propagator}) and (\ref{SfpD}).  The gap equation can also
describe real processes, e.g., Bremsstrahlung.  Moreover, as we shall see, a solution of the analogous equation in QCD provides information about dynamical chiral
symmetry breaking and also quark confinement.  

\begin{figure}[t]
\centerline{\includegraphics[width=0.33\textwidth,angle=-90]{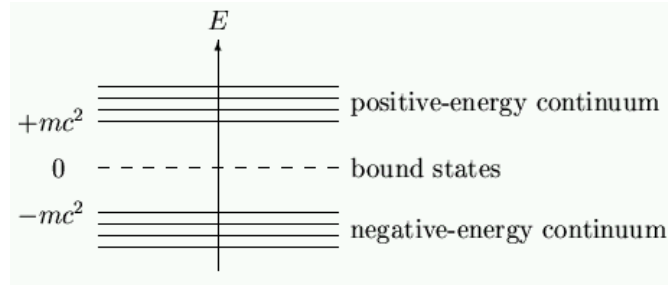}}
\caption{\label{diracgap} The Dirac spectrum of a free particle in a box.  (In this figure, $c$ is retained.)  In Dirac's hole picture, the negative energy states are fully occupied by negative energy electrons.  This is the \emph{Dirac sea} ground state.  In the usual scattering theory, bound states appear with an energy less-than the sum of the masses of the constituents and are thus indicated here as appearing with negative energy.}
\end{figure}

It is natural to ask why Eq.\,(\ref{GapEqQED}) is called the \emph{gap equation}.  The name may be traced to the particle-hole interpretation of the Dirac equation's solutions.  The Dirac equation for a free fermion in a box admits infinitely many solutions with positive energy and an equal number with negative energy.  As illustrated in Fig.\,\ref{diracgap}, the negative energy solutions are separated from the positive energy solutions by a gap whose width is $ 2 m$, where $m$ is the fermion's mass.  This is the \emph{mass gap}.  In order to avoid the catastrophe of positive energy fermions cascading into the negative energy levels, Dirac formulated the \emph{hole theory}, which postulates that all the negative energy levels are already occupied by fermions.  Then, in accordance with the Pauli exclusion principle, the states are not accessible to positive energy fermions.  The negative energy fermions form the Dirac sea.  A fermion at the surface of the sea can be excited to a positive energy level if it receives an energy transfer $> 2 m$; i.e., an energy deposit that is sufficient to bridge the mass gap.  Should that occur, the system has one energy $+|E|$ fermion with charge $-|e|$ and a \emph{hole} in the Dirac sea, which is registered as the absence of a fermion with charge $-|e|$ and energy $-E$.  This absence may equally be interpreted by an observer as the presence of a object with charge $+|e|$ and energy $+E$; namely, an antifermion.  It is plain that the mass gap vanishes for a massless fermion.  However, as we shall see, there are interactions for which a self-consistent solution of Eq.\,(\ref{GapEqQED}) generates an effective mass for the fermion even though the perturbative (or Dirac equation) mass vanishes.  In this instance the fermion DSE describes the dynamical generation of a mass gap.

Dynamical mass generation, which is also called dynamical chiral symmetry breaking (DCSB), is a keystone of hadron physics.  However, in order to understand DCSB one must first come to terms with explicit chiral symmetry breaking.  Consider then the DSE for the quark self-energy in QCD:
\begin{equation} 
\label{GapEq} -i \,\Sigma(p) =  g_0^2 \, \int\! \frac{d^4 \ell}{(2\pi)^4}\, 
iD^{\mu\nu}(p-\ell)\, \frac{i}{2} \lambda^a \gamma_\mu \, iS(\ell)\, 
 i\Gamma_\nu^a(\ell,p) \,, 
\end{equation} 
where the flavour label is suppressed.  The form of this equation is the same as the gap equation in QED, Eq.~(\ref{GapEqQED}), except for the following:
\begin{itemize}
\item colour (Gell-Mann) matrices: $\{ \lambda^a; a=1,\ldots,8\}$, appear 
at the fermion-gauge-boson vertex;
\item $D^{\mu\nu}(\ell)$ represents the colour-diagonal connected gluon two-point function;
\item and $\Gamma_\nu^a(\ell,\ell^\prime)$ is the proper quark-gluon vertex.
\end{itemize}

The one-loop (leading perturbative order) contribution to a quark's self-energy is obtained by evaluating the rhs of Eq.\,(\ref{GapEq}) with the free/noninteracting quark and gluon propagators, and the quark-gluon vertex: 
\begin{equation} 
\Gamma_\nu^{a\,(0)}(\ell,\ell^\prime) = \frac{1}{2} \lambda^a \gamma_\nu\,,
\end{equation} 
which yields
\begin{eqnarray} 
\nonumber -i \,\Sigma^{(2)}(p) & = & - g_0^2 \, \int\! \frac{d^4 k}{(2\pi)^4}\, 
\left(-g^{\mu\nu} + (1-\xi_0) \frac{k^\mu k^\nu}{k^2+i\eta^+}\right) 
\, \frac{1}{k^2 + i\eta^+}\, \\ 
& & \times  \,  \frac{i}{2} \lambda^a \gamma_\mu \, \frac{1}{\not\! k + \not\!\! p - 
m_0 + i\eta^+}\, \frac{i}{2} \lambda^a \gamma_\mu \,. \label{Sigma2a} 
\end{eqnarray} 

Equation (\ref{Sigma2a}) can be re-expressed as
\begin{eqnarray} 
\nonumber \lefteqn{-i \,\Sigma^{(2)}(p)
=  - g_0^2\,C_2(R) \, \int\! \frac{d^4 k}{(2\pi)^4}\, 
\frac{1}{(k+p)^2 - m_0^2 + i\eta^+}\,\frac{1}{k^2+i\eta^+}}\\ 
\nonumber & & \!\! \times \left\{
\gamma^\mu \,(\not\! k + \not\!\! p + m_0) \,\gamma_\mu - (1 - \xi_0)\, 
(\not\! k - \not\!\! p + m_0) - 2\, (1-\xi_0)\, \frac{ k\cdot p \not\! k }{k^2 + 
i\eta^+} 
\right\},\\ 
& & 
\end{eqnarray} 
where we have used 
\begin{equation}
\frac{1}{2} \lambda^a \, \frac{1}{2} \lambda^a  = C_2(R) \,
\mbox{\boldmath $I$}_c\,;\; C_2(R) = \frac{N_c^2-1}{2N_c}, 
\end{equation}
with $N_c$ the number of colours ($N_c=3$ in QCD), and $\mbox{\boldmath
$I$}_c$ is the identity matrix in colour space.  Note now that 
\begin{equation}
2\,k\cdot p = [ (k+p)^2 - m_0^2 ] - [ k^2 ] - [ p^2 - m_0^2] 
\end{equation}
and hence 
\begin{eqnarray} 
\nonumber -i \,\Sigma^{(2)}(p) & = & - g_0^2 \, C_2(R)\,\int\! \frac{d^4 
k}{(2\pi)^4}\, 
\frac{1}{(k+p)^2 - m_0^2 + i\eta^+}\,\frac{1}{k^2+i\eta^+}\,\\ 
\nonumber && \left\{\rule{0ex}{3.5ex} 
\gamma^\mu \,(\not\! k + \not\!\! p + m_0) \,\gamma_\mu + (1 - \xi_0)\, 
(\not\!\! p  - m_0)  \right. \\ 
\nonumber && + \,  (1-\xi_0)\, (p^2 - m_0^2)\,
\frac{\not\! k }{k^2 + i\eta^+} \\
&& \left. 
- \, (1-\xi_0)\, [(k+p)^2 - m_0^2] \, \frac{\not\! k}{k^2 + i\eta^+}  
\rule{0ex}{3.5ex} \right\}.
\label{Sigma2b} 
\end{eqnarray} 

Let's focus now on the last term in Eq.\,(\ref{Sigma2b}): 
\begin{eqnarray} 
\nonumber && \int\! \frac{d^4 k}{(2\pi)^4}\, \frac{1}{(k+p)^2 - m_0^2 +  
i\eta^+}\,\frac{1}{k^2+i\eta^+}\, [(k+p)^2 - m_0^2]\, \frac{\not\! k}{k^2 + 
i\eta^+} \\ 
& = & \int\! \frac{d^4 k}{(2\pi)^4}\, \frac{1}{k^2+i\eta^+}\, 
\frac{\not\! k }{k^2 + i\eta^+} = 0 \label{oddbutdivergent} 
\end{eqnarray} 
because the integrand is odd under $k \to -k$, and so this term in Eq.~(\ref{Sigma2b}) vanishes.  

The second term in Eq.\,(\ref{Sigma2b}):
\begin{eqnarray} 
\nonumber & & (1 - \xi_0)\, \mbox{$(\not\!\! p - m_0)$} 
\int\! \frac{d^4 k}{(2\pi)^4}\, \frac{1}{(k+p)^2 - m_0^2 + 
i\eta^+}\,\frac{1}{k^2+i\eta^+}\,, \label{secondterm} 
\end{eqnarray} 
is more interesting.  To understand why, we will consider the behaviour of the integrand at large $k^2$: 
\begin{equation} 
\frac{1}{(k+p)^2 - m_0^2 + i\eta^+}\,\frac{1}{k^2+i\eta^+} \stackrel{k^2\to \pm 
\infty}{\sim} \frac{1}{(k^2-m_0^2+ i\eta^+)\, (k^2+ i \eta^+)}\,. 
\end{equation} 

The integrand has poles in the second and fourth quadrants of the complex-$k_0$-plane but vanishes on any circle of radius $R\to \infty$ in this plane.  That means one may rotate the contour anticlockwise to find
\begin{eqnarray} 
\nonumber \lefteqn{\int_{0}^{\infty}\! dk^0 \, \frac{1}{(k^2-m_0^2+ 
i\eta^+)\,(k^2+ i \eta^+)}}\\ 
\nonumber & = & \int_0^{i\infty} \!dk^0 \, \frac{1}{([k^0]^2 - \vec{k}^2 - 
m_0^2 + i \eta^+) 
 ([k^0]^2 - \vec{k}^2+ i \eta^+)} \\ 
& \stackrel{k^0 \to i k_4}{=} & i\int_0^{\infty} \!d k_4 \, \frac{1}{(-k_4^2 - 
\vec{k}^2 -m_0^2)\,(-k_4^2 - \vec{k}^2 )} \,.  \label{WickRotated}
\end{eqnarray} 
Performing a similar analysis of the $\int_{-\infty}^0$ part, one obtains the 
complete result: 
\begin{eqnarray} 
\nonumber \lefteqn{\int\!\frac{d^4k}{(2\pi)^4} \, \frac{1}{(k^2-m_0^2+
i\eta^+)\,(k^2+ i  \eta^+)}}\\
& = & 
i\,  \int\!\frac{d^3 k}{(2\pi)^3}\,\int_{-\infty}^\infty \!\frac{dk_4}{2\pi} \, 
\frac{1}{(-\vec{k}^2 - k_4^2-m_0^2)\, (-\vec{k}^2 - k_4^2)}\,. \label{WickRotated2}
\end{eqnarray} 
These two steps constitute what is called a \textit{Wick rotation}.  

The integral on the rhs in Eq.\,(\ref{WickRotated2}) is defined in a four-dimensional Euclidean space; i.e., $k^2:= k_1^2 + k_2^2+k_3^2 + k_4^2 \geq 0$, with $k^2$ nonnegative.  A general vector in this space can be written in the form:
\begin{equation} 
(k) = |k|\,(\cos \phi\,\sin\theta\,\sin\beta,\sin \phi\,\sin\theta\,\sin\beta, 
\cos\theta\,\sin\beta,\cos\beta)\,;
\end{equation} 
i.e., using hyperspherical coordinates, and clearly $k^2 = |k|^2$.  Using these coordinates the four-vector measure factor is
\begin{eqnarray} 
\nonumber \lefteqn{\int d_E^4k\, f(k_1,\ldots,k_4) }\\
\nonumber &=& \frac{1}{2}\,\int_0^\infty \! dk^2 \, k^2\,  
\int_0^\pi \! d\beta\,sin^2\beta \,\int_0^\pi \! d\theta \sin\theta \, 
\int_0^{2\pi}\! d\phi\, f(k,\beta,\theta,\phi)\,. \\
\end{eqnarray}
Integration is straightforward because of the non-negative metric.

We now return to Eq.~(\ref{secondterm}) wherein, making use of the material just introduced, the large $k^2$ behaviour of the integral can be determined via
\begin{eqnarray} 
\nonumber && \int\! \frac{d^4 k}{(2\pi)^4}\, 
 \frac{1}{(k+p)^2 - m_0^2 + i\eta^+}\,\frac{1}{k^2+i\eta^+} \approx \frac{i}{16\pi^2}\int_0^\infty\! dk^2 \frac{1}{(k^2+m_0^2)}\\ 
\nonumber & = & \frac{i}{16\pi^2}
\lim_{\Lambda\to\infty}\,\int_0^{\Lambda^2}\! dx \, \frac{1}{x+m_0^2} \\  
& = &  \frac{i}{16\pi^2}\lim_{\Lambda\to\infty} \, \ln\left(1 + 
\Lambda^2/m_0^2\right) \to \infty\,. \label{uvdivergence} 
\end{eqnarray} 
This is why it is interesting; viz., after all this work, the result is meaningless: the one-loop contribution to the quark's self-energy is divergent!

Such ``ultraviolet'' divergences, and others which are more complicated, arise in perturbation theory whenever loops appear.\footnote{The others include ``infrared'' divergences associated with the gluons' masslessness; e.g., consider what would happen in Eq.\ (\ref{uvdivergence}) with $m_0\to 0$.}  In a \textit{renormalisable} quantum field theory there exists a well-defined set of rules that can be used to render perturbation theory sensible.  First, however, one must \textit{regularise} the theory; i.e., introduce a cutoff or use some other means to make finite every integral that appears.  Then each step in the calculation of an observable is rigorously understood.  Renormalisation follows; i.e, the absorption of divergences, and the
redefinition of couplings and masses, so that finally one arrives at ${S}$-matrix amplitudes which are finite and physically meaningful.  The {regularisation} procedure must preserve the Ward-Takahashi identities (the Slavnov-Taylor identities in QCD) because they are crucial in proving that a theory can sensibly be renormalised.  A theory is called renormalisable if, and only if, the number of different types of divergent integral is finite.  Then only a finite number of masses and couplings need to be renormalised; i.e., \textit{a priori} the theory has only a finite number of
undetermined parameters that must subsequently be fixed through comparison with experiments.

Herein we will not explain the procedure.  For those interested, Ref.\,[\refcite{pt84}] is a pedagogical introduction and illustration: all the steps for many calculations are presented and explained.  Nevertheless, we present the one-loop result in the momentum subtraction scheme for the renormalised self-energy:
\begin{equation}
\label{SigmaMom}
\Sigma_R^{(2)}(\not\!\! p) = \Sigma_{VR}^{(2)}(p^2) \,\not\!\! p + \Sigma_{SR}^{(2)}(p^2) \, \mbox{\boldmath $I$}_{\rm D}\,;
\end{equation}
\begin{eqnarray} 
\nonumber \lefteqn{\Sigma_{VR}^{(2)}(p^2;\zeta^2) = \frac{\alpha(\zeta)}{\pi} \, 
\xi(\zeta) \, \frac{1}{4}\, C_2(R)\, \left\{ 
-m(\zeta)^2\left(\frac{1}{p^2}+\frac{1}{\zeta^2}\right) \right. } \\ 
\nonumber & & + \left. \left( 1 - \frac{m(\zeta)^4}{p^4}\right) \ln \left( 1 - 
\frac{p^2}{m(\zeta)^2}\right) - \left( 1 - \frac{m(\zeta)^4}{\zeta^4}\right) \ln 
\left( 1 + \frac{ \zeta^2}{m(\zeta)^2}\right) \right\}\,, \\ \\
\nonumber\lefteqn{ \Sigma_{SR}^{(2)}(p^2;\zeta^2) = m(\zeta) \, 
\frac{\alpha(\zeta)}{\pi}\, 
\frac{1}{4}\, C_2(R)\, \left\{ \rule{0em}{3.0ex} -[3+\xi(\zeta)] \right.} \\
\nonumber && \times  \left[ \left( 1 - \frac{m(\zeta)^2}{p^2}\right) 
\ln\left( 1 - \frac{p^2}{m(\zeta)^2}\right) \left. - \left( 1 + 
\frac{m(\zeta)^2}{\zeta^2}\right) 
\ln\left( 1 + \frac{\zeta^2}{m(\zeta)^2}\right) \right]\right\},\\ \label{Bp2QCD} 
\end{eqnarray} 
where the renormalised quantities depend on the point at which the renormalisation was conducted; e.g., $\alpha(\zeta)$ is the running coupling and $m(\zeta)$ is the running quark mass, and both are evaluated at the renormalisation scale $\zeta$.  NB.\ $\zeta^2$ is a spacelike point.

QCD is asymptotically free.  Hence, at some large spacelike $p^2=-\zeta^2$ the quark propagator is exactly the free propagator \textit{except} that the bare mass is replaced by the renormalised mass.  At one-loop order, the vector part of the dressed self-energy is proportional to the running gauge parameter.  In Landau gauge, that parameter is zero.  Hence, in this gauge the vector part of the renormalised dressed-quark self-energy vanishes at one-loop order in perturbation theory.  The same is true for a charged fermion in QED.   

The scalar part of the dressed-quark self-energy is proportional to the renormalised current-quark mass.  This is true at one-loop order, and indeed at every finite order in perturbation theory.  Hence, if the current-quark mass vanishes, then $\Sigma_{SR}\equiv 0$ in perturbation theory.  That means if one starts with a chirally symmetric theory, then in perturbation theory one also ends up with a chirally symmetric theory: the fermion DSE cannot generate a gap if there is no bare-mass seed in the first place.  Thus DCSB is \label{impossible} impossible in perturbation theory.  

\section{Hadron Physics}
\label{hadronphysics}
\setcounter{equation}{0}
Hadron physics is a key part of the international effort in basic science.  The Thomas Jefferson National Accelerator Facility (JLab) and the Relativistic Heavy Ion Collider (RHIC) are essential facilities for pursuing long term goals in a world-wide effort. Progress in this field is gauged via the successful completion of precision measurements of fundamental properties of the proton, neutron and simple nuclei, for comparison with theoretical calculations to provide a quantitative understanding of their quark substructure.

The proton and neutron (collectively termed \emph{nucleons}) are fermions.  They are characterised by two static properties: an electric charge and a magnetic moment.  In Dirac's theory of pointlike relativistic fermions the magnetic moment is
\begin{equation}
\mu_D = \frac{e}{2 M}\,.
\end{equation}
The proton's magnetic moment was discovered in 1933 by Otto Stern, who was awarded the Nobel Prize in 1943 for this work.\cite{nobelStern}  He found, however, that 
\begin{equation}
\mu_p = (1 + \mbox{\underline{$1.79$}}) \, \mu_D\,.
\end{equation}
This was the first indication that the proton is not a point particle.  Of course, we now explain the proton as a bound state of quarks and gluons, a composition which is seeded and determined by the properties of three valence quarks.  These quarks and gluons are the elementary quanta of QCD.

Important experiments at JLab measure hadron and nuclear form factors via electron scattering.  The electron current is known from QED:
\begin{eqnarray}
\label{jmue}
j_\mu(P^\prime,P)  &=& ie\,\bar u_e(P^\prime)\, { \Lambda_\mu(Q,P)} \,u_e(P)\,, \; Q=P^\prime-P  \\
& = &  i e \,\bar u_e(P^\prime)\,{ \gamma_\mu (-1)} \, u_e(P)\,, \label{eborn}
\end{eqnarray}
where $u_e$, $\bar u_e$ are Dirac spinors for a real (on-mass-shell) electron.  Equation (\ref{eborn}) is the Born approximation result, in which the dressed-electron-photon vertex is just $(- \gamma_\mu)$ and where the negative sign merely indicates the sign of the electron charge.  For the nucleons, Eq.\,(\ref{jmue}) is written
\begin{eqnarray}
\lefteqn{ J_\mu^N(P^\prime,P)  = ie\,\bar u_N(P^\prime)\, { \Lambda_\mu(Q,P)} \,u_N(P) } \\
& = &  i e \,\bar u_N(P^\prime)\,\left( \gamma_\mu F_1^N(Q^2) +
\frac{1}{2M}\, \sigma_{\mu\nu}\,Q_\nu\,F_2^N(Q^2)\right) u_N(P) \label{jmuN}
\end{eqnarray}
where $u_N$, $\bar u_N$ are on-shell nucleon spinors, and $F_{1}^N(Q^2)$ is the Dirac form factor and $F_2^N(Q^2)$ is the Pauli form factor.  The so-called Sachs form factors are defined via
\begin{equation}
\label{defGEGM}
G_E(Q^2)  =  F_1(Q^2) - \frac{Q^2}{4 M^2} F_2(Q^2)\,,\; 
G_M(Q^2)  =  F_1(Q^2) + F_2(Q^2)\,.
\end{equation}
In the Breit frame in the nonrelativistic limit, the three-dimensional Fourier transform of $G_E(Q^2)$ provides the electric-charge-density distribution within nucleon, while that of $G_M(Q^2)$ gives the magnetic-current-density distribution.  This explains their names: $G_E^N(Q^2)$ is the nucleon's electric form factor and $G_M^N(Q^2)$ is the magnetic form factor.  It is apparent via a comparison between Eqs.\,(\ref{eborn}) and (\ref{jmuN}) that $F_2\equiv 0$ for a point particle, in which case $G_E = G_M$.  This means, of course, that if the neutron is a point particle then it has neither an electric nor a magnetic form factor.

We know of six quarks in the Standard Model of particle physics: $u$ (up), $d$ (down), $s$ (strange), $c$ (charm), $b$ (bottom) and $t$ (top).  The first three are most important in hadron physics.  A central goal of nuclear physics is to understand the structure and properties of protons and neutrons, and ultimately atomic nuclei, in terms of the quarks and gluons of QCD.  So, why don't we just go ahead and do it?  One of the answers is \emph{confinement}: no quark or gluon has ever been seen in isolation.  Another is dynamical chiral symmetry breaking; e.g., the masses of the $u$, $d$ quarks in perturbative QCD provide no explanation for $\simeq 98$\% of the proton's mass.  One therefore has to ask, with quarks and gluons are we dealing with the right degrees of freedom?

\begin{figure}[t]
\centerline{\includegraphics[width=0.7\textwidth,angle=270]{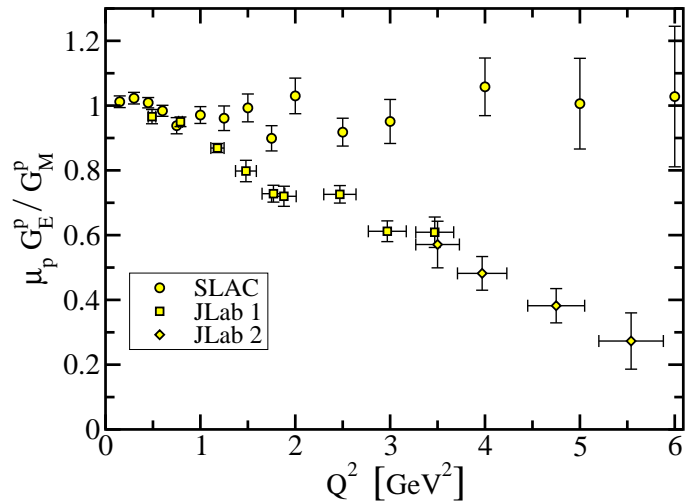}}
\caption{\label{gepgmpdata} Proton form factor ratio: $\mu_p\, G_E^p(Q^2)/G_M^p(Q^2)$. The data are obtained via Rosenbluth separation, \textit{circles} - Ref.\,[\protect\refcite{walker}]; and polarisation transfer, \textit{squares} - Ref.\,[\protect\refcite{jones}]; \textit{diamonds} - Ref.\,[\protect\refcite{gayou}].}
\end{figure}

The search for patterns in the hadron spectrum adds emphasis to this question.  Let's consider the proton, for example.  It has a mass of approximately $1\,$GeV.  Suppose it to be composed of three (two $u$ and one $d$) \emph{constituent}-quarks, as in the \emph{eightfold way} classification of hadrons into groups on the basis of their symmetry properties \cite{gellmannNobel}.  A first guess would place the mass of these constituents at $\sim 350\,$MeV.  In the same approach, the pseudoscalar $\pi$-meson is composed of a constituent-quark and a constituent-antiquark.  It should therefore have a mass of $\sim 700\,$MeV.  However, its true mass is $\sim 140\,$MeV!  On the other hand, the mass of the vector $\rho$-meson is correctly estimated in this way: $m_\rho = 770\,$MeV.  Such mismatches are repeated in the spectrum.  \label{eightfold} 

Furthermore, modern high-luminosity experimental facilities, such as JLab, that employ large momentum transfer reactions are providing remarkable and intriguing new information on nucleon structure.\cite{gao,leeburkert}  For an example one need only look so far as the discrepancy between the ratio of electromagnetic proton form factors, $\mu_p G_E^p(Q^2)/G_M^p(Q^2)$, extracted via the Rosenbluth separation method\,\cite{walker} and that inferred from polarisation transfer experiments.\cite{jones,roygayou,gayou,arrington,qattan}  This discrepancy is marked for $Q^2\gsim 2\,$GeV$^2$ and grows with increasing $Q^2$.  Before the JLab data were analysed it was assumed that $\mu_p \,G_E^p(Q^2)/G_M^p(Q^2)=1$ based on the seemingly sensible argument that the distribution of quark charge and the distribution of quark current should be the same.  However, if the available JLab results turn out to provide the true measure of this ratio, then we must dramatically rethink that picture.

An immediate question to ask is where does modern hadron physics theory stand on this issue?  A theoretical understanding might begin with a calculation of the proton's Poincar\'e covariant wave function.  (Remember, the discrepancy is marked at larger momentum transfer; viz., on the relativistic domain.)  One might think that is not a problem.  After all, the wave functions of few electron atoms can be calculated very reliably.  However, there are some key differences.  One is that the ``potential'' between light-quarks is \textbf{completely} \textbf{unknown} throughout $\simeq 98$\% of the proton's volume.  Moreover, as we shall see, a reliable description of the proton's wave function will require an accurate treatment of virtual particle effects, which are a quintessential part of relativistic quantum field theory.  In fact, a computation of the proton's wave function requires the \emph{ab initio} solution of a fully-fledged relativistic quantum field theory, but that is yet far beyond the capacity of modern physics and mathematics.

\subsection{Aspects of QCD}
\label{formulationofqcd}
The theory we must explore is quantum chromodynamics (QCD).  The action is expressed through a local Lagrangian density; viz.,
\begin{eqnarray}
\nonumber S[A_\mu^a, q, \bar{q}\,] & = & \int d^4x\,\bigg\{ \bar q(x)\left[ \gamma_\mu D_\mu + M \right]q(x)\\
&&  + \frac{1}{4} F_{\mu\nu}^a(x) F_{\mu\nu}^a(x) + \frac{1}{2\xi} \, \partial_\mu A_\mu^a(x) \, \partial_\nu A_\nu^a(x) \bigg\},
\label{Sqcd}
\end{eqnarray}
in which the first term of the second line is the chromomagnetic field strength tensor
\begin{equation}
F_{\mu\nu}^a(x)  = \partial_\mu A_\nu^a(x) -\partial_\nu A_\mu^a(x) + g f^{abc} A_\mu^b(x) A_\nu^c(x),
\end{equation}
where $\{f^{abc}:a,b,c=1,\ldots,8\}$ are the structure constants of $SU(3)$, and the second term is a covariant gauge fixing term -- almost identical to that in QED, Eq.\,(\ref{Sqed}) -- with $\xi$ the gauge fixing parameter.  The first term in Eq.\,(\ref{Sqcd}) involves the covariant derivative:
\begin{equation}
\displaystyle D_\mu = \partial_\mu - i g \frac{\lambda^a}{2} A^a_\mu(x)\,,
\end{equation}
where $\{ \mbox{\small $\frac{1}{2}$}\lambda^a;a=1,\ldots,8\}$ are the generators of $SU(3)$ for the fundamental representation; and, in addition, the current-quark mass matrix
\begin{equation}
M= \left(\begin{array}{cccc}
m_u & 0 & 0 & \ldots \\
0 & m_d & 0 & \ldots \\
0 & 0 & m_s & \ldots \\
\vdots & \vdots & \vdots & 
\end{array}\right)\,,
\end{equation}
wherein we have only indicated the light-quark elements.  

Understanding observables that have been and will be measured using the Continuous Electron Beam Accelerator Facility (CEBAF) at JLab means knowing all that the quantum field theory based on Eq.\,(\ref{Sqcd}) predicts.  As we have emphasised, perturbation theory is inadequate to that task because confinement, DCSB, and the formation and structure of bound states are all essentially nonperturbative phenomena.  We will subsequently provide an overview of a nonperturbative approach to exploring strong QCD in the continuum.\footnote{Almost all nonperturbative studies in relativistic quantum field theory employ a Euclidean Metric.  (Remember the Wick Rotation?!) We have used a Euclidean metric in writing Eq.\,(\protect\ref{Sqcd}).  \ref{Appendix1} provides some background and describes our Euclidean conventions.}

QCD is a local, renormalisable, non-Abelian gauge theory, in which each flavour of quark comes in three colours and there are eight gauge bosons, called gluons.  It is a peculiar feature of non-Abelian gauge theories that the gauge bosons each carry the gauge charge, \emph{colour} in this case, and hence self-interact.  This is the key difference between QED, an Abelian gauge theory wherein the photons are neutral, and QCD.  The gluon self-interaction is primarily responsible for the marked difference between the running coupling in QCD and that in QED; namely, that the running coupling in QCD decreases relatively rapidly with increasing momentum transfer -- the theory is asymptotically free, whereas the QED coupling increases very slowly with growing momentum transfer.  (See, e.g., Sec.\,1.2 in Ref.\,[\refcite{cdrANU}].)

\begin{table}[t]
\tbl{Comparison between various hadron mass ratios, involving the $\pi$, $\rho$, $\sigma$ and $a_1$ mesons, the nucleon -- $N$, and the first radial excitations of: the $\pi$ $=\,\pi_1$; and the $\rho$ $=\,\rho_1$.  The individual masses can be found in Ref.\,[\protect\refcite{pdg}].}
{\normalsize
\begin{tabular*}{\hsize} {
l@{\extracolsep{0ptplus1fil}}%
l@{\extracolsep{0ptplus1fil}}%
l@{\extracolsep{0ptplus1fil}}%
}
{ $\bullet$} $\displaystyle \frac{m_\rho^2}{m_\pi^2} = 30$ & 
{ $\bullet$} $\displaystyle \frac{m_{a_1}^2}{m_{\sigma}^2} = 2.1$ & 
{ ?} Hyper{f}ine Splitting\\[2ex]
{ $\bullet$} $\displaystyle \frac{m_{\pi_1}^2}{m_\pi^2} = 86$ &
{ $\bullet$} $\displaystyle \frac{m_{\rho_1}^2}{m_{\rho}^2} = 3.5$ &
{ ?} Excitation Energy \\[2ex]
{ $\bullet$} $\displaystyle \frac{m_N}{m_\pi} \approx 7$ &
{ $\bullet$} $\displaystyle \frac{m_N}{m_\rho} = \frac{5}{4} \approx
\frac{3}{2} $ & { ?} Quark Counting\\
\end{tabular*}\label{tableSpectrum}}
\end{table}

In Table~\ref{tableSpectrum} we return to the hadron spectrum and focus on some of its features.  In a constituent-quark model the $J^{PC}=1^{--}$ $\rho$-meson is obtained from the $0^{-+}$ $\pi$-meson by a spin flip, yielding a vector meson state in which the spins of the constituent-quark and -antiquark are aligned.  The same procedure would yield the $1^{++}$ $a_1$-meson from the $0^{++}$ $\sigma$-meson.  Thus the difference between the masses of $\rho$ and $\pi$, and the $a_1$ and $\sigma$ would appear to owe to an hyperfine interaction.  As the Table asks in row one, why is this interaction so much greater in the $\pi$ channel?  Another questions is raised in row two: why is the radial excitation energy in the pseudoscalar channel so much greater than that in the vector channel?  And row three asks the question we posed on page~\pageref{eightfold}: why doesn't constituent-quark counting work for the $\pi$?  Additional questions can be posed.  The range of an interaction is inversely proportional to the mass of the boson that mediates the force.  The nucleon-nucleon interaction has a long-range component generated by the $\pi$.  The fact that the pion is so much lighter than all other hadrons composed of $u$- and $d$-quarks is crucially important in nuclear physics.   If this were not the case; viz., were the pion roughly as massive as all like-constituted hadrons, then the domain of stable nuclei would be much reduced.  In such a universe the Coulomb force would prevent the formation of elements like Fe, and planets such as ours and we, ourselves, would not exist.

\subsection{Emergent Phenomena}
\label{emergentQCD}
A true understanding of the visible universe thus requires that we learn just what it is about QCD which enables the formation of an unnaturally light pseudoscalar meson from two rather massive constituents.  The correct understanding of hadron observables must explain why the pion is light but the $\rho$-meson and the nucleon are heavy.  The keys to this puzzle are QCD's \emph{emergent phenomena}: \textbf{confinement} and \textbf{dynamical chiral symmetry breaking}.  Confinement is the feature that no matter how hard one strikes a hadron, it never breaks apart into quarks and/or gluons that ultimately reach a detector alone.  DCSB is signalled by the very unnatural pattern of bound state masses, something that we have partly illustrated with Table~\ref{tableSpectrum} and the associated discussion.  Neither of these phenomena is apparent in QCD's action and yet they are the dominant determining characteristics of real-world QCD.  Attaining an understanding of these phenomena is one of the greatest intellectual challenges in physics.

In order to come to grips with DCSB it is first necessary to know the meaning of chiral symmetry.  It is a fact that, at the Lagrangian level, local gauge theories with massless fermions posses chiral symmetry.   Consider then helicity, which may be viewed as the projection of an object's spin, $\vec{j}$, onto its direction of motion, $\vec p\,$; viz., $\lambda \propto \vec{j}\cdot \vec p$.  For massless particles, helicity is a Lorentz invariant \emph{spin observable}.  Plainly, it is either parallel or anti-parallel to the direction of motion.  

In the Dirac basis, $\gamma_5$ is the chirality operator and we may represent a positive helicity (right-handed) fermion via \begin{equation}
q_+(x) = \frac{1}{2} \left(\mbox{\boldmath $I$}_{\rm D} + \gamma_5 \right) q(x) =: P_+ \,q(x)
\end{equation}
and a left-handed fermion through
\begin{equation}
q_-(x) = \frac{1}{2} \left(\mbox{\boldmath $I$}_{\rm D} - \gamma_5 \right) q(x) =: P_-\, q(x)\,.
\end{equation}
A global chiral transformation is enacted by\footnote{For this illustrative purpose it is not necessary to consider complications that arise in connection with $U(1)$ chiral anomalies, which appear via quantisation.}
\begin{equation}
\label{globalchiral} 
q(x) \to q(x)^\prime = {\rm e}^{i \gamma_5 \theta} q(x)\,,\;
\bar q(x) \to \bar q(x)^\prime = \bar q(x)\, {\rm e}^{i \gamma_5 \theta} ,
\end{equation}
and with the choice $\theta=\pi/2$ it is evident that this transformation maps $q_+ \to q_+$ and $q_- \to - q_-$.  Hence, a theory that is invariant under chiral transformations can only contain interactions that are insensitive to a particle's helicity.  

Consider now a composite local pseudoscalar: $\bar q(x) i \gamma_5 q(x)$.  According to Eq.\,(\ref{globalchiral}), a chiral rotation through an angle $\theta = \pi/4$ effects the transformation
\begin{equation}
\bar q(x) i \gamma_5 q(x) \to - \bar q(x) \mbox{\boldmath $I$}_{\rm D}\, q(x)\,;
\end{equation}
i.e., it turns a pseudoscalar into a scalar.  Thus the spectrum of a theory invariant under chiral transformations should exhibit degenerate parity doublets.  Is such a prediction borne out in the hadron spectrum?  Let's check:\cite{pdg}
\begin{equation}
\begin{array}{lcr}
N(J^P= \frac{1}{2}^+,m=938) & \;\mbox{cf.} \;& N(J^P= \frac{1}{2}^-,m=1535) \\
\pi(0^-,140) & \mbox{cf.} & \sigma(0^+,600)\\
\rho(1^-,770) & \mbox{cf.} & a_1(1^+,1260)
\end{array}
\end{equation}
Quite clearly, it is not: the difference in masses between parity partners is very large, which forces a conclusion that chiral symmetry is badly broken.  Since the current-quark mass term is the only piece of the QCD Lagrangian that breaks chiral symmetry, this appears to suggest that the quarks are quite massive.  The conundrum reappears again: how can the pion be so light if the quarks are so heavy?

The extraordinary phenomena of confinement and DCSB can be identified with properties of dressed-quark and -gluon propagators.  These two-point functions describe the in-medium propagation characteristics of QCD's elementary excitations.  Here the medium is QCD's ground state; viz., the interacting vacuum.  

The propagation of a photon through a dense electron gas is a well-known example from solid state physics of the effect that a medium can have on particle propagation.  Such a photon acquires a Debye mass: $m_D^2 \propto k_F^2$, where $k_F$ is the Fermi momentum of the electron gas, so that the photon propagator is modified:
\begin{equation}
\frac{1}{q^2} \rightarrow \frac{1}{q^2+m_D^2}\,.
\end{equation}
The appearance of this dynamically generated mass leads to a screening of  electromagnetic interactions within the gas; namely, interactions are material only between particles separated by $r \lsim r_D:=1/m_D$.  

Similar but more dramatic changes occur in the quark and gluon propagators.  They acquire momentum-dependent mass functions, an outcome which fundamentally alters the spectral properties of these elementary excitations.  

A mass term in the QCD Lagrangian explicitly breaks chiral symmetry.  The effect can be discussed in terms of the quark propagator.  It is sufficient to consider that of a noninteracting fermion of mass $m$:
\begin{equation}
S(p) = \frac{- i \gamma\cdot p + m}{p^2 + m^2}.
\end{equation}
On this propagator, the chiral rotation of Eq.\,(\ref{globalchiral}) is effected through
\begin{equation}
S(p) \rightarrow {\rm e}^{i \gamma_5 \theta} S(p) {\rm e}^{i \gamma_5 \theta} 
= \frac{- i \gamma\cdot p }{p^2 + m^2} + {\rm e}^{2i \gamma_5 \theta} \frac{ m}{p^2 + m^2} .
\end{equation}
It is therefore clear that the symmetry violation is proportional to the current-quark mass and hence that the theory is chirally symmetric for $m=0$.   Another way of looking at this is to consider the fermion condensate:
\begin{equation}
\label{condensatesimple}
\langle \bar q q \rangle = \,- \,{\rm tr} \int \frac{d^4 p }{(2\pi)^4} \, S(p) \propto 
\,- \int \frac{d^4 p }{(2\pi)^4} \, \frac{m}{p^2 + m^2} .
\end{equation}
This is a quantity that can rigorously be defined in quantum field theory\,\cite{kurtcondensate} and whose strength measures the violation of chiral symmetry.  It is a standard \emph{order parameter} for chiral symmetry breaking, playing a role analogous to that of the magnetisation in a ferromagnet.

\begin{figure}[t]

\vspace*{-7em}

\centerline{\includegraphics[clip,height=1.0\textwidth]{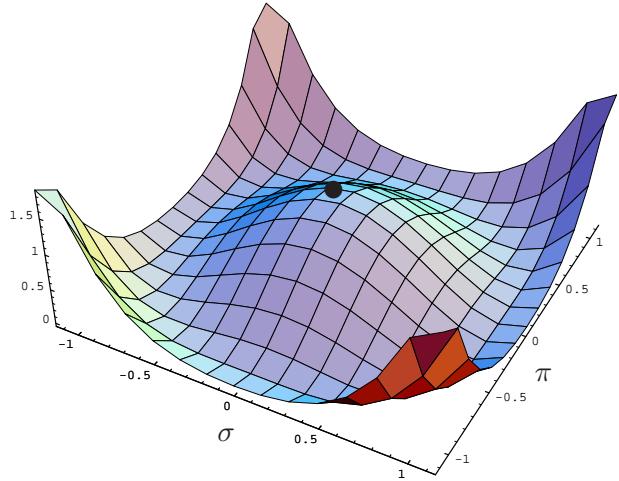}}
\vspace*{-19em}

\hspace*{16.3em}%
{\LARGE $\bullet$}\vspace*{5.5em}

\hspace*{24em} {\large $\pi$}\vspace*{1em}

\hspace*{12em} {\large $\sigma$}\vspace*{2em}

\caption{\label{mexicanhat} A rotationally invariant but unstable extremum of the Hamiltonian obtained with the potential in Eq.\,(\protect\ref{Emexican}). (Adapted from Ref.\,[\protect\refcite{cdrANU}].)}

\end{figure}

This connects immediately to dynamical symmetry breaking.  Consider a point-particle in the rotationally invariant potential 
\begin{equation}
\label{Emexican}
V(\sigma,\pi) = (\sigma^2+\pi^2-1)^2,
\end{equation}
which is illustrated in Fig. \ref{mexicanhat}.  The figure depicts a state wherein the particle is stationary at an extremum of the action.  That state is rotationally invariant but unstable.  On the other hand, in the ground state of the system the particle is stationary at any point $(\sigma,\pi)$ in the trough of the potential, for which $\sigma^2+\pi^2=1$.  There are infinitely many (an uncountable infinity of) such vacua, $|\theta\rangle$, which are related, one to another, by rotations in the $(\sigma,\pi)$-plane.  The vacua are degenerate but not rotationally invariant and hence, in general, $\langle \theta | \sigma | \theta\rangle \neq 0\neq\langle \theta | \pi | \theta\rangle$.  In this case the rotational invariance of the Hamiltonian is not exhibited in any single ground state: the symmetry is dynamically broken with interactions being responsible for $\langle \theta | \sigma | \theta\rangle \neq 0\neq\langle \theta | \pi | \theta\rangle$.

The connection between dynamics and symmetries is now in plain view.  The elementary excitations of QCD's action are absent from the strong interaction spectrum: neither a quark nor a gluon ever reaches a detector alone.  This is the physics of confinement.  Chirality is the projection of a particle's spin onto its direction of motion.  It is a Lorentz invariant for massless quarks.  To classical QCD interactions, left-handed and right-handed quarks are indistinguishable.  This symmetry has implications for the spectrum that do not appear to be realised.  That is DCSB.  Our challenge is to understand the emergence of confinement and DCSB from the QCD Lagrangian, and therefrom describe their impact on the strong interaction spectrum and hadron dynamics.  These two phenomena need not be separate.  They are likely both manifestations of the same mechanism.  That mechanism must be elucidated.  It is certainly nonperturbative.

\section{Nonperturbative Tool in the Continuum}
\label{dsea}
\setcounter{equation}{0}
In Sec.\,\ref{DSEsection} we introduced the Dyson-Schwinger equations (DSEs).  They can provide a nonperturbative tool for the study of continuum strong QCD.  At the simplest level the DSEs provide a generating tool for perturbation theory.  Since QCD is asymptotically free, that means that any model-dependence in the application of these methods can be restricted to the infrared or, equivalently, the long-range domain.  In this mode, the DSEs provide a means by which to use nonperturbative strong interaction phenomena to map out, e.g., the behaviour at long range of the interaction between light-quarks.  A nonperturbative solution of the DSEs enables the study of: hadrons as composites of dressed-quarks and -gluons; the phenomena of confinement and DCSB; and therefrom an articulation of any connection between them.  The solutions of the DSEs are Schwinger functions and because all cross-sections can be constructed from such $n$-point functions the DSEs can be used to make predictions for real-world experiments.  One of the merits in this is that any assumptions employed, or guesses made, can be tested, verified and improved, or rejected in favour of more promising alternatives.  The modern application of these methods is described in Refs.\,[\refcite{bastirev},\refcite{cdrwien},\refcite{reinhardrev},\refcite{pieterrev}].

Let's return to the dressed-quark propagator, which is given by the solution of QCD's gap equation.  In Minkowski space that is Eq.\,(\ref{GapEq}), but in Euclidean space the gap equation takes the form:
\begin{eqnarray}
\label{Sqdse} 
\lefteqn{S(p)^{-1} = i\gamma\cdot p \,+ m + \Sigma(p)\,,}\\
\Sigma(p) &= & \int^\Lambda\!\!\! {d^4\ell\over (2\pi)^4} \, g^2\,D_{\mu\nu}(p-\ell)\, \gamma_\mu\frac{\lambda^a}{2} \frac{1}{i\gamma\cdot \ell A(\ell^2) + B(\ell^2)} \, \Gamma_\nu^a(\ell,p). \label{qdse} 
\end{eqnarray}
At zero temperature and chemical potential the most general Poincar\'e covariant solution of this gap equation involves two scalar functions.  There are three common expressions:
\begin{equation}
\label{Sgeneral}
S(p) = \frac{1}{i\gamma\cdot p \, A(p^2) + B(p^2)} = \frac{Z(p^2)}{i \gamma\cdot p + M(p^2)} = -i \gamma\cdot p\, \sigma_V(p^2) + \sigma_S(p^2)\,,
\end{equation}
which are equivalent to Eq.\,(\ref{SigmaMom}).  In the second form, $Z(p^2)$ is called the wave-function renormalisation and $M(p^2)$ is the dressed-quark mass function.  

A weak coupling expansion of the DSEs produces every diagram in perturbation theory, and we reproduced the one-loop result in Eq.\,(\ref{Bp2QCD}).  The general result in perturbation theory can be summarised via 
\begin{equation}
B_{\rm pert}(p^2) = m \left( 1 - \frac{\alpha}{\pi} \ln \left[\frac{p^2}{m^2}\right] + \ldots \right),
\end{equation}
where the ellipsis denotes terms of higher order in $\alpha$ that involve $(\ln [p^2/m^2])^2$ and $(\ln\ln [p^2/m^2])$, etc.  However, at arbitrarily large finite order it is always true that 
\begin{equation}
\lim_{m\to 0} B_{\rm pert}(p^2) \equiv 0.
\end{equation}
This restates the remark made after Eq.\,(\ref{Bp2QCD}) on page~\pageref{impossible}.  Our question is whether this conclusion can ever be avoided; namely, are there circumstances under which it is possible to obtain a nonzero dressed-quark mass function in the chiral limit; viz., for $m\to 0$?

\subsection{Dynamical Mass Generation}
\label{dynamicalmass}
To begin the search for an answer, consider Eqs.\,(\ref{Sqdse}), (\ref{qdse}) with the following model forms for the dressed-gluon propagator and quark-gluon vertex:\footnote{The form for the gluon two-point function implements a four-dimensional-cutoff version of the Nambu--Jona-Lasinio model, which has long been used to model QCD at low energies, e.g., Refs.\,[\protect\refcite{weisenjl},\protect\refcite{klevanskynjl},\protect\refcite{ebertnjl}].}
\begin{eqnarray}
\label{Dnjl}
g^2 D_{\mu\nu}(p-\ell) & = &  \delta_{\mu\nu}\, 
\frac{1}{m_G^2}\,\theta(\Lambda^2-\ell^2)\,,\\
\label{Gnjl}
\Gamma_\nu^a(k,p) & = & \gamma_\mu \frac{\lambda^a}{2}\,,
\end{eqnarray}
wherein $m_G$ is some gluon mass-scale and $\Lambda$ serves as a cutoff.  The model thus obtained is not renormalisable so that the regularisation scale $\Lambda$, upon which all calculated quantities depend, plays a dynamical role.  In this case the gap equation is
\begin{eqnarray} 
\nonumber \lefteqn{ 
i\gamma\cdot p\, A(p^2) + B(p^2)= i \gamma\cdot p + m_0 }\\ 
 &  & + 
\frac{4}{3}\, \frac{1}{m_G^2}\,\int \frac{d^4\ell}{(2\pi)^4}\, 
\theta(\Lambda^2 -\ell^2)\, \gamma_\mu \,\frac{-i\gamma\cdot \ell A(\ell^2) + 
B(\ell^2)}{\ell^2 A^2(\ell^2) + B^2(\ell^2)}\, \gamma_\mu\,,  \label{njlgap2} 
\end{eqnarray} 
wherein we employ $m_0$ to represent the mass that explicitly breaks chiral symmetry. 

If one multiplies Eq.\,(\ref{njlgap2}) by $(-i\gamma\cdot p)$ and subsequently evaluates a trace over spinor (Dirac) indices, then one finds
\begin{equation} 
\label{Aeqnnjl}
 p^2 \, A(p^2)= p^2 + \frac{8}{3}\, \frac{1}{m_G^2}\,\int 
\frac{d^4\ell}{(2\pi)^4}\,\theta(\Lambda^2 
-\ell^2)\,p\cdot\ell\,\frac{A(\ell^2)}{\ell^2 A^2(\ell^2) + B^2(\ell^2)} \,.
\end{equation} 
It is straightforward to show that $\int d^4\ell p\cdot \ell = 0$; i.e., the angular integral in Eq.\,(\ref{Aeqnnjl}) vanishes, from which it follows that
\begin{equation} 
\label{Aisonenjl} A(p^2) \equiv 1\,. 
\end{equation} 
This owes to the fact that models of the Nambu--Jona-Lasinio type are defined via a four-fermion contact interaction in configuration space, which entails momentum-independence of the interaction and therefore also of the gap equation's solution in momentum space.  

If, on the other hand, one multiplies Eq.\,(\ref{njlgap2}) by $\mbox{\boldmath $I$}_{\rm D}$, uses Eq.\,(\ref{Aisonenjl}) and subsequently evaluates a trace over Dirac indices, then \begin{equation} 
B(p^2) = m_0 + \frac{16}{3}\, \frac{1}{m_G^2}\,\int 
\frac{d^4\ell}{(2\pi)^4}\,\theta(\Lambda^2 -\ell^2)\,\frac{B(\ell^2)}{\ell^2 + 
B^2(\ell^2)}\,. \label{BMnjl} 
\end{equation} 
Since the integrand here is $p^2$-independent then a solution at one value of $p^2$ must be the solution at all values; viz., any nonzero solution must be of the form \begin{equation}
B(p^2) = {\rm constant}=M\,.
\end{equation}
Using this result, Eq.\,(\ref{BMnjl}) becomes 
\begin{eqnarray} 
M & = &  m_0 + M\,\frac{1}{3\pi^2} \, \frac{1}{m_G^2}\, { C}(M^2,\Lambda^2)\,,\\ 
\label{CMLamda} { C}(M^2,\Lambda^2) & = & \Lambda^2 - M^2 \ln\left[1+\Lambda^2/M^2\right]
\,. 
\end{eqnarray} 
Recall now that $\Lambda$ defines the mass-scale in a nonrenormalisable model.  Hence we can set $\Lambda \equiv 1$ and hereafter merely interpret all other mass-scales as being expressed in units of $\Lambda$, whereupon the gap equation becomes 
\begin{equation} 
\label{pgapnjl} M = m_0 + M\,\frac{1}{3\pi^2} \, \frac{1}{m_G^2}\, { 
C}(M^2,1)\,. 
\end{equation} 

Let us consider Eq.\,(\ref{pgapnjl}) in the chiral limit: $m_0=0$,
\begin{equation}
\label{pgapnjl0}
M = M\,\frac{1}{3\pi^2} \, \frac{1}{m_G^2}\, { 
C}(M^2,1)\,.
\end{equation}
One solution is obviously $M\equiv 0$.  This is the result that connects smoothly with perturbation theory: one starts with no mass, and no mass is generated.  In this instance the theory is said to realise chiral symmetry in the Wigner-Weyl mode.  

Suppose, however, that $M\neq 0$ in Eq.\,(\ref{pgapnjl0}).  That is possible if, and only if, the following equation has a solution:
\begin{equation}
1= \frac{1}{3\pi^2} \, \frac{1}{m_G^2}\,  { C}(M^2,1)\,.
\end{equation}
It is plain from Eq.\,(\ref{CMLamda}) that $C(M^2,1)$ is a monotonically decreasing function of $M$ whose maximum value occurs at $M=0$: $ {C}(0,1)=1$.  Consequently, $\exists M \neq 0$ solution if, and only if, 
\begin{equation}
\label{critmG1}
\frac{1}{3\pi^2} \, \frac{1}{m_G^2} > 1\,.
\end{equation}
It is thus apparent that there is always a domain of values for the gluon mass-scale, $m_G$, for which a nontrivial solution of the gap equation can be found.  If we suppose $\Lambda \sim 1\,$GeV, which is a scale that is typical of hadron physics, then $\exists M \neq 0$ solution for 
\begin{equation}
\label{critmG2}
m_G^2 < \frac{\Lambda^2}{3\pi^2} \simeq (0.2\,{\rm GeV})^2.
\end{equation}

This result, derived in a straightforward manner, is astonishing!  It reveals the power of a nonperturbative solution to nonlinear equations.  Although we started with a model of massless fermions, the interaction alone has provided the fermions with mass.  This is dynamical chiral symmetry breaking; namely, the generation of mass \textit{from nothing}.  When this happens chiral symmetry is said to be realised in the Nambu-Goldstone mode.  It is clear from Eqs.\,(\ref{critmG1}), (\ref{critmG2}) that DCSB is guaranteed to be possible so long as the interaction exceeds a particular minimal strength.  

\begin{figure}[t]
\centerline{%
\includegraphics[clip,width=0.75\textwidth]{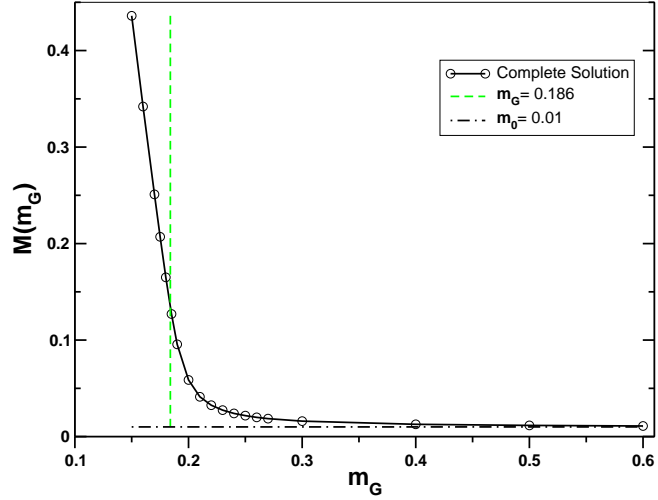}
}

\caption{\label{NJLDCSB} With a bare mass $m=0.01$, the mass gap obtained using the interaction specified by Eqs.\,(\ref{Dnjl}), (\ref{Gnjl}) -- solid curve with circles.  (All dimensioned quantities measured in units of $\Lambda$, the regularisation scale.)}
\end{figure}

In Fig.\,\ref{NJLDCSB} we depict the $m_G$-dependence of the nontrivial self-consistent solution of Eq.\,(\ref{pgapnjl}) obtained with a current-quark mass $m = 0.01$ (measured in units of $\Lambda$).  The vertical line marks 
\begin{equation}
m_G^{\rm cr} = \frac{1}{\surd{3}\,\pi}\,;
\end{equation}
namely, the value of $m_G$ above which a DCSB solution of Eq.\,(\ref{pgapnjl0}) is impossible.  Evidently, for $m_G > m_G^{\rm cr}$, $M(m_G) \approx m_0$ and the self-consistent solution is well approximated by the perturbative result.  However, a transition takes place for $m_G \simeq m_G^{\rm cr}$, and for $m_G < m_G^{\rm cr}$ the dynamical mass is much greater than the bare mass, with $M$ increasing rapidly as $m_G$ is reduced and the effective strength of the interaction is thereby increased.

\subsection{Dynamical Mass and Confinement}
\label{confinement}
One aspect of quark confinement is the absence from the strong interaction spectrum of free-particle-like quarks.  What does the model of Sec.\,\ref{dynamicalmass} have to contribute in this connection?  Well, whether one works within the domain of the model on which DCSB takes place, or not, the quark propagator always has the form
\begin{equation}
S^{\rm NJL}(p)=\frac{1}{i\gamma\cdot p \, [A(p^2)=1] + [B(p^2)=M]} =  \frac{- i\gamma\cdot p  + M}{p^2 + M^2}\,,
\end{equation}
where $M$ is constant.  This expression has a pole at $p^2 + M^2 = 0$ and thus is always effectively the propagator for a noninteracting fermion with mass $M$.  Hence, while it is generally true that models of the Nambu--Jona-Lasinio type support DCSB, they do not exhibit confinement.  

Consider an alternative,\cite{mn83} defined via Eq.\,(\ref{Gnjl}) and 
\begin{equation}
\label{mnprop} g^2 D_{\mu\nu}(k) = (2\pi)^4\, \check{G}\,\delta^4(k) \left[\delta_{\mu\nu} - \frac{k_\mu k_\nu}{k^2}\right].
\end{equation}
Here $\check{G}$ defines the model's mass-scale and the interaction is a $\delta$-function in momentum space, which may be compared with models of the Nambu--Jona-Lasinio type wherein the interaction is instead a $\delta$-function in configuration space.  In this instance the gap equation is 
\begin{equation} 
i\gamma\cdot p\, A(p^2) + B(p^2)  =  i \gamma\cdot p + m_0 + \check{G}\,\gamma_\mu\, 
\frac{-i\gamma\cdot p\, A(p^2) + B(p^2)}{p^2 A^2(p^2) + B^2(p^2)}\, 
\gamma_\mu\,,\label{mngap1} 
\end{equation} 
which yields the following coupled nonlinear algebraic equations: 
\begin{eqnarray} 
A(p^2) & = & 1 + 2 \, \frac{A(p^2)}{p^2 A^2(p^2) + B^2(p^2)} \, ,\label{AMN}\\ 
B(p^2) & = & m_0 + 4\, \frac{B(p^2)}{p^2 A^2(p^2) + B^2(p^2)} \,. \label{BMN}
\end{eqnarray} 
Equation (\ref{mngap1}) yields an ultraviolet finite model and hence there is no regularisation mass-scale.  In this instance we can therefore refer all dimensioned quantities to the model's mass-scale and set $\check{G}=1$. 

Consider the chiral limit of Eq.\,(\ref{BMN}):
\begin{equation} 
B(p^2)  =   4\, \frac{B(p^2)}{p^2 A^2(p^2) + B^2(p^2)}\,. \label{BMN0} 
\end{equation} 
Obviously, like Eq.\,(\ref{pgapnjl0}), this equation admits a trivial solution $B(p^2) \equiv 0$ that is smoothly connected to the perturbative result, but is there another? 
The existence of a $B\not\equiv 0$ solution; i.e., a solution that dynamically 
breaks chiral symmetry, requires (in units of $\check{G}$)
\begin{equation} 
p^2 A^2(p^2) + B^2(p^2) = 4\,.
\end{equation} 
Suppose this identity to be satisfied, then its substitution into Eq.~(\ref{AMN}) gives \begin{equation} 
A(p^2) - 1 = \frac{1}{2}\,A(p^2) \; \Rightarrow \; A(p^2) \equiv 2\,, 
\end{equation} 
which in turn entails 
\begin{equation} 
B(p^2) = 2\,\sqrt{1 - p^2}\,. 
\end{equation} 

A complete chiral-limit solution is composed subject to the physical requirement that the quark self-energy is real on the spacelike momentum domain, and hence 
\begin{eqnarray} 
\label{AMNres}
A(p^2) &= & \left\{ \begin{array}{ll} 
2\,;\; & p^2\leq 1\\ 
\frac{1}{2}\left( 1 + \sqrt{1+8/p^2} \right)\,;\; & p^2>1 \end{array} \right.\\ 
\label{BMNres}
B(p^2) &= & \left\{ \begin{array}{ll} 
\sqrt{1-p^2}\,;\; & p^2\leq 1\\ 
0\,; & p^2>1 \,.\end{array} \right. 
\end{eqnarray} 
In this case both scalar functions characterising the dressed-quark propagator differ significantly from their free-particle forms and are momentum dependent.  (As we will see, this is also true in QCD.)  It is noteworthy that the magnitude of the model's mass-scale plays no role in the appearance of this DCSB solution of the gap equation.  Thus, in models of the Munczek-Nemirovsky type, the interaction is always strong enough to support the generation of mass from nothing.

The DCSB solution of Eqs.\,(\ref{AMNres}), (\ref{BMNres}) is defined and continuous for all $p^2$, including timelike momenta, $p^2<0$.  It gives a dressed-quark propagator whose denominator
\begin{equation} 
p^2\,A^2(p^2) + B^2(p^2) > 0\,, \; \forall \, p^2\,. 
\end{equation} 
This is a novel and remarkable result, which means that the propagator does not exhibit any free-particle-like poles!  This feature can be interpreted as a realisation of quark confinement.  

The Munczek-Nemirovosky interaction has taken a massless quark and turned it into something which at timelike momenta bears little resemblance to the perturbative quark.  It does that for all nonzero values of the model's mass-scale.  In this model one exemplifies an intriguing possibility that all models with quark-confinement necessarily exhibit DCSB.  It is obvious from Sec.\,\ref{dynamicalmass} that the converse is certainly not true.

\begin{figure}[t]
\centerline{%
\includegraphics[clip,width=0.75\textwidth]{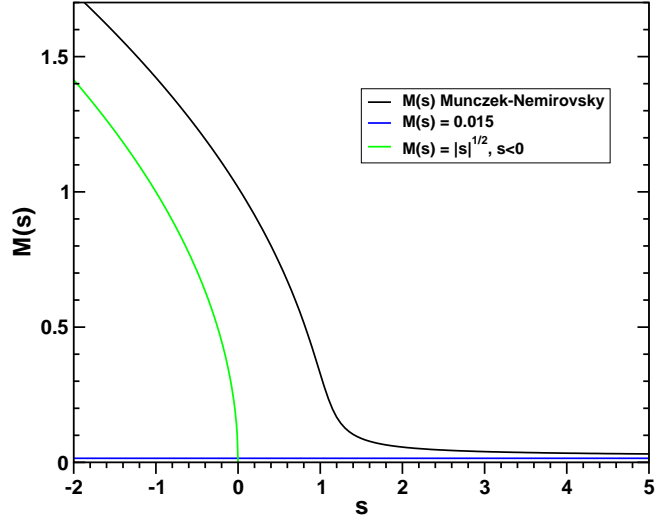}
}

\caption{\label{MNmassfn} With a bare mass $m_0=0.015$, the dressed-quark mass function \mbox{$M(s=p^2)$} obtained using the interaction specified by Eqs.\,(\ref{Gnjl}), (\ref{mnprop}) -- solid curve.  A free-particle-like mass-pole would occur at that value of $s$ for which the solid and green curves intersect.  As the figure suggests, this never happens in models of the Munczek-Nemirovsky type.  (All dimensioned quantities measured in units of $\check{G}$, the model's mass-scale.)}
\end{figure}

In the chirally asymmetric case the gap equation yields 
\begin{eqnarray} 
A(p^2) & = & \frac{2 \,B(p^2)}{ m+ B(p^2) }\,,\\ 
B(p^2) & = & m_0 + \frac{4\, [m + B(p^2)]^2}{B(p^2) ([m+B(p^2)]^2 + 4 p^2)}\,. 
\end{eqnarray} 
The second of these is a quartic equation for $B(p^2)$.  It can be solved algebraically.  There are four solutions, obtained in closed form, only one of which possesses the physically sensible ultraviolet spacelike behaviour: $B(p^2) \to m$ as \mbox{$p^2\to \infty$}.  The physical solution is depicted in Fig.\,\ref{MNmassfn}.  At large spacelike momenta, $M(s=p^2)\to m_0^+$ and a perturbative analysis is reliable.  That is never the case for $s\lsim 1$ (in units of $\check{G}$), on which domain $M(s)\gg m_0$ and the difference $M(s) - |s|$ is always nonzero, a feature that is consistent with confinement.

\begin{figure}[t]
\centerline{%
\includegraphics[clip,width=0.75\textwidth]{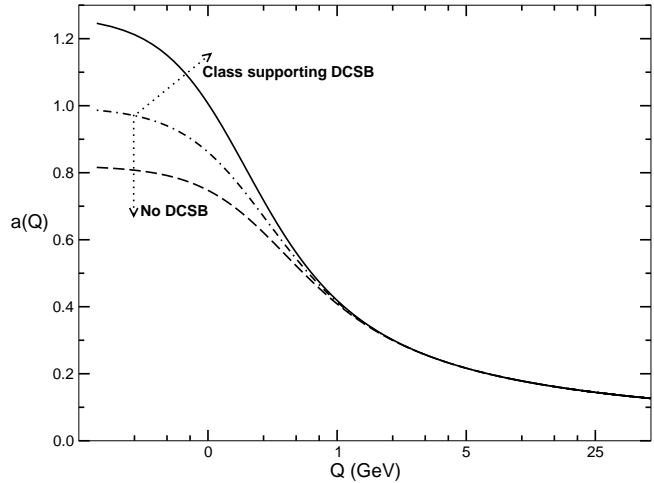}
}\vspace*{-1ex}

\caption{\label{generalinteraction} Two classes of effective interaction in the gap equation.  A theory will exhibit dynamical chiral symmetry breaking if, and only if, $a(Q=0) \geq 1$.  Here the ultraviolet behaviour of the interactions is constrained to be that of QCD.}
\end{figure}

We have illustrated two variations on the theme of dynamical mass generation via the gap equation.  A general class of models for asymptotically free theories may be discussed in terms of an effective interaction
\begin{equation}
g^2 D(Q^2) = 4\pi \frac{a(Q^2)}{Q^2}\,,
\end{equation}
where $a(Q^2)$ is such that $g^2 D(Q^2)$ evolves according to QCD's renormalisation group in the ultraviolet.  This situation is depicted in Fig.\,\ref{generalinteraction}.  The types of effective interaction fall into two classes.  In those for which $a(0) < 1$, the only solution of the gap equation in the chiral limit is \mbox{$M(s) \equiv 0$}.  Whereas, when $a(0)\geq 1$, $M(s)\neq 0$ is possible in the chiral limit and, indeed, corresponds to the energetically favoured ground state.  

It is the right point to recall the conundrums described in Sec.\,\ref{hadronphysics}. We have seen how confinement and DCSB can emerge in the nonperturbative solution of a theory's Dyson-Schwinger equations.  Therefore theories whose classical Lagrangian possesses no mass-scale can, in fact, via DCSB, behave like theories with large quark masses and can also exhibit confinement.  This understanding is a step toward a resolution of the riddles posed by strong interaction physics.  

\section{Meson Properties}
\label{mesonproperties}
\setcounter{equation}{0}
\subsection{Coloured Two- and Three-point Schwinger Functions}
\label{coloured}
We now begin a consideration of QCD proper.  The preceding discussion leads to the key question: what is the behaviour of the kernel of QCD's gap equation?  That kernel is constituted by the contraction of the dressed-gluon propagator and the dressed-quark-gluon vertex:
\begin{equation}
D_{\mu\nu}(p-q) \, \Gamma_\nu(q)\,.
\end{equation}

\begin{figure}[t]
\centerline{%
\includegraphics[clip,width=0.75\textwidth]{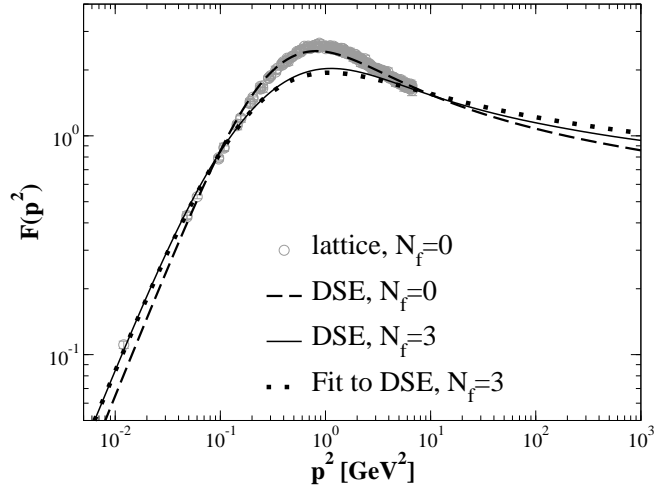}
}

\caption{\label{gluoncf} Solid curve -- $F(p^2)$ in Eq.\,(\protect\ref{gluonZ}) obtained through the solution of one particular truncation of the coupled ghost-gluon-quark DSEs with $N_f=3$ flavours of massless quark.  Dashed curve -- DSE solution with the same truncation but omitting coupling to the quark DSE.  Open circles -- quenched lattice-QCD results.  NB.\ There is sufficient lattice data at intermediate spacelike $p^2$ for the open circles to appear as a grey band.  Within the lattice errors, the quenched DSE and lattice results are indistinguishable.  (Adapted from Ref.\,[\protect\refcite{alkoferdetmold}].)}
\end{figure}

In Landau gauge the two-point gluon Schwinger function can be expressed
\begin{equation}
\label{gluonZ}
D_{\mu\nu}(p)= \left(\delta_{\mu\nu} - \frac{p_\mu p_\nu}{p^2}\right) \frac{F(p^2)}{p^2}\,,
\end{equation}
where $F(p^2)$ involves the vacuum polarisation discussed in Sec.\,\ref{photonvacuum}.  The modern DSE perspective on $F(p^2)$ is reviewed in Ref.\,[\refcite{reinhardrev}] and the predictions described therein were verified in contemporary simulations of lattice-regularised QCD.\cite{latticegluon}  The agreement is illustrated in Fig.\,\ref{gluoncf}.  

The DSE result depicted in Fig.\,\ref{gluoncf} describes a gluon two-point function that is suppressed at small $p^2$; i.e., in the infrared.  This deviation from expectations based on perturbation theory becomes apparent at $p^2 \simeq 1\,$GeV$^2$.  A mass-scale of this magnitude has long been anticipated as characteristic of nonperturbative gauge-sector dynamics.  Its origin is fundamentally the same as that of $\Lambda_{\rm QCD}$, which appears in perturbation theory.  This phenomenon whereby the value of a dimensionless quantity becomes essentially linked to a dynamically generated mass-scale is sometimes called \emph{dimensional transmutation}.

\begin{figure}[t]
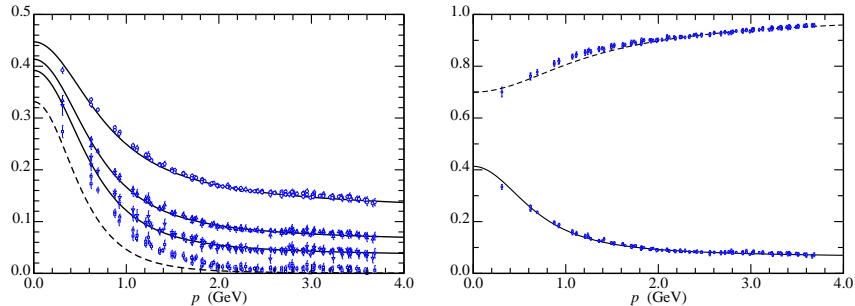

\centerline{%
\resizebox{0.47\textwidth}{!}{\includegraphics{fig1mb.eps}}\hspace*{1em}
\resizebox{0.47\textwidth}{!}{\includegraphics{fig3mb.eps}}
}

\caption{\label{quarkcf} \textit{Left Panel} -- Dashed-curve: gap 
equation's solution in the chiral limit; solid curves: solutions for $M(p^2)$ 
obtained using the current-quark masses in Eq.\ (\protect\ref{amvalues}).  Data, upper three sets: lattice results for $M(p^2)$ in GeV at $am$ values in Eq.\,(\protect\ref{amvalues}); lower points (boxes): linear extrapolation of these results to $a m =0$.\protect\cite{bowman2} \textit{Right Panel} -- Dashed curve, $Z(p^2)$, and solid curve, $M(p^2)$ calculated from the gap equation with $m(\zeta)=55\,$MeV.  Data, quenched lattice-QCD results for $M(p^2)$ and $Z(p^2)$ obtained with $am = 0.036$.\protect\cite{bowman2}  (Adapted from Ref.\,[\protect\refcite{bhagwat}].)}
\end{figure}

The dressed-quark two-point function has the form presented in Eq.\,(\ref{Sgeneral}), and we know that for a free particle $Z(p^2)=1$ and $M(p^2) = m_{\rm current}$.  On the other hand, the behaviour of these functions in QCD is a longstanding prediction of DSE studies,\cite{cdragw} and could have been anticipated from Refs.\,[\refcite{lane},\refcite{politzer}].  These DSE predictions, too, are confirmed in numerical simulations of lattice-QCD,\cite{bowman2,bowman} and the conditions have been explored under which pointwise agreement between DSE results and lattice simulations may be obtained.\cite{alkoferdetmold,bhagwat,bhagwat2}.  This agreement is illustrated in Fig.\,\ref{quarkcf}, wherein the nonzero current-quark masses are 
\begin{equation} 
\label{amvalues} 
\begin{array}{l|lll} 
a\,m_{\rm lattice} ~&~ 0.018 ~&~ 0.036 ~&~ 0.072 \\
\hline m(\zeta) ~({\rm GeV}) &~
0.030 ~&~ 0.055 ~&~ 0.110 
\end{array}\,.
\end{equation} 
The top row here lists the values used in the quenched-QCD lattice simulations, with $a$ the lattice spacing so that $ a\, m$ is dimensionless,\cite{bowman2} and the second row provides the matched current-quark masses used in the DSE study, with the renormalisation scale $\zeta = 19\,$GeV.\cite{bhagwat}

Figure \ref{quarkcf} confirms that DCSB is a reality in QCD.  At ultraviolet momenta the magnitude of the mass function is determined by the current-quark mass.  In the infrared, however, for light-quarks, $M(p^2)$ is orders-of-magnitude larger.  The DSE analysis alone, and its correlation of lattice data, indicates that the mass function is nonzero and retains its magnitude in the chiral limit.  

What \label{confpage} about confinement?  We have already mentioned that this phenomenon might be expressed in the analyticity properties of the dressed propagators.  In fact it is sufficient for confinement that dressed propagators for coloured excitations not possess a Lehmann representation, since this is associated with a violation of reflection positivity.  An excitation connected with a propagator that violates reflection positivity cannot appear in the Hilbert space of physical states; viz., it won't propagate to a detector.  These notions may be traced from Refs.\,[\refcite{entire1,entire2,stingl,krein}] and are described in Refs.\,[\refcite{cdragw,bastirev,cdrwien,reinhardrev}].  (See also, e.g., the discussion of the reconstruction theorem in Ref.\,[\refcite{glimm}].)  Any Schwinger function that exhibits an inflexion point cannot be expressed through a Lehmann representation.  An inspection of the DSE and lattice results for $F(p^2)/p^2$ and $\sigma_S(p^2)$ suggests strongly that the dressed-gluon and -quark propagators display an inflexion point.  Moreover, in QCD there are DSE studies which suggest that quark confinement and DCSB both owe to the same dynamical mechanism,\cite{dsenonzeroT,dsenonzeromu} and therefore that one does not appear without the other.

In connection with Fig.\,\ref{generalinteraction} in Sec.\,\ref{confinement} we described a class of effective interaction in the gap equation that can generate DCSB.  The dressed-gluon propagator obtained from $F(p^2)$ in Fig.\,\ref{gluoncf}, combined with Eq.\,(\ref{Gnjl}), does not yield an interaction that is a member of that class.\cite{cdrwien,hawes} How then is it possible that Ref.\,[\refcite{bhagwat}] unified the gluon and quark two-point functions?  The problem was anticipated in Ref.\,[\refcite{hawes}] and an answer suggested; namely, the dressed-quark-gluon vertex must exhibit an enhancement in the infrared.  This is precisely the means employed in Ref.\,[\refcite{bhagwat}].  The exact nature of the enhancement and its origin in QCD is the subject of contemporary research.\cite{bhagwat2,jisvertex,bhagwatvertex}

\subsection{Colour-singlet Schwinger Functions: Bound States}
\label{colourless}
At this point it is apparent that a semi-quantitatively reliable picture of the key propagators and vertices is established in QCD.  What about bound states?  Without them, of course, a direct comparison with experiment is impossible.  Bound states appear as pole contributions to colour-singlet Schwinger functions and this observation may be viewed as the origin of the Bethe-Salpeter equation.  The Bethe-Salpeter equation (BSE) has an history that predates QCD but we will not go into that.  It can be traced, e.g., from Ref.\,[\refcite{iz80}], Chap.\,10.  

The DSE for the dressed-quark-gluon vertex can be viewed as a BSE, as can that for the dressed-quark-photon vertex.  The latter is a colour singlet vertex and its lowest mass pole-contribution is the $\rho$-meson.\cite{marisphotonvertex}  That fact underlies the success of vector meson dominance phenomenology.  

The axial-vector vertex is of primary interest to hadron physics.  It may be obtained as the solution of the inhomogeneous Bethe-Salpeter equation
\begin{equation}
\label{avbse}
\left[\Gamma_{5\mu}(k;P)\right]_{tu}
 =  Z_2 \left[\gamma_5\gamma_\mu\right]_{tu} + \int^\Lambda_q
[S(q_+) \Gamma_{5\mu}(q;P) S(q_-)]_{sr} K_{tu}^{rs}(q,k;P)\,,
\end{equation}
where $q_\pm = q \pm P/2$ and the colour-, Dirac- and flavour-matrix structure of the elements in the equation is denoted by the indices $r,s,t,u$.  In Eq.\,(\ref{avbse}), $K(q,k;P)$ is the fully-amputated quark-antiquark scattering kernel.  It is one-particle-irreducible and hence, by definition, does not contain quark-antiquark to single gauge-boson annihilation diagrams, such as would describe the leptonic decay of the pion, nor diagrams that become disconnected by cutting one quark and one antiquark line.  If one knows the form of $K$ then one completely understands the nature of the interaction between quarks in QCD.

In addition to $K$, the kernel of Eq.\,(\ref{avbse}) also contains the dressed-quark propagator.  That is obtained from the gap equation, which in QCD is written 
\begin{equation}
\label{gendse} S(p)^{-1} = Z_2 \,(i\gamma\cdot p + m^{\rm bare}) +\, Z_1
\int^\Lambda_q \, g^2 D_{\mu\nu}(p-q) \frac{\lambda^a}{2}\gamma_\mu S(q)
\Gamma^a_\nu(q;p) \,.
\end{equation}
We have seen this equation before but here it is written \textit{properly}; viz., Eq.\,(\ref{gendse}) is the renormalised DSE for the dressed-quark propagator.  Therein $D_{\mu\nu}(k)$ is the renormalised dressed-gluon propagator, $\Gamma^a_\nu(q;p)$ is the renormalised dressed-quark-gluon vertex, $m^{\rm bare}$ is the $\Lambda$-dependent current-quark bare mass that appears in the QCD Lagrangian and \mbox{$\int^\Lambda_q := \int^\Lambda d^4 q/(2\pi)^4$} represents a \textit{Poincar\'e-invariant} regularisation of the integral, with $\Lambda$ the regularisation mass-scale.  In addition, $Z_1(\zeta^2,\Lambda^2)$ and $Z_2(\zeta^2,\Lambda^2)$ are the quark-gluon-vertex and quark wave function renormalisation constants, which depend on the renormalisation point, $\zeta$, and the regularisation mass-scale.  The solution of Eq.\,(\ref{gendse}) is obtained subject to the renormalisation condition
\begin{equation}
\label{renormS} \left.S(p)^{-1}\right|_{p^2=\zeta^2} = i\gamma\cdot p +
m(\zeta)\,,
\end{equation}
where $m(\zeta)$ is the renormalised mass: 
\begin{equation}
Z_2(\zeta^2,\Lambda^2) \, m^{\rm bare}(\Lambda) = Z_4(\zeta^2,\Lambda^2) \, m(\zeta)\,,
\end{equation}
with $Z_4$ the Lagrangian mass renormalisation constant.  In QCD the chiral limit is unambiguously defined by
\begin{equation}
\label{limchiral}
Z_2(\zeta^2,\Lambda^2) \, m^{\rm bare}(\Lambda) \equiv 0 \,, \forall \Lambda \gg \zeta \,,
\end{equation}
which is equivalent to stating that the renormalisation-point-invariant current-quark mass vanishes; i.e., $\hat m = 0$.

\subsubsection{Model-independent results}
\label{modelindependent}
We have made much of chiral symmetry in the preceding discussion.  In quantum field theory, chiral symmetry and the pattern by which it is broken is expressed via the chiral Ward-Takahashi identity:
\begin{equation}
\label{avwtim}
P_\mu \Gamma_{5\mu}^H(k;P)  = \check{S}(k_+)^{-1} i \gamma_5\frac{T^H}{2}
+  i \gamma_5\frac{T^H}{2} \check{S}(k_-)^{-1}
- i\,\{ {M}^\zeta ,\Gamma_5^H(k;P) \} ,
\end{equation}
where the pseudoscalar vertex is given by
\begin{equation}
\label{genpve}
\left[\Gamma_{5}^H(k;P)\right]_{tu} =
Z_4\,\left[\gamma_5 \frac{T^H}{2}\right]_{tu} \,+
\int^\Lambda_q \,
\left[ \chi_5^H(q;P)\right]_{sr}
K^{rs}_{tu}(q,k;P)\,,
\end{equation}
with $\chi_5^H(q;P) = \check{S}(q_+) \Gamma_{5}^H(q;P) \check{S}(q_-)$, $\check{S}= {\rm diag}[S_u,S_d,S_s,\ldots]$ and $M ^\zeta = {\rm diag}[m_u(\zeta),m_d(\zeta),m_s(\zeta),\ldots]$.  

We have written Eqs.\,(\ref{avwtim}), (\ref{genpve}) for the case of a flavour-nonsinglet vertex in a theory with $N_f$ quark flavours.  The matrices $T^H$ are constructed from the generators of $SU(N_f)$ with, e.g., \mbox{$T^{\pi^+}=\mbox{\small $\frac{1}{2}$} (\lambda^1+i\lambda^2)$} providing for the flavour content of a positively charged pion.  Writing the equations in this manner is straightforward.  However, a unified description of light- and heavy-quark systems is not.  Truncations and approximations that are reliable in one sector need not be valid in the other.

The axial-vector Ward-Takahashi identity relates the solution of a BSE to that of the gap equation.  If the identity is always to be satisfied and in a model-independent manner, as it must be in order to preserve an essential symmetry of the strong interaction and its breaking pattern, then the kernels of the gap and Bethe-Salpeter equations must be intimately related.  Any truncation or approximation of these equations must preserve that relation.  This is an extremely tight constraint.  Perturbation theory is one systematic truncation that, order by order, guarantees Eq.\,(\ref{avwtim}).  However, as we have emphasised, perturbation theory is inadequate in the face of QCD's emergent phenomena.  Something else is needed. 

Happily, at least one systematic, nonperturbative and symmetry preserving truncation of the DSEs exists.\cite{bhagwatvertex,munczek,truncscheme}  This makes it possible to prove Goldstone's theorem in QCD.\cite{mrt98} Namely, when chiral symmetry is dynamically broken: the axial-vector vertex, Eq.\,(\ref{avbse}), is dominated by the pion pole for $(P^2\sim 0)$ and the homogeneous, isovector, pseudoscalar BSE has a massless ($P^2 = 0$) solution.  The converse is also true, so that DCSB is a sufficient and necessary condition for the appearance of a massless pseudoscalar bound state of dynamically-massive constituents which dominates the axial-vector vertex for infrared total momenta.  

Furthermore, from the axial-vector Ward-Takahashi identity and the existence of a systematic, nonperturbative symmetry-preserving truncation, one can prove the following identity involving the mass-squared of a pseudoscalar meson:\cite{mrt98}
\begin{equation}
\label{gengmor}
f_H \, m_H^2 = 
\rho_H(\zeta) {M}_H^\zeta,
\end{equation}
where ${M}_H^\zeta = m_{q_1}(\zeta) + m_{q_2}(\zeta)$ is the sum of the
current-quark masses of the meson's constituents;
\begin{equation}
\label{fH}
f_H \, P_\mu = Z_2 {\rm tr} \int_q^\Lambda \! \sfrac{1}{2} (T^H)^T \gamma_5
\gamma_\mu \check{S}(q_+)\, \Gamma^H(q;P)\, \check{S}(q_-)\,,
\end{equation}
where $(\cdot)^{\rm T}$ indicates matrix transpose, and
\begin{equation}
\label{qbqH}
\rho_H(\zeta) =   Z_4\, {\rm tr}\int_q^\Lambda \!
\sfrac{1}{2}(T^H)^T \gamma_5 \check{S}(q_+) \,\Gamma^H(q;P)\, \check{S}(q_-) \,.
\end{equation}
The renormalisation constants in Eqs.\,(\ref{fH}), (\ref{qbqH}) play a pivotal role.  Indeed, the expressions would be meaningless without them.  They serve to guarantee that the quantities described are gauge invariant and finite as the regularisation scale is removed to infinity, which is the final step in any calculation.  Moreover, $Z_2$ in Eq.\,(\ref{fH}) and $Z_4$ in Eq.\,(\ref{qbqH}) ensure that both $f_H$ and the product $\rho_H(\zeta) M_H^\zeta$ are renormalisation point independent, which is an absolute necessity for any observable quantity.

Taking note that in a Poincar\'e invariant theory a pseudoscalar meson Bethe-Salpeter amplitude assumes the form
\begin{eqnarray}
\nonumber
\lefteqn{i \Gamma_{H}^j(k;P) = T^H \gamma_5
\left[ i E_H(k;P)  + \gamma\cdot P \, F_H(k;P)\right.} \\
 & &
\left.+ \, \gamma\cdot k \,k\cdot P\, G_H(k;P)+
 \sigma_{\mu\nu}\,k_\mu P_\nu \,H_H(k;P) \right],
\label{genpvv}
\end{eqnarray}
then, in the chiral limit, one can also prove that 
\begin{eqnarray}
\label{bwti}
f_H E_H(k;0)  &= &  B(k^2)\,, \\
 F_R(k;0) +  2 \, f_H F_H(k;0)  & = & A(k^2)\,,
 \label{fwti}\\
G_R(k;0) +  2 \,f_H G_H(k;0)    & = & 2 A^\prime(k^2)\,,
\label{gwti}\\
\label{hwti}
H_R(k;0) +  2 \,f_H H_H(k;0)    & = & 0\,.
\end{eqnarray}
The functions $F_R$, $G_R$, $H_R$ are associated with terms in the axial-vector vertex that are regular in the neighbourhood of $P^2 +m_H^2 = 0$ and do not vanish at $P_\mu = 0$.  These four identities are quark-level Goldberger-Treiman relations for the pion.  They are exact in QCD and are a pointwise expression of Goldstone's theorem.  These identities relate the pseudoscalar meson Bethe-Salpeter amplitude directly to the dressed-quark propagator.  Equation (\ref{bwti}) explains why DCSB and the appearance of a Goldstone mode are so intimately connected, and Eqs.\,(\ref{fwti})-(\ref{hwti}) entail that in general a pseudoscalar meson Bethe-Salpeter amplitude has what might be called pseudovector components; namely: $F_H$, $G_H$, $H_H$.  It is the latter which, in a covariant treatment, guarantee that the electromagnetic pion form factor behaves as $1/Q^2$ at large spacelike momentum transfer.\cite{mrpion}

Equation (\ref{gengmor}) and its corollaries are of fundamental importance in QCD.  To exemplify let's focus first on the chiral limit behaviour of Eq.\,(\ref{qbqH})
whereat, using Eqs.\,(\ref{genpvv}), (\ref{bwti})-(\ref{hwti}), one finds readily
\begin{equation}
f_H^0\,\rho^0_H(\zeta) = 
Z_4(\zeta,\Lambda)\, N_c\, {\rm tr}_{\rm D} \int^\Lambda_q  S_{\hat m =0}(q) = - \langle \bar q q \rangle_\zeta^0  \,, \label{qbq0}
\end{equation}
where $f_H^0$ is the chiral limit value from Eq.\,(\ref{fH}), which is nonzero when chiral symmetry is dynamically broken.  Equation (\ref{qbq0}) is unique as the expression for the chiral limit \textit{vacuum quark condensate}, and is the true definition of the order parameter first described in Eq.\,(\ref{condensatesimple}).  It thus follows from Eqs.\,(\ref{gengmor}), (\ref{qbq0}) that in the neighbourhood of the chiral limit
\begin{equation}
\label{gmor}
(f_H^0)^2 \, m_H^2 = - \, M_H^\zeta\, \langle \bar q q \rangle_\zeta^0 +
{\rm O}(\hat M^2)\,.
\end{equation}
Hence what is commonly known as the Gell-Mann--Oakes--Renner relation is a \textit{corollary} of Eq.\,(\ref{gengmor}).

Let's now consider another extreme; viz., when one of the constituents is a heavy quark, a domain on which Eq.\,(\ref{gengmor}) is equally valid.  In this case Eq.\,(\ref{fH}) yields the model-independent result\cite{marismisha} 
\begin{equation}
\label{fHheavy}
f_H \propto \frac{1}{\sqrt{M_H}}\,;
\end{equation}
i.e., it reproduces a well-known consequence of heavy-quark symmetry.\cite{neubert93} A similar analysis of Eq.\ (\ref{qbqH}) gives a new result\cite{marisAdelaide,mishaSVY}
\begin{equation}
\label{qbqHheavy}
- \langle \bar q q \rangle^H_\zeta = \mbox{constant} +
  O\left(\frac{1}{m_H}\right) \mbox{~for~} \frac{1}{m_H} \sim 0\,.
\end{equation}
Combining Eqs.\,(\ref{fHheavy}), (\ref{qbqHheavy}), one finds\cite{marisAdelaide,mishaSVY}
\begin{equation}
m_H \propto \hat m_f \;\; \mbox{for} \;\; \frac{1}{\hat m_f} \sim 0\,,
\end{equation}
where $ \hat m_f$ is the renormalisation-group-invariant current-quark mass of the flavour-nonsinglet pseudoscalar meson's heaviest constituent.  This is the result one would have anticipated from constituent-quark models but here we have indicated a direct proof in QCD. 

Pseudoscalar mesons hold a special place in QCD and there are three states, composed of $u$,$d$ quarks, in the hadron spectrum with masses below $2\,$GeV:\cite{pdg} $\pi(140)$; $\pi(1300)$; and $\pi(1800)$.  Of these, the pion [$\pi(140)$] is naturally well known and much studied.  The other two are observed, e.g., as resonances in the coherent production of three pion final states via pion-nucleus collisions \cite{experiment}.  In the context of a model constituent-quark Hamiltonian, these mesons are often viewed as the first three members of a $Q\bar Q$ $n\, ^1\!S_0$ trajectory, where $n$ is the principal quantum number; i.e., the $\pi(140)$ is viewed as the $S$-wave ground state and the others are its first two radial excitations.  By this reasoning the properties of the $\pi(1300)$ and $\pi(1800)$ are likely to be sensitive to details of the long-range part of the quark-quark interaction because the constituent-quark wave functions will possess material support at large interquark separation.  Hence the development of an understanding of their properties may provide information about light-quark confinement, which complements that obtained via angular momentum excitations\,\cite{a1b1}.  As we have emphasised, the development of an understanding of confinement is one of the greatest intellectual challenges in physics.

We have already seen that Eq.\,(\ref{gengmor}) is a powerful result.  That is further emphasised by the fact that it is applicable here, too.\cite{krassnigg1,krassnigg2}  The result holds at each pole common to the pseudoscalar and axial-vector vertices and therefore it also impacts upon the properties of non ground state pseudoscalar mesons.  Let's work with a label $n\geq 0$ for the pseudoscalar mesons: $\pi_n$, with $n=0$ denoting the ground state, $n=1$ the state with the next lowest mass, and so on.  Plainly, by assumption, $m_{\pi_{n\neq 0}}>m_{\pi_0}$ , and hence $m_{\pi_{n\neq 0}} > 0$ in the chiral limit.  Moreover, the ultraviolet behaviour of the quark-antiquark scattering kernel in QCD guarantees that Eq.\,(\ref{qbqH}) is cutoff independent.  Thus
\begin{equation}
0 < \rho_{\pi_{n}}^0(\zeta):= \lim_{\hat m\to 0} \rho_{\pi_{n}}(\zeta) < \infty \,, \; \forall \, n\,.
\end{equation}
Hence, it is a necessary consequence of chiral symmetry and its dynamical breaking in QCD; viz., Eq.\,(\ref{gengmor}), that
\begin{equation}
\label{fpinzero}
f_{\pi_n}^0 \equiv 0\,, \forall \, n\geq 1\,.
\end{equation}
This result is consistent with Refs.\,[\refcite{dominguez}], as appreciated in Ref.\,[\refcite{volkov}]. It means that in the presence of DCSB all pseudoscalar mesons except the ground state decouple from the weak interaction.  NB.\ Away from the chiral limit the quantities $f_{\pi_n}$ alternate in sign; i.e., they are positive for even $n$ but negative for odd $n$.  This is because they are the residues of poles in a \textit{vertex} that, considered as a function of $P^2$, is continuous and does not vanish between adjacent bound state poles.

This argument is legitimate in any theory that has a valid chiral limit.  It is logically possible that such a theory does not exhibit DCSB; i.e., realises chiral symmetry in the Wigner-Weyl mode, as was illustrated in Sec.\,\ref{dynamicalmass} and is the case in QCD above the critical temperature for chiral symmetry restoration.\cite{bastirev}  Equation (\ref{gengmor}) is still valid in the Wigner phase.  However, its implications are different; namely, in the Wigner phase, one has 
\begin{equation}
\label{Bm0}
B^W(0,\zeta^2)\propto m(\zeta) \propto\hat m\,;
\end{equation}
i.e., the mass function and constituent-quark mass vanish in the chiral limit.  This result is accessible in perturbation theory.  Equation (\ref{bwti}) applies if there is a massless bound state in the chiral limit.  Suppose such a bound state persists in the absence of DCSB.\footnote{If that is false then considering this particular case is unnecessary.  However, it is true at the transition temperature in QCD \cite{bastirev}.} It then follows from Eqs.\,(\ref{bwti}) and (\ref{Bm0}) that \begin{equation}
f^W_{\pi_0} \propto \hat m\,.
\end{equation}
In this case the leptonic decay constant of the ground state pseudoscalar meson also vanishes in the chiral limit, and hence all pseudoscalar mesons are blind to the weak interaction.

It is always true that 
\begin{equation}
f_{\pi_0}\,\rho_{\pi_{0}}(\zeta) \stackrel{\hat m\approx 0}{\propto} - \langle \bar q q \rangle_\zeta^0 \,.
\end{equation}
In the Wigner phase\,\cite{kurtcondensate}, $ \langle \bar q q \rangle^{0\,W}_\zeta \propto \hat m^3$.  Hence, via Eq.\,(\ref{gengmor}), if a rigorously chirally symmetric theory possesses a massless pseudoscalar bound state then\footnote{cf.\ The case of DCSB; i.e., the Nambu phase, wherein $m^N_{\pi_o} \stackrel{\hat m\approx 0}{\propto} \surd \hat m$, Eq.\,(\ref{gmor}).}
\begin{equation}
m^W_{\pi_0} \stackrel{\hat m\approx 0}{\propto} \hat m \,.
\end{equation}
In this case there is also a degenerate scalar meson partner whose mass behaves in the same manner.

In the presence of DCSB the ground state neutral pseudoscalar meson decays predominantly into two photons, a process connected with the Abelian anomaly.  Given that $f_{\pi_n \neq 0} \equiv 0$ in the chiral limit, it is natural to ask whether $\pi_{n\neq 0} \to \gamma\gamma$ is similarly affected.  Since rainbow-ladder is the leading order in a symmetry preserving truncation, it can be used to provide a model-independent analysis of this process.  In this instance, the two-photon coupling for all $u,d$ pseudoscalar mesons, including $n=0$, is described by the renormalised triangle diagrams
\begin{eqnarray}
\nonumber T^3_{5\mu\nu\rho}(k_1,k_2) &=& {\rm tr}\int_\ell^M \check{S}(\ell_{0+}) \, \Gamma^3_{5\rho}(\ell_{0+},\ell_{-0}) \, \check{S}(\ell_{-0}) \\%
 & \times&  \, i \check{Q}\Gamma_\mu(\ell_{-0},\ell) \, \check{S}(\ell) \, i \check{Q}\Gamma_\nu(\ell,\ell_{0+})\,,\label{Tmnr}\\
\nonumber T^3_{5\mu\nu}(k_1,k_2) &=& {\rm tr}\int_\ell^M \check{S}(\ell_{0+}) \, \Gamma^3_{5}(\ell_{0+},\ell_{-0}) \, \check{S}(\ell_{-0}) \\
 &\times&  \, i \check{Q}\Gamma_\mu(\ell_{-0},\ell) \, \check{S}(\ell) \, i \check{Q}\Gamma_\nu(\ell,\ell_{0+})\,,\label{Pmnr}
\end{eqnarray}
where $\ell_{\alpha\beta}=\ell+\alpha k_1+\beta k_2$, the electric charge matrix $\check{Q}={\rm diag}[e_u,e_d]=e\,{\rm diag}[2/3,-1/3]$, $\check{S}= {\rm diag}[S_u,S_d]$ and 
\begin{equation}
\label{photonvertex}
\left[\Gamma_{\mu}(k;P)\right]_{tu} = Z_2 \left[\gamma_\mu  \right]_{tu}\\
+ \int^\Lambda_q
[\check{S}(q_+) \Gamma_{\mu}(q;P) \check{S}(q_-)]_{sr} K_{tu}^{rs}(q,k;P) 
\end{equation}
is the renormalised dressed-quark-photon vertex.  (We assume subsequently that $m_u(\zeta)=m(\zeta)=m_d(\zeta)$ so that $S_u =S_d$.)

\begin{figure}[t]
\centerline{%
\includegraphics[width=0.95\textwidth]{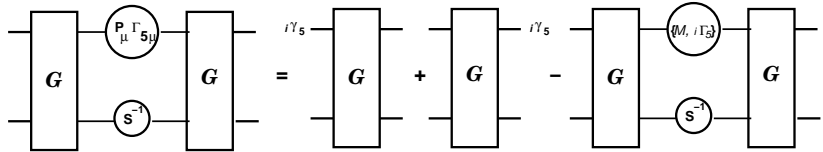}}
\caption{\label{bicudoavwti} This axial-vector Ward-Takahashi identity is an analogue of Eq.\,(\protect\ref{avwtim}).  It is valid if, and only if: the dressed-quark propagator, $S$, is obtained from Eq.\,(\ref{rainbowdse}); the axial-vector vertex, $\Gamma_{5\mu}$, is obtained from Eq.\,(\protect\ref{avbse}) with a ladder-like kernel, $K$, that ensures Eq.\,(\ref{avwtim}); the pseudoscalar vertex is constructed analogously; and the unamputated renormalised quark-antiquark scattering matrix: $G= (SS) + (SS)K(SS) + (SS)K(SS)K(SS)+ [\ldots]$, is constructed from the elements just described.  (Adapted from Ref.\,[\protect\refcite{krassnigg2}].)}
\end{figure}

The dressed-quark propagators in Eqs.\,(\ref{Tmnr}) -- (\ref{photonvertex}) are understood to be calculated using a rainbow-truncation gap equation, and $K$ is a ladder-like quark-antiquark scattering kernel that ensures Eq.\,(\ref{avwtim}) is satisfied.  At this point, nothing more need be said of these elements.  The results we describe are independent of the details discussed in Sec.\,(\ref{practical}).  

Under the conditions just described, one may derive\,\cite{bicudo} the Ward-Takahashi identity for the quark-antiquark scattering matrix, $G$, depicted in Fig.\,\ref{bicudoavwti}.  Using this identity it can be shown\cite{lc03}
\begin{equation}
\label{anomaly}
P_\rho T^3_{5\mu\nu\rho}(k_1,k_2) + 2 i m(\zeta) \, T^3_{5\mu\nu}(k_1,k_2) = \frac{\alpha}{2 \pi}\, \varepsilon_{\mu\nu\rho\sigma} k_{1\rho} k_{2\sigma}\,,
\end{equation}
where $\alpha= e^2/(4\pi)$.  This is an explicit demonstration that the triangle-diagram representation of the axial-vector--two-photon coupling calculated in the rainbow-ladder truncation is a necessary and sufficient pairing to preserve the Abelian anomaly.

In general the coupling of an axial-vector current to two photons is described by a six-point Schwinger function, to which Eq.\,(\ref{Tmnr}) is an approximation.  The same is true of the pseudoscalar--two-photon coupling and its connection with Eq.\,(\ref{Pmnr}).  Equation~(\ref{anomaly}) is valid for any and all values of $P^2=(k_1+k_2)^2$.  It is an exact statement of a divergence relation between these two six-point Schwinger functions, which is preserved by the rainbow-ladder truncation, and any systematic improvement thereof.  We can now report corollaries of Eq.\,(\ref{anomaly}) that have important implications for the properties of pseudoscalar bound states.

It follows from the discussion presented above that, unless there is a reason for the residue to vanish, every isovector pseudoscalar meson appears as a pole contribution to the axial-vector and pseudoscalar vertices:\cite{mrt98}
\begin{eqnarray}
\left. \Gamma_{5 \mu}^j(k;P)\right|_{P^2+m_{\pi_n}^2 \approx 0}&=&   \frac{f_{\pi_n} \, P_\mu}{P^2 + 
m_{\pi_n}^2} \Gamma_{\pi_n}^j(k;P) +\; \Gamma_{5 \mu}^{j\,{\rm reg}}(k;P) \,, \label{genavv} \\
\left. i\Gamma_{5 }^j(k;P)\right|_{P^2+m_{\pi_n}^2 \approx 0}
&=&   \frac{\rho_{\pi_n}(\zeta) }{P^2 + 
m_{\pi_n}^2} \Gamma_{\pi_n}^j(k;P) + \; i\Gamma_{5 }^{j\,{\rm reg}}(k;P) \,, \label{genpv} 
\end{eqnarray}
where the residues are given in Eqs.\,(\ref{fH}), (\ref{qbqH}); viz., each vertex may be expressed as a simple pole plus terms regular in the neighbourhood of this pole, with $\Gamma_{\pi_n}^j(k;P)$ being the bound state's canonically normalised Bethe-Salpeter amplitude.  

If one inserts Eqs.\,(\ref{genavv}) and (\ref{genpv}) into Eq.\,(\ref{anomaly}), and uses Eq.\,(\ref{gengmor}), one finds that in the neighbourhood of each electric-charge-neutral pseudoscalar-meson bound-state pole  
\begin{equation}
P_\rho T_{5\mu\nu\rho}^{3\,{\rm reg}}(k_1,k_2) + 2 i m(\zeta) \, T_{5\mu\nu}^{3\,{\rm reg}}(k_1,k_2)
  + f_{\pi_n} \,T^{\pi_n^0}_{\mu\nu}(k_1,k_2) = \frac{\alpha}{2 \pi} i\varepsilon_{\mu\nu\rho\sigma} k_{1\rho} k_{2\sigma}\,.
\label{reganomaly}
\end{equation}
In this equation, $T^{3\,{\rm reg}}(k_1,k_2)$ are nonresonant or \emph{continuum} contributions to the relevant Schwinger functions, whose form is concretely illustrated upon substitution of $\Gamma_{5 \mu}^{j\,{\rm reg}}(k;P)$ and $\Gamma_{5}^{j\,{\rm reg}}(k;P)$ into Eqs.\,(\ref{Tmnr}) and (\ref{Pmnr}), respectively.  Moreover, $T^{\pi_n^0}_{\mu\nu}$ is the six-point Schwinger function describing the bound state contribution, which in rainbow-ladder truncation is realised as   
\begin{eqnarray}
\nonumber T^{\pi_n^0}_{\mu\nu}(k_1,k_2)  &=& {\rm tr}\int_\ell^{M\to\infty} \!\! \check{ S}(\ell_{0+}) \, \Gamma_{\pi_n^0}(\ell_{-\frac{1}{2}\frac{1}{2}};P) \, \check{ S}(\ell_{-0}) \\%
&\times&  \, i\check{Q}\Gamma_\mu(\ell_{-0},\ell) \, \check{S}(\ell) \, i \check{ Q}\Gamma_\nu(\ell,\ell_{0+})\, .
\label{Tpingg}
\end{eqnarray}
This Schwinger function describes the direct coupling of a pseudoscalar meson to two photons.  The support properties of the bound state Bethe-Salpeter amplitude guarantee that the renormalised Schwinger function is finite so that the regularising parameter can be removed; i.e., $M\to \infty$.

We note that owing to the $O(4)$ (Euclidean Lorentz) transformation properties of each term on the lhs in Eq.\,(\ref{anomaly}), one may write
\begin{eqnarray}
P_\rho T_{5\mu\nu\rho}^{3\,{\rm reg}}(k_1,k_2) & = & \frac{\alpha}{\pi} i\varepsilon_{\mu\nu\rho\sigma} k_{1\rho} k_{2\sigma}\,A^{3\,{\rm reg}}(k_1,k_2) \,,\; \\
T_{5\mu\nu}^{3\,{\rm reg}}(k_1,k_2) & = & \frac{\alpha}{\pi} i\varepsilon_{\mu\nu\rho\sigma} k_{1\rho} k_{2\sigma}\,P^{3\,{\rm reg}}(k_1,k_2)\,,\; \\
T^{\pi_n^0}_{\mu\nu}(k_1,k_2) & = & \frac{\alpha}{\pi} i\varepsilon_{\mu\nu\rho\sigma} k_{1\rho} k_{2\sigma}\,G^{\pi_n^0}(k_1,k_2)\,, \; \label{TGdef}
\end{eqnarray}
so that Eq.\,(\ref{anomaly}) can be compactly expressed as
\begin{equation}
\label{reganomaly0}
A^{3\,{\rm reg}}(k_1,k_2) + 2 i m(\zeta) P^{3\,{\rm reg}}(k_1,k_2) + f_{\pi_n} G^{\pi_n^0}(k_1,k_2) = \frac{1}{2}.
\end{equation}
It follows from Eqs.\,(\ref{fpinzero}), (\ref{reganomaly}) that in the chiral limit all pseudoscalar mesons, \emph{except} the Goldstone mode, decouple from the divergence of the axial-vector--two-photon vertex.  

In the chiral limit the pole associated with the ground state pion appears at $P^2=0$ and thus  
\begin{equation}
\left. P_\rho T_{5\mu\nu\rho}^{3}(k_1,k_2)\right|_{P^2\neq 0}
  = \left.P_\rho T_{5\mu\nu\rho}^{3\,{\rm reg}}(k_1,k_2)\right|_{P^2\neq 0} = \frac{\alpha}{2 \pi}\, i\varepsilon_{\mu\nu\rho\sigma} k_{1\rho} k_{2\sigma}\,;
\end{equation}
namely, outside the neighbourhood of the ground state pole the regular (or continuum) part of the divergence of the axial-vector vertex saturates the anomaly in the divergence of the axial-vector--two-photon coupling.  On the other hand, in the neighbourhood of $P^2=0$ 
\begin{eqnarray} 
\left. A^{3\,{\rm reg}}(k_1,k_2) \right|_{ P^2\simeq 0} + f_{\pi_0} \,G^{\pi_0}(k_1,k_2)
& =& \frac{1}{2}\,; \label{anomalypion}
\end{eqnarray}
i.e., on this domain the contribution to the axial-vector--two-photon coupling from the regular part of the divergence of the axial-vector vertex combines with the direct $\pi_0^0 \gamma \gamma$ vertex to fulfill the anomaly.  This fact was illustrated in Ref.\,[\refcite{mrpion}] by direct calculation: Eqs.\,(\ref{bwti}) -- (\ref{hwti}) are an essential part of that demonstration.

If one defines 
\begin{equation}
\label{TpiG}
\check{T}_{\pi_n^0}(P^2,Q^2) = \left. G^{\pi_n^0}(k_1,k_2) \right|_{k_1^2=Q^2=k_2^2},
\end{equation}
in which case $ P^2= 2(k_1\cdot k_2+Q^2)$, then the physical width of the neutral ground state pion is determined by
\begin{equation}
g_{\pi_0^0 \gamma\gamma}:= \check{T}_{\pi_0^0}(-m_{\pi_0^0}^2,0) ;
\end{equation}
viz., the second term on the l.h.s.\ of Eq.\,(\ref{anomalypion}) evaluated at the on-shell points.  This result is not useful unless one has a means of estimating the contribution from the first term; viz., $A^{3\,{\rm reg}}(k_1,k_2)$.  However, that is readily done.  A consideration\,\cite{mrt98} of the structure of the regular piece in Eq.\,(\ref{genavv}) indicates that the impact of this continuum term on the $\pi_0^0 \gamma\gamma$ coupling is modulated by the magnitude of the pion's mass, which is small for realistic $u$ and $d$ current-quark masses and vanishes in the chiral limit.  One therefore expects this term to contribute very little and anticipates from Eq.\,(\ref{anomalypion}) that 
\begin{equation}
\label{anomalycouple}
g_{\pi_0^0 \gamma\gamma} = \frac{1}{2} \frac{1}{f_{\pi_0}}
\end{equation}
is a good approximation.  This is verified in explicit calculations; e.g., in Ref.\,[\refcite{maristandy4}], which evaluates the triangle diagrams described herein, the first term on the l.h.s.\ modifies the result in Eq.\,(\ref{anomalycouple}) by less than 2\%.  

There is no reason to expect an analogous result for pseudoscalar mesons other than the $\pi(140)$; i.e., the states which we denote by $n\geq 1$.  Indeed, as all known such pseudoscalar mesons have experimentally determined masses that are greater than $1\,$GeV, the reasoning used above suggests that the presence of the continuum terms, $A^{3\,{\rm reg}}(k_1,k_2)$ and $P^{3\,{\rm reg}}(k_1,k_2)$, must materially impact upon the value of $g_{\pi_1^0 \gamma\gamma}$.  This was shown to be true in Ref.\,[\refcite{krassnigg2}]; e.g., $g_{\pi_1^0 \gamma\gamma} = - 0.13\,g_{\pi_0^0 \gamma\gamma}$. 

On the kinematically accessible $Q^2$ domains, $\check{T}_{\pi_0^0}(-m_{\pi_0}^2,Q^2)>0$ and $\check{T}_{\pi_1^0}(-m_{\pi_1}^2,Q^2)<0$.  It is anticipated that this pattern of signs will repeat, as described after Eq.\,(\ref{fpinzero}); i.e., on the kinematically accessible domains $\check{T}_{\pi_n^0}(-m_{\pi_n}^2,Q^2)$ is positive for even $n$ and negative for odd $n$.

We have stated that the rainbow-ladder truncation preserves the one-loop renormalisation group properties of QCD.  It follows that Eq.\,(\ref{Tpingg}) should reproduce the leading large-$Q^2$ behaviour of the $\gamma^\ast(Q) \pi_n(P) \gamma^\ast(Q)$ transition form factor inferred from perturbative QCD.  That is true.\cite{kekez}  

Reference~[\refcite{krassnigg2}] considered the general case; i.e., arbitrary $n$, and proved 
\begin{equation}
\check{T}_{\pi_n^0}(-m_{\pi_n}^2,Q^2) \stackrel{Q^2\gg \Lambda_{\rm QCD}^2}{=} \frac{4\pi^2}{3}
\left[ \frac{f_{\pi_n}}{Q^2} + F_n^{(2)}(-m_{\pi_n}^2) 
\frac{\ln^{\gamma} Q^2/\omega_{\pi_n}^2}{Q^4}
\right] ,
\label{UVnot0}
\end{equation}
where: $\gamma$ is an anomalous dimension, which cannot be determined accurately in rainbow-ladder truncation; $\omega_{\pi_n}$ is a mass-scale that gauges the momentum space width of the pseudoscalar meson; and $F_n^{(2)}(-m_{\pi_n}^2)$ is a structure-dependent constant, similar but unrelated to $f_{\pi_n}$.\footnote{With the interaction of Eq.\,(\ref{gk2}), $F_1^{(2)}(-m_{\pi_n}^2) \simeq -\langle\bar q q\rangle^0$, and it is generally nonzero in the chiral limit.}   It is now plain that $\forall\, n\geq 1$ 
\begin{equation}
\lim_{\hat m\to 0} \check{T}_{\pi_n^0}(-m_{\pi_n}^2,Q^2) 
 \stackrel{Q^2\gg \Lambda_{\rm QCD}^2}{=} \frac{4\pi^2}{3}\left. F^{(2)}_{n }(-m_{\pi_n}^2)\frac{\ln^{\gamma} Q^2/\omega_{\pi_n}^2}{Q^4}\right|_{\hat m=0} ;
\label{UVchiralnot0}
\end{equation}
namely, in the chiral limit the leading-order power-law in the transition form factor for excited state pseudoscalar mesons is O$(1/Q^4)$.  This is a novel, nonperturbative and model-independent result.

\begin{figure}[t]
\begin{center}
\includegraphics[clip,width=0.75\textwidth]{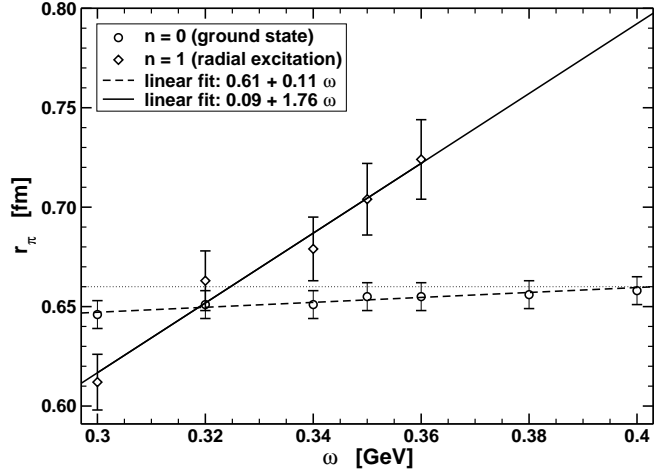}
\caption{\label{fig:emradii} Evolution of ground and first excited state pseudoscalar mesons' electromagnetic charge radii with the scale parameter $\omega$ in Eq.\,(\protect\ref{gk2}): $r_a=1/\omega$ gauges the range of the confining interaction between light quarks.  \textit{Dotted line}: $r_\pi=0.66\,$fm, which indicates the experimental value of the ground state's radius.  The radius is evaluated numerically from the electromagnetic form factor.  That is the primary source of the theory error depicted in the figure, which corresponds to a relative error  $\lesssim 1$\% for $n=0$ and $\lesssim 3$\% for $n=1$.  (Adapted from Ref.\,[\protect\refcite{krassnigg2}].)}
\end{center}
\end{figure}

Reference~[\refcite{krassnigg2}] also provides quantitative illustrations of these and other results relating to the electromagnetic properties of ground- and excited-state pseudoscalar mesons.  For example, the two photon width of the $\pi_1$-meson is estimated
\begin{equation}
\label{Gpiggbest}
\Gamma_{\pi_1\gamma\gamma} \simeq 240\,{\rm eV} \simeq 30\,\Gamma_{\pi_0\gamma\gamma}\,,
\end{equation} 
as is the electromagnetic radius of the charge states
\begin{equation}
r_{\pi_1} \simeq 1.4 \,r_{\pi_0}\,.
\end{equation}

These results were obtained with the model rainbow-ladder interaction described below, in Sec.\,\ref{practical}.  However, the calculation of the latter was used to verify that the properties of excited states are indeed a sensitive probe of the interaction between light-quarks at long-range.  This is shown in Fig.\,\ref{fig:emradii}.  As with so many properties of the ground state Goldstone mode, its radius is protected against rapid evolution by DCSB.  However, the charge radius of the first excited state changes swiftly with increasing $\omega$, with the ratio $r_{\pi_1}/r_{\pi_0}$ varying from $0.9$ -- $1.2$.  This outcome can readily be interpreted.  The length-scale $r_a := 1/\omega$ is a measure of the range of strong attraction: magnifying $r_a$ increases the active range of the confining piece of the interaction.  This effectively strengthens the confinement force and that compresses the bound state, as one observes in Fig.\,\ref{fig:emradii}: $r_{\pi_1}$ decreases quickly with decreasing $\omega$  (increasing $r_a$).  While it is natural to suppose $r_{\pi_1}>r_{\pi_0}$; namely, that a radial excitation is larger than the associated ground state, the calculations of Ref.\,[\refcite{krassnigg2}] illustrated that with the ground state pseudoscalar meson's properties constrained by Goldstone's theorem and its pointwise consequences, Eqs.\,(\ref{bwti}) -- (\ref{hwti}), it is possible in quantum field theory for a confining interaction to compress the excited state with the consequence that $r_{\pi_1}<r_{\pi_0}$.  

A related result for the evolution of the mass was observed in Ref.\,[\refcite{krassnigg1}]; namely, the mass of the first excited state dropped rapidly with increasing $r_a$.  (However, in this case DCSB guarantees $m_{\pi_1}>m_{\pi_0}$.)  On the $\omega$-domain illustrated in Fig.\,\ref{fig:emradii}, the mass of the ground state obtained with nonzero current-quark mass varied by only 3\% while that of the first excited state changed by 14\%.  It is natural to expect that an increase in the strength of the confinement force should increase the magnitude of the binding energy and hence reduce the mass.  That is precisely what occurs.    

\subsubsection{Practical, predictive tool}
\label{practical}
We have stated that there is a systematic, nonperturbative symmetry-preserving truncation of the DSEs.  The leading-order is provided by the renormalisation-group-improved rainbow-ladder truncation, which has been used widely; e.g., Refs.\,[\refcite{jainmunczek,mr97,klabucar}], and references thereto.  A practical renormalisation-group-improved rainbow-ladder truncation preserves the one-loop ultraviolet behaviour of perturbative QCD.  However, a model assumption is required for the behaviour of the kernel in the infrared; viz., on the domain $Q^2 \lsim 1\,$GeV$^2$, which corresponds to length-scales $\gsim 0.2\,$fm.  This is the confinement domain whereupon little is truly known about the interaction between light-quarks.  That information is, after all, what we seek!  

The systematic application of a single model to an extensive range of JLab-related phenomena is pursued in Refs.\,[\refcite{marisphotonvertex,maristandy4,maristandy1,maristandy3,marisji,%
marisbicudo,maristandy5,mariscotanch,wright05}].  The model interaction is expressed via
\begin{equation}
\frac{\check{G}(Q^2)}{Q^2} = \frac{4\pi^2}{\omega^6} D\, Q^2 {\rm
e}^{-Q^2/\omega^2} + \, \frac{ 8\pi^2\, \gamma_m } { \ln\left[\tau + \left(1 +
Q^2/\Lambda_{\rm QCD}^2\right)^2\right]} \, {\check F}(Q^2) \,, \label{gk2}
\end{equation}
wherein ${\check F}(Q^2)= [1 - \exp(-Q^2/[4 m_t^2])]/Q^2$, $m_t$ $=$ $0.5\,$GeV; $\tau={\rm e}^2-1$; $\gamma_m = 12/25$; and $\Lambda_{\rm QCD} = \Lambda^{(4)}_{\overline{\rm MS}}=0.234\,$GeV. (NB.\ Eq.\,(\protect\ref{gk2}) gives $\alpha(m_Z^2)= 0.126$.)  This simple form represents the interaction strength as a sum of two terms: the second ensures that perturbative behaviour is preserved at short-range; and the first provides for the possibility of enhancement in $K$ and the gap equation's kernel at long-range.\cite{bhagwat}  The true parameters in Eq.\,(\ref{gk2}) are $D$ and $\omega$, which together determine the integrated infrared strength of the rainbow-ladder kernel; i.e., the so-called interaction tension,\cite{cdrwien} $\sigma^\Delta$.  However, we emphasise that they are not independent:\cite{maristandy1} in fitting to a selection of observables, a change in one is compensated by altering the other; e.g., on the domain $\omega\in[0.3,0.5]\,$GeV, the fitted observables are approximately constant along the trajectory\cite{raya3}
\begin{equation}
\label{omegaD}
\omega \,D = (0.72\,{\rm GeV})^3 =: m_g^3.
\end{equation}
This correlation: a reduction in $D$ compensating an increase in $\omega$, acts to keep a fixed value of the interaction tension.  Equation (\ref{gk2}) is thus a one-parameter model.  NB.\ As we saw in connection with Fig.\,\ref{gluoncf}, this value of $m_g$ is typical of the mass-scale associated with nonperturbative gluon dynamics.

\begin{figure}[t]
\centerline{%
\includegraphics[clip,width=0.75\textwidth]{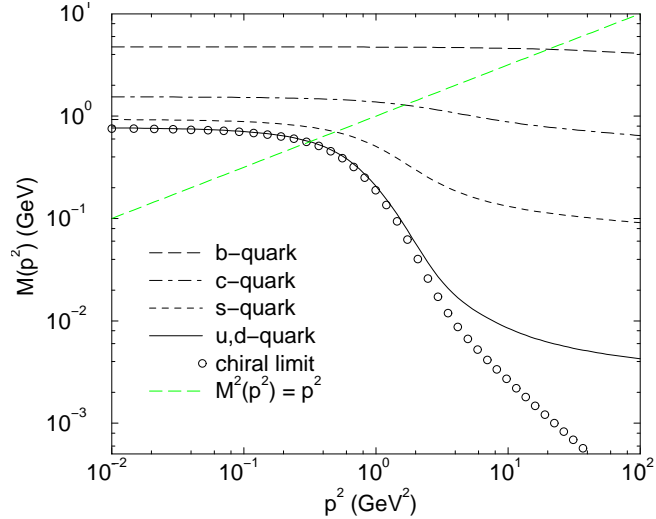}
}
\caption{\label{Mpsqplot}  Quark mass function obtained by solving the complex formed by Eqs.\,(\protect\ref{renormS}), (\protect\ref{gk2}), (\protect\ref{rainbowdse}), (\protect\ref{rainbowmasses}). NB.\ A nonzero solution is obtained in the chiral limit.  This solution is power law suppressed in the ultraviolet\,\protect\cite{lane,politzer} and essentially nonperturbative.  Thus when the $u,d$-quark solution melds with the chiral limit solution one has entered the domain on which all effects are primarily nonperturbative.  The transition takes place for $m_g \lsim s \lsim 3\,m_g$. (Adapted from Ref.\,[\protect\refcite{mishaSVY}].)}
\end{figure}

With the interaction in Eq.\,(\ref{gk2}), one obtains the rainbow-truncation gap equation:
\begin{equation}
S(p)^{-1} = Z_2 \,(i\gamma\cdot p + m^{\rm bare}) + \int^\Lambda_q\! {\check G}((p-q)^2) D_{\mu\nu}^{\rm free}(p-q) \frac{\lambda^a}{2}\gamma_\mu S(q) \frac{\lambda^a}{2}\gamma_\nu , \label{rainbowdse} 
\end{equation} 
wherein $D_{\mu\nu}^{\rm free}(\ell)$ is the free gauge boson propagator; namely, the Euclidean space version of Eq.\,(\ref{photonfree}).  The self-consistent solution obtained with current-quark masses (in GeV):
\begin{equation}
\label{rainbowmasses}
\begin{array}{l|c cccc}
f ~&~ \mbox{chiral} & u,d & s & c & b \\\hline
m_f(\zeta) ~&~ 0.0 & 0.0037 & 0.082 & 0.97 & 4.1\\
\end{array}
\end{equation}
with $\zeta = 19\,$GeV, is depicted in Fig.\,\ref{Mpsqplot}.  

The mass function is a renormalisation group invariant and can be used to define a Euclidean constituent-quark mass; viz.,\footnote{The model produces dressed-quarks, which are confined in the sense that their propagator does not exhibit a true Minkowski space mass pole; i.e., there is no solution of \mbox{$s + M(s)^2 = 0$}.}
\begin{equation}
\label{CQM}
(M^E)^2 := s \,|\, s = M(s)^2 .
\end{equation}
For the solutions depicted in Fig.\,\ref{Mpsqplot}, one finds (in GeV)
\begin{equation}
\label{rainbowCQmasses}
\begin{array}{l|ccccc}
f ~&~ \mbox{chiral} & u,d & s & c & b \\\hline
\rule{0em}{2.5ex} M^E_f ~&~ 0.42 & 0.42 & 0.56 & 1.57 & 4.68
\end{array}
\end{equation}
The ratio $\check L_f:=M^E_f/m_f(\zeta)$ is a measure of the impact of the DCSB mechanism on a particular flavour of quark.  A comparison between Eqs.\,(\ref{rainbowmasses}) and (\ref{rainbowCQmasses}) shows that for quark flavours with $\hat m_f \ll m_g$ the effect of DCSB on their propagation characteristics is very great $\forall \,s\lsim m_g$.  The domain on which the impact is large diminishes rapidly as the current-quark mass increases.

\begin{figure}[t]
\centerline{\includegraphics[clip,width=0.75\textwidth]{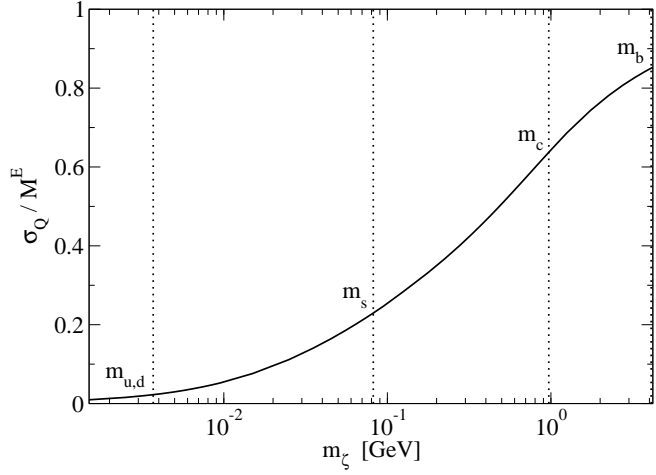}}
\caption{\label{fig:quark_ratio} Ratio $\sigma_Q/M^E_Q$ [Eqs.\,(\protect\ref{CQM}) \& (\protect\ref{sMEQ})] as a function of current-quark mass.  It is a measure of the current-quark-mass-dependence of dynamical chiral symmetry breaking: a zero value indicates complete dominance of dynamical over explicit chiral symmetry breaking; and a value of one, the opposite.  The vertical dotted lines correspond to the $u=d$, $s$, $c$ and $b$ current-quark masses listed in Eq.\,\protect\ref{rainbowmasses}. (Adapted from Ref.\,[\protect\refcite{wright05}].)}
\end{figure}

Another way of looking at this is via a constituent-quark $\sigma$-term:\cite{wright05,sigmaterms}
\begin{equation}
\label{sMEQ}
\sigma_f := m_f(\zeta) \, \frac{\partial  M_f^E}{\partial m_f(\zeta)}\,.
\end{equation}
This renormalisation-group-invariant can be determined from solutions of the gap equation, Eq.\,(\ref{rainbowdse}), and is a keen probe of the impact of explicit chiral symmetry breaking on the mass function.  The ratio $\sigma_{f}/M^E_f$ plotted in Fig.\,\ref{fig:quark_ratio} therefore measures the effect of explicit chiral symmetry breaking on the dressed-quark mass-function compared with the sum of the effects of explicit and dynamical chiral symmetry breaking.  It is evident in the figure that this ratio vanishes for light-quarks because the magnitude of their constituent-mass owes primarily to dynamical chiral symmetry breaking, while for heavy-quarks it approaches one.  A useful illustration of the chiral limit behaviour is provided by the model in Sec.\,\ref{confinement}, which yields algebraic results in the neighbourhood of the chiral limit:\cite{changlei} $M^E_0=\mbox{\small $\frac{1}{\surd 2}$}\, \check G$, $\sigma_0 = \mbox{\small $\frac{3}{2}$}\,m$, and hence $\mbox{\small $\frac{\sigma_{0}}{M^E_0}$} = \mbox{\small $\frac{3}{\surd 2}$} \mbox{\small $\frac{m}{\check G}$}$.  With the parameter values in Ref.\,[\refcite{mn83}], this gives an estimate $\mbox{\small $\frac{\sigma_{0}}{M^E_0}$} \approx 0.04$ for $u=d$ quarks.

A value for the vacuum quark condensate, Eq.\,(\ref{qbq0}), may be obtained from the chiral limit solution depicted in Fig.\,\ref{Mpsqplot}
\begin{equation}
 -\langle \bar q q \rangle_\zeta^0 =
\lim_{\Lambda\to \infty} Z_4(\zeta,\Lambda)\, N_c\, {\rm tr}_{\rm D} \int^\Lambda_q  S_{\hat m =0}(q) 
=(0.275\,{\rm GeV})^3\,.
\end{equation}
Since QCD is multiplicatively renormalisable, then
\begin{equation}
\langle \bar q q \rangle_{\zeta^\prime}^0 = Z_m(\zeta^\prime,\zeta) \, \langle \bar q q \rangle_\zeta^0\,.
\end{equation}
Applying one-loop evolution to define the vacuum condensate at a typical hadronic scale, one therefore obtains\cite{mr97}
\begin{equation}
\label{qbq1}
- \langle \bar q q \rangle_{\zeta=1\,{\rm GeV}}^0 = (0.24\,{\rm GeV})^3.
\end{equation}
This condensate might be interpreted as measuring the density of quark-antiquark pairs in the vacuum of chiral QCD, in which connection the result in Eq.\,(\ref{qbq1}) corresponds to $\rho_{\bar q q} = 1.8\,{\rm fm}^{-3}$.  Were we to assume that this vacuum could be viewed as a medium of close-packed spheres, then each would occupy a volume $V_{\bar q q} = 0.55\,{\rm fm}^3$, which corresponds to a radius $r_{\bar q q} = 0.51\,$fm.  For comparison, the measured electromagnetic charge radius of a pion can thus be written $r_\pi= 0.77 r_{\bar q q}$ and that of a proton, $r_p = 0.58\, r_{\bar q q}$.  It is therefore clear that a veracious understanding of the length-scale defined by the chiral limit vacuum quark condensate is a keystone for bridging the gap between theory and experiment. 

The renormalisation-group-improved rainbow-ladder truncation obtained with Eq.\,(\ref{gk2}) has been employed very successfully.  For example, it predicted\,\cite{marisphotonvertex,maristandy3} the electromagnetic pion form factor measured\,\cite{Volmer2000} at JLab,\footnote{The result is perhaps too good given that pion-loop contributions were omitted.  Such effects, which appear beyond rainbow-ladder truncation, are important at small momentum transfer.  For example, they add $\lsim 15$\% to the rainbow-ladder result for $r_\pi$.\protect\cite{benderpion}} and over an illustrative basket of thirty-one calculated quantities, tabulated in Ref.\,[\refcite{pieterrev}], its agreement with data has an average relative error of $1.6$\% and standard-deviation of $15$\%.

\begin{figure}[t]
\centerline{%
\includegraphics[clip,width=0.80\textwidth]{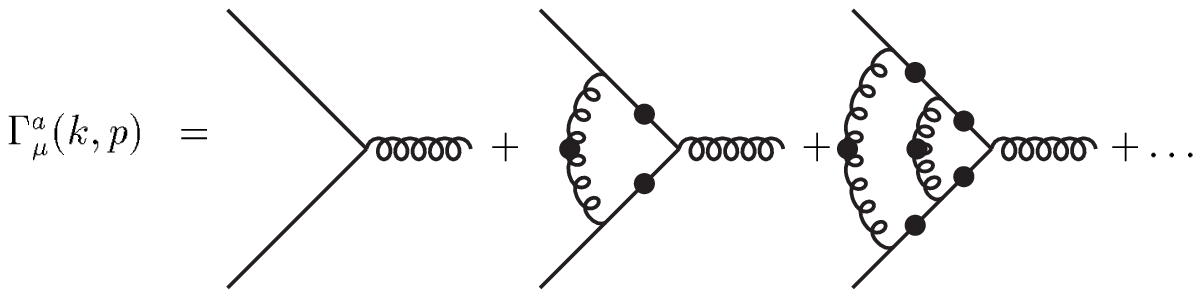}}
\caption{\label{detmoldvertex}  Integral equation for a planar dressed-quark-gluon vertex.  The ``springs'' between dressed-quark lines indicate the dressed-gluon propagator.  The first diagram depicts the rainbow-truncation; the second adds a single gluon rung; etc.  While this series apparently neglects the three-gluon vertex, its effect has satisfactorily been modelled via the \textit{Ansatz} $g^2 \to -\,\check{C} g^2$.\protect\cite{bhagwatvertex}  The change of sign is an important consequence of the non-Abelian nature of QCD.  (Adapted from Ref.\,[\protect\refcite{detmold}].)}
\end{figure}

\subsubsection{Beyond rainbow-ladder}
\label{beyond}
\begin{figure}[t]
\centerline{%
\includegraphics[width=0.80\textwidth]{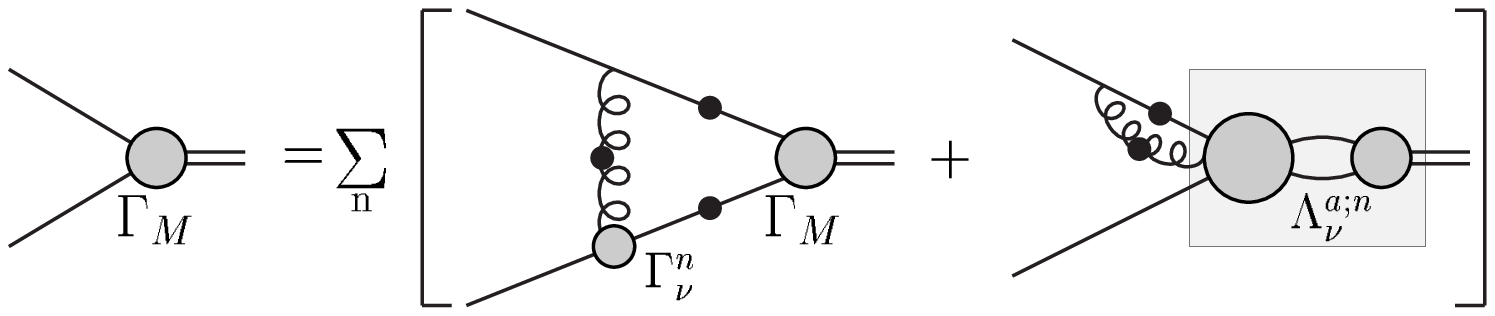}}
\centerline{%
\includegraphics[width=0.80\textwidth]{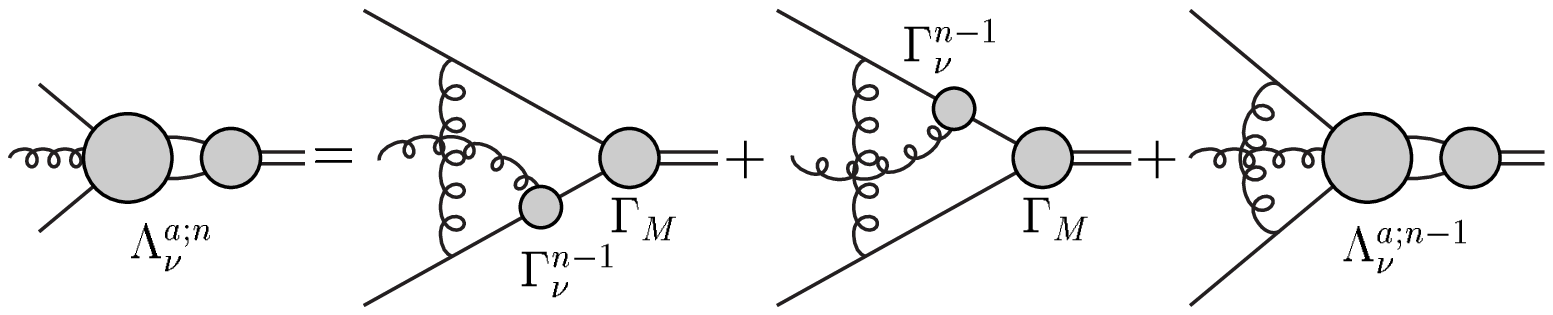}}
\caption{\label{detmoldkernel}  Implicit definition of the integral equation for a symmetry preserving Bethe-Salpeter kernel that is consistent with the vertex depicted in Fig.\,\protect\ref{detmoldvertex}.  Beyond rainbow-truncation, symmetry preservation requires that this kernel be nonplanar even though the vertex in the gap equation is planar.  (Adapted from Ref.\,[\protect\refcite{detmold}].)}
\end{figure}

We have stated that rainbow-ladder is the leading order in a nonperturbative, systematic and symmetry preserving truncation of the DSEs.  This is explained and illustrated well in Refs.\,[\refcite{bhagwatvertex,truncscheme,detmold}].  Figure \ref{detmoldvertex} depicts a natural extension of the vertex \textit{Ansatz} in Eq.\,(\ref{Gnjl}).  It modifies the gap equation's kernel.  In order to preserve the axial-vector Ward-Takahashi identity, Eq.\,(\ref{avwtim}), the Bethe-Salpeter kernel, $K$, must also be modified.  That is systematically accomplished for this vertex via the procedure indicated in Fig.\,\ref{detmoldkernel}, as detailed in Refs.\,[\refcite{bhagwatvertex,detmold}]. 

The model interaction of Eq.\,({\ref{mnprop}}) provides an excellent tool with which to illustrate the procedure.  When used for the ``spring'' in the diagrams of Figs.\,\ref{detmoldvertex}, \ref{detmoldkernel}, it gives the results in Table~\ref{pirhores}.  The first striking observation is that $m_\pi=0$ in the chiral limit for every value of $n$.  Plainly, the truncation procedure preserves the Ward-Takahashi identity, Eq.\,(\ref{avwtim}).  In addition, all of the $\rho$-$\pi$ mass splitting is present in the chiral limit.  This answers the questions raised on page~\pageref{eightfold} and in connection with Table~\ref{tableSpectrum}: the remarkably large difference between the $\pi$ and $\rho$ masses owes to DCSB, which forces $m_\pi$ to be unnaturally small.  It is also apparent that $m_\pi$ is not very sensitive to the order of the truncation.  This is another corollary of DCSB for the Goldstone mode; namely, the cancellations which ensure $m_\pi = 0$ in the chiral limit are still quite effective for small nonzero current-quark masses.  Finally, $m_\rho$ does exhibit some sensitivity to the order of the truncation.  The rainbow-ladder truncation is accurate to $20$\%; the one-loop correction, to $13$\%; and the two-loop result, to $4$\%.  At this point one has become very sensitive to details of the model and is thus in a position to make a quantitatively accurate map of the interaction between light-quarks at long-range via comparison with experiment.  These results explain the level of accuracy attained with a rainbow-ladder truncation based on Eq.\,(\ref{gk2}).

\begin{table}[t]
\tbl{Calculated $\pi$- and $\rho$-meson masses, in GeV, quoted with $\check{G}=0.65\,$GeV in which case $m=0.016 \check{G} = 10\,$MeV.  $n$ is the number of interaction rungs retained in dressing the quark-gluon vertex, see Fig.\,(\protect\ref{detmoldvertex}), and hence the order of the vertex-consistent
Bethe-Salpeter kernel.  NB.\ $n=0$ corresponds to the rainbow-ladder truncation, in which case $m_\rho = \sqrt{2}\, \check{G}$ for $m=0$.}
{\normalsize
\begin{tabular*}
{\hsize} {l@{\extracolsep{0ptplus1fil}}
|c@{\extracolsep{0ptplus1fil}}c@{\extracolsep{0ptplus1fil}}
c@{\extracolsep{0ptplus1fil}}c@{\extracolsep{0ptplus1fil}}}
%
 & $M_H^{n=0}$ & $M_H^{n=1}$ & $M_H^{n=2}$ & $M_H^{n=\infty}$\\\hline
$\pi$, $m=0$ & 0 & 0 & 0 & 0\\
$\pi$, $m=0.011$ & 0.147 & 0.135 & 0.139 & 0.138\\\hline
$\rho$, $m=0$ & 0.920 & 0.648 & 0.782 & 0.754\\
$\rho$, $m=0.011$ & 0.936 & 0.667 & 0.798 &0.770
\end{tabular*}\label{pirhores}}
\end{table}

\begin{figure}[t] 
\centerline{\includegraphics[clip,width=0.75\textwidth]{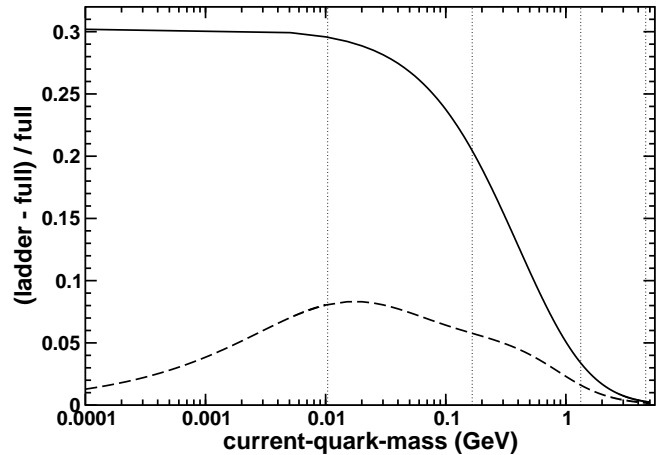}}
\caption{\label{ladderfull} Evolution with current-quark mass of the relative difference between the meson mass calculated in the rainbow-ladder truncation and the exact value.  Solid lines: vector meson trajectories; and dashed-lines; pseudoscalar meson trajectories.  The dotted vertical lines mark the $u=d$, $s$, $c$ and $b$ current-quark masses. (Adapted from Ref.\,[\protect\refcite{bhagwatvertex}].}
\end{figure}

Reference [\protect\refcite{bhagwatvertex}] provides a discussion of the transition between the light-quark and heavy-quark sectors, and the relative strength of the corrections to rainbow-ladder truncation as this transition is made.  Herein we only illustrate the results with Fig.\,\ref{ladderfull}, from which it is apparent that with increasing current-quark mass the contributions from nonplanar diagrams and vertex corrections are suppressed.   Naturally, they must still be included in precision spectroscopic calculations.  As usual, for small current-quark masses, owing to the effects of DCSB, the pseudoscalar channel is a little different.  However, the trend in this channel becomes the same as that in the vector channel for current-quark masses above $\sim 2\,m_s$.  

A contemporary challenge is to generalise the procedure described in this section to mesons composed of constituents with different current-quark masses.  Trajectories for such states are presented in Ref.\,[\protect\refcite{mariscairns}] but only on a limited mass domain.  It is important to expand this domain, e.g., in order to increase the capacity for description and prediction, and to make further contact with results that are exact in the limit of one infinitely heavy constituent.

\section{Baryon Properties}
\label{baryon}
\setcounter{equation}{0}
Significant progress has been made with the study of mesons.  While that is good, it does not directly impact on the important challenge of baryons.  Mesons fall within the class of two-body problems.  They are the simplest bound states for theory.  However, the absence of meson targets poses significant difficulties for the experimental verification of predictions such as those reported above.  On the other hand, it is relatively straightforward to construct a proton target but, as a three-body problem in relativistic quantum field theory, here the difficulty is for theory.  With this problem the current expertise is approximately at the level it was for mesons ten years ago; namely, model building and phenomenology, making as much use as possible of the results and constraints outlined above.  

A natural primary aim is to provide a true picture of the proton's electromagnetic form factors and therefrom a reliable determination of the ratio depicted in Fig.\,\ref{gepgmpdata}.  On the domain of momentum transfer for which there is apparently a discrepancy between the experimental results; namely, $Q^2 > M^2$, where $M$ is the nucleon's mass, a veracious understanding of these and other contemporary data require a Poincar\'e covariant description of the nucleon.  This is apparent in applications of relativistic quantum mechanics; e.g., Refs.~[\refcite{millerfrank,boffi,fuda,stan,bruno}].  A different tack follows the formulation of a Poincar\'e covariant Faddeev equation\,\cite{regfe,hugofe}.  Its foundation is understood through the observation that the same interaction which describes colour-singlet mesons also generates quark-quark (diquark) correlations in the colour-$\bar 3$ (antitriplet) channel\,\cite{regdq}.  While diquarks do not survive as asymptotic states;\cite{bhagwatvertex,truncscheme,detmold,njldiquark} i.e., they do not appear in the strong interaction spectrum, the attraction between quarks in this channel grounds a picture of baryons in which two quarks are always correlated as a colour-$\bar 3$ diquark pseudoparticle, and binding is effected by the iterated exchange of roles between the bystander and diquark-participant quarks. 

A first numerical study of this Faddeev equation for the nucleon was reported 
in Ref.\,[\refcite{cjbfe}], and following that there have been numerous more 
extensive analyses; e.g., Refs.~[\refcite{bentzfe,oettelfe,hechtfe,birsefe}].  It has become apparent that the dominant correlations for ground state octet and decuplet baryons are scalar and axial-vector diquarks, primarily because the associated mass-scales are smaller than the masses of these baryons\,\cite{cjbsep,marisdq} and the positive parity of the correlations matches that of the baryons.  Both scalar and axial-vector diquarks provide attraction in the Faddeev equation; e.g., a scalar diquark alone provides for a bound octet baryon and including axial-vector correlations reduces that baryon's mass.

With the retention of axial-vector diquark correlations a quantitative description of baryon properties is attainable.  Indeed, the 
formulation of Ref.\,[\refcite{oettelfe}] employs confined quarks, and confined
pointlike-scalar and -axial-vector diquark correlations, to obtain a spectrum 
of octet and decuplet baryons in which the rms-deviation between the calculated 
mass and experiment is only $2$\%.  The model also reproduces nucleon form 
factors over a large range of momentum transfer\,\cite{oettel2}, and its 
descriptive success in that application is typical of such Poincar\'e covariant 
treatments; e.g., Refs.\,[\refcite{jacquesA,jacquesmyriad,cdrqciv,nedm}]. 
 
However, these successes might themselves indicate a flaw in the application of the Faddeev equation to the nucleon.  For example, in the context of spectroscopy, studies using the Cloudy Bag Model (CBM)\,\cite{tonyCBM} indicate that the dressed-nucleon's mass receives a negative contribution of as much as $300$-$400\,$MeV from pion self-energy corrections; i.e.,\,\cite{tonyANU,bruceCBM} $\delta M_+ = -300\,$ to $-400\,$MeV.  Furthermore, a perturbative study, using the Faddeev equation, of the mass shift induced by pointlike-$\pi$ exchange between the quark and diquark constituents of the nucleon obtains\,\cite{ishii} $\delta M_+ = -150$ to $-300\,$MeV.  Unameliorated these mutually consistent results would much diminish the value of the $2$\% spectroscopic accuracy obtained using only quark and diquark degrees of freedom. 

In addition to masses, pseudoscalar meson loops make important contributions to many other baryon properties; e.g., to charge and magnetic radii, and magnetic moments.\cite{radiiCh,young}  These effects must not be overlooked because the size and qualitative impact of meson contributions provide material constraints on the development of a realistic quark-diquark picture of the nucleon, and its interpretation and application.  

\subsection{Faddeev Equation}
\label{faddeev}
\setcounter{equation}{0}
For quarks in the fundamental representation of colour-$SU(3)$:
\begin{equation} 
3_c \otimes 3_c \otimes 3_c = (\bar 3_c \oplus 6_c) \otimes 3_c = 1_c \oplus 
8_c^\prime \oplus 8_c \oplus 10_c\,,
\end{equation} 
and hence any two quarks in a colour-singlet three-quark bound state must constitute a relative colour-antitriplet.  This fact enables the derivation of a Faddeev equation for the bound state contribution to the three quark scattering kernel\,\cite{regfe} because the same kernel that describes mesons so well\,\cite{pieterrev} is also attractive for quark-quark scattering in the colour-$\bar 3$ channel.  

In this truncation of the three-body problem the interactions between two selected quarks are added to yield a quark-quark scattering matrix, which is then approximated as a sum over all possible diquark pseudoparticle terms \,\cite{cjbfe}: Dirac-scalar $+$ -axial-vector $+ [\ldots]$.  The Faddeev equation thus obtained describes the baryon as a composite of a dressed-quark and nonpointlike diquark with an iterated exchange of roles between the bystander and diquark-participant quarks.  The baryon is consequently represented by a Faddeev amplitude: 
\begin{equation} 
\Psi = \Psi_1 + \Psi_2 + \Psi_3 \,, 
\end{equation} 
where the subscript identifies the bystander quark and, e.g., $\Psi_{1,2}$ are obtained from $\Psi_3$ by a correlated, cyclic permutation of all the quark 
labels.  

The Faddeev equation is simplified further by retaining only the lightest diquark correlations in the representation of the quark-quark scattering matrix.  A simple Goldstone-theorem-preserving, rainbow-ladder DSE-model\,\cite{cjbsep} yields the following diquark pseudoparticle masses (isospin symmetry is assumed): 
\begin{equation} 
\label{dqmass} 
\begin{array}{l|cccc} 
(qq)_{J^P}           & (ud)_{0^+} & (us)_{0^+}  & (uu)_{1^+}& (us)_{1^+}\\ 
 m_{qq}\,({\rm GeV}) & 0.74       & 0.88        & 0.95      & 1.05 \\\hline 
(qq)_{J^P}           & (ss)_{1^+} & (uu)_{1^-} & (us)_{1^-} & (ss)_{1^-}  \\ 
 m_{qq}\,({\rm GeV}) & 1.13 & 1.47 & 1.53 & 1.64 
\end{array} 
\end{equation} 
The mass ordering is characteristic and model-independent (cf.\ Refs.\,[\refcite{marisdq,gunner}], lattice-QCD estimates\,\cite{hess} and 
studies of the spin-flavour dependence of parton distributions\,\cite{Close:br}), and indicates that a study of the $N$ and $\Delta$ must retain at least the scalar and pseudovector $(uu)$- and $(ud)$-correlations if it is to be accurate.  NB.\ The spin-$3/2$ $\Delta$ is inaccessible unless pseudovector correlations are retained.

The simplest realistic representation of the Faddeev amplitude for the spin- and isospin-$1/2$ nucleon is therefore 
\begin{equation} 
\label{Psi} \Psi_3(p_i,\alpha_i,\tau_i) = \check{N}_3^{0^+} + \check{N}_3^{1^+};
\end{equation} 
namely, a sum of scalar and axial-vector diquark correlations, with $(p_i,\alpha_i,\tau_i)$ the momentum, spin and isospin labels of the quarks constituting the bound state, and $P=p_1+p_2+p_3$ the system's total momentum.\footnote{NB.\ Hereafter we assume isospin symmetry of the strong interaction; i.e., the $u$- and $d$-quarks are indistinguishable but for their electric charge.  This simplifies the form of the Faddeev amplitudes.}  For the $\Delta$, since it is not possible to combine an isospin-$0$ diquark with an isospin-$1/2$ quark to obtain isospin-$3/2$, the spin- and isospin-$3/2$ $\Delta$ contains only an axial-vector diquark component
\begin{equation}
\label{PsiD} \Psi^\Delta_3(p_i,\alpha_i,\tau_i) = \check{D}_3^{1^+}.
\end{equation} 

The scalar diquark piece in Eq.\,(\ref{Psi}) is 
\begin{eqnarray} 
\check{N}_3^{0^+}(p_i,\alpha_i,\tau_i)&=& [\Gamma^{0^+}(\sfrac{1}{2}p_{[12]};K)]_{\alpha_1 
\alpha_2}^{\tau_1 \tau_2}\, \Delta^{0^+}(K) \,[\mathbf{S}(\ell;P) u(P)]_{\alpha_3}^{\tau_3}\,,%
\label{calS} 
\end{eqnarray} 
where: the spinor satisfies Eq.\,(\ref{DiracN}), and it is also a spinor in isospin space with $\varphi_+= {\rm col}(1,0)$ for the proton and $\varphi_-= {\rm col}(0,1)$ for the neutron; $K= p_1+p_2=: p_{\{12\}}$, $p_{[12]}= p_1 - p_2$, $\ell := (-p_{\{12\}} + 2 p_3)/3$; $\Delta^{0^+}$ is a pseudoparticle propagator for the scalar diquark formed from quarks $1$ and $2$, and $\Gamma^{0^+}\!$ is a Bethe-Salpeter-like amplitude describing their relative momentum correlation; and ${\mathbf S}$, a $4\times 4$ Dirac matrix, describes the relative quark-diquark momentum correlation.  (${\mathbf
S}$, $\Gamma^{0^+}$ and $\Delta^{0^+}$ are discussed in Sect.\,\ref{completing}.)  The necessary colour antisymmetry of $\Psi_3$ is implicit in $\Gamma^{J^P}\!\!$, with the 
Levi-Civita tensor, $\epsilon_{c_1 c_2 c_3}$, expressed via the antisymmetric Gell-Mann matrices; viz., defining 
\begin{equation} 
\{H^1=i\lambda^7,H^2=-i\lambda^5,H^3=i\lambda^2\}\,, 
\end{equation} 
then $\epsilon_{c_1 c_2 c_3}= (H^{c_3})_{c_1 c_2}$.  [See Eqs.\,(\ref{Gamma0p}), (\ref{Gamma1p}).]

The axial-vector component in Eq.\,(\ref{Psi}) is
\begin{equation} 
\check{N}^{1^+}(p_i,\alpha_i,\tau_i) =  [{\tt t}^i\,\Gamma_\mu^{1^+}(\sfrac{1}{2}p_{[12]};K)]_{\alpha_1 
\alpha_2}^{\tau_1 \tau_2}\,\Delta_{\mu\nu}^{1^+}(K)\, 
[{\mathbf A}^{i}_\nu(\ell;P) u(P)]_{\alpha_3}^{\tau_3}\,,
\label{calA} 
\end{equation} 
where the symmetric isospin-triplet matrices are 
\begin{equation} 
{\tt t}^+ = \frac{1}{\surd 2}(\tau^0+\tau^3) \,,\; 
{\tt t}^0 = \tau^1\,,\; 
{\tt t}^- = \frac{1}{\surd 2}(\tau^0-\tau^3)\,, 
\end{equation} 
and the other elements in Eq.\,(\ref{calA}) are straightforward generalisations of those in Eq.\,(\ref{calS}). 

The general form of the Faddeev amplitude for the spin- and isospin-$3/2$ $\Delta$ is complicated.  However, isospin symmetry means one can focus on the $\Delta^{++}$ with it's simple flavour structure, because all the charge states are degenerate, and consider 
\begin{equation}
{\mathbf D}_3^{1^+}= [{\tt t}^+ \Gamma^{1^+}_\mu(\sfrac{1}{2}p_{[12]};K)]_{\alpha_1 \alpha_2}^{\tau_1 \tau_2}
\, \Delta_{\mu\nu}^{1^+}(K) \, [{\mathbf D}_{\nu\rho}(\ell;P)u_\rho(P)\, \varphi_+]_{\alpha_3}^{\tau_3}\,, \label{DeltaAmpA} 
\end{equation} 
where $u_\rho(P)$ is a Rarita-Schwinger spinor, Eq.\,(\ref{rarita}).

The general forms of the matrices ${\mathbf S}(\ell;P)$, ${\mathbf A}^i_\nu(\ell;P)$ and ${\mathbf D}_{\nu\rho}(\ell;P)$, which describe the momentum space correlation between the quark and diquark in the nucleon and the $\Delta$, respectively, are described in Ref.\,[\refcite{oettelfe}].  The requirement that ${\mathbf S}(\ell;P)$ represent a positive energy nucleon; namely, that it be an eigenfunction of $\Lambda_+(P)$, Eq.\,(\ref{Lplus}), entails
\begin{equation}
\label{Sexp} 
{\mathbf S}(\ell;P) = s_1(\ell;P)\,\mbox{\boldmath $I$}_{\rm D} + \left(i\gamma\cdot \hat\ell - \hat\ell \cdot \hat P\, \mbox{\boldmath $I$}_{\rm D}\right)\,s_2(\ell;P)\,, 
\end{equation} 
where $\hat \ell^2=1$, $\hat P^2= - 1$.  In the nucleon rest frame, $s_{1,2}$ describe, respectively, the upper, lower component of the bound-state nucleon's spinor.  Placing the same constraint on the axial-vector component, one has
\begin{equation}
\label{Aexp}
 {\mathbf A}^i_\nu(\ell;P) = \sum_{n=1}^6 \, p_n^i(\ell;P)\,\gamma_5\,A^n_{\nu}(\ell;P)\,,\; i=+,0,-\,,
\end{equation}
where ($ \hat \ell^\perp_\nu = \hat \ell_\nu + \hat \ell\cdot\hat P\, \hat P_\nu$, $ \gamma^\perp_\nu = \gamma_\nu + \gamma\cdot\hat P\, \hat P_\nu$)
\begin{equation}
\begin{array}{lll}
A^1_\nu= \gamma\cdot \hat \ell^\perp\, \hat P_\nu \,,\; &
A^2_\nu= -i \hat P_\nu \,,\; &
A^3_\nu= \gamma\cdot\hat \ell^\perp\,\hat \ell^\perp\,,\\
A^4_\nu= i \,\hat \ell_\mu^\perp\,,\; &
A^5_\nu= \gamma^\perp_\nu - A^3_\nu \,,\; &
A^6_\nu= i \gamma^\perp_\nu \gamma\cdot\hat \ell^\perp - A^4_\nu\,.
\end{array}
\end{equation}
Finally, requiring also that ${\mathbf D}_{\nu\rho}(\ell;P)$ be an eigenfunction of $\Lambda_+(P)$, one obtains
\begin{equation}
{\mathbf D}_{\nu\rho}(\ell;P) = {\mathbf S}^\Delta(\ell;P) \, \delta_{\nu\rho} + \gamma_5{\mathbf A}_\nu^\Delta(\ell;P) \,\ell^\perp_\rho \,,
\end{equation}
with ${\mathbf S}^\Delta$ and ${\mathbf A}^\Delta_\nu$ given by obvious analogues of Eqs.\,(\ref{Sexp}) and (\ref{Aexp}), respectively.

We are now in a position to write the Faddeev equation satisfied by $\Psi_3$:
\begin{equation} 
 \left[ \begin{array}{r} 
{\mathbf S}(k;P)\, u(P)\\ 
{\mathbf A}^i_\mu(k;P)\, u(P) 
\end{array}\right]\\ 
 = -\,4\,\int\frac{d^4\ell}{(2\pi)^4}\,{\mathbf M}(k,\ell;P) 
\left[ 
\begin{array}{r} 
{\mathbf S}(\ell;P)\, u(P)\\ 
{\mathbf A}^j_\nu(\ell;P)\, u(P) 
\end{array}\right] .
\label{FEone} 
\end{equation} 
The kernel in Eq.~(\ref{FEone}) is 
\begin{equation} 
\label{calM} {\mathbf M}(k,\ell;P) = \left[\begin{array}{cc} 
{\mathbf M}_{00} & ({\mathbf M}_{01})^j_\nu \\ 
({\mathbf M}_{10})^i_\mu & ({\mathbf M}_{11})^{ij}_{\mu\nu}\rule{0mm}{3ex} 
\end{array} 
\right] 
\end{equation} 
with 
\begin{equation} 
\label{calM00}
 {\mathbf M}_{00} = \Gamma^{0^+}\!(k_q-\ell_{qq}/2;\ell_{qq})\, 
S^{\rm T}(\ell_{qq}-k_q) \,\bar\Gamma^{0^+}\!(\ell_q-k_{qq}/2;-k_{qq})\, 
S(\ell_q)\,\Delta^{0^+}(\ell_{qq}) \,, 
\end{equation} 
where: $\ell_q=\ell+P/3$, $k_q=k+P/3$, $\ell_{qq}=-\ell+ 2P/3$, 
$k_{qq}=-k+2P/3$ and the superscript ``T'' denotes matrix transpose; and
\begin{eqnarray}
\nonumber
\lefteqn{({\mathbf M}_{01})^j_\nu= {\tt t}^j \,
\Gamma_\mu^{1^+}\!(k_q-\ell_{qq}/2;\ell_{qq})} \\
&& \times 
S^{\rm T}(\ell_{qq}-k_q)\,\bar\Gamma^{0^+}\!(\ell_q-k_{qq}/2;-k_{qq})\, 
S(\ell_q)\,\Delta^{1^+}_{\mu\nu}(\ell_{qq}) \,, \label{calM01} \\ 
\nonumber \lefteqn{({\mathbf M}_{10})^i_\mu = 
\Gamma^{0^+}\!(k_q-\ell_{qq}/2;\ell_{qq})\, 
}\\ 
&&\times S^{\rm T}(\ell_{qq}-k_q)\,{\tt t}^i\, \bar\Gamma_\mu^{1^+}\!(\ell_q-k_{qq}/2;-k_{qq})\, 
S(\ell_q)\,\Delta^{0^+}(\ell_{qq}) \,,\\ 
\nonumber \lefteqn{({\mathbf M}_{11})^{ij}_{\mu\nu} = {\tt t}^j\, 
\Gamma_\rho^{1^+}\!(k_q-\ell_{qq}/2;\ell_{qq})}\\ 
&&\times \, S^{\rm T}(\ell_{qq}-k_q)\,{\tt t}^i\, \bar\Gamma^{1^+}_\mu\!(\ell_q-k_{qq}/2;-k_{qq})\, 
S(\ell_q)\,\Delta^{1^+}_{\rho\nu}(\ell_{qq}) \,. \label{calM11} 
\end{eqnarray} 

The $\Delta$'s Faddeev equation is
\begin{eqnarray} 
{\mathbf D}_{\lambda\rho}(k;P)\,u_\rho(P) & = & 4\int\frac{d^4\ell}{(2\pi)^4}\,{\mathbf
M}^\Delta_{\lambda\mu}(k,\ell;P) \,{\mathbf D}_{\mu\sigma}(\ell;P)\,u_\sigma(P)\,, \label{FEDelta} 
\end{eqnarray} 
with
\begin{eqnarray}
\nonumber\lefteqn{ {\mathbf M}^\Delta_{\lambda\mu} = {\tt t}^+ 
\Gamma_\sigma^{1^+}\!(k_q-\ell_{qq}/2;\ell_{qq})}\\
&& \times\, 
 S^{\rm T}\!(\ell_{qq}-k_q)\, {\tt t}^+\bar\Gamma^{1^+}_\lambda\!(\ell_q-k_{qq}/2;-k_{qq})\, 
S(\ell_q)\,\Delta^{1^+}_{\sigma\mu}\!(\ell_{qq}). \label{MDelta}
\end{eqnarray}

\begin{figure}[t]
\centerline{%
\includegraphics[clip,width=0.70\textwidth]{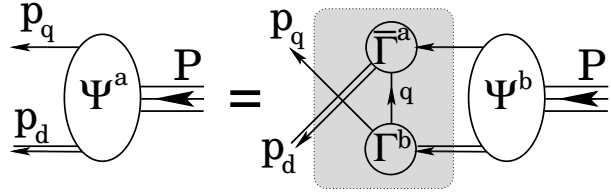}}
\caption{\label{faddeevpic} Pictorial representation of the Faddeev equation, Eq.\,(\ref{FEone}).  A nucleon of four-momentum $P$ is constituted from a dressed-quark (single line, momentum $p_q=k_q$) and dressed-diquark (double line, momentum $p_d=k_{qq}$).  Binding is effected by an iterated exchange of roles between the bystander and diquark-participant quarks, which is described by the kernels $\mathbf M$ in Eqs.\,(\protect\ref{calM00})-(\protect\ref{calM11}), (\protect\ref{MDelta}).  The exchange takes place within the shaded region.}
\end{figure}

The Faddeev equation is illustrated in Fig.\,\ref{faddeevpic}.  It is a linear, homogeneous matrix equation whose solution yields the Poincar\'e covariant Faddeev amplitude, which describes the relative motion between the quark and diquark within the baryon.  Orbital angular momentum is not a Poincar\'e invariant.  However, if absent in a particular frame, it will almost inevitably appear in another frame related via a Poincar\'e transformation.  Nonzero quark orbital angular momentum is the necessary outcome of a Poincar\'e covariant description.  This is why the covariant Faddeev amplitude is a matrix-valued function with a rich structure that, in the baryons' rest frame, corresponds to a relativistic wave function with $s$-wave, $p$-wave and even $d$-wave components. (Details can be found in Ref.\,[\refcite{oettelthesis}], Sec.\,2.4.)  

\subsubsection{Faddeev equation kernels}
\label{completing} 
To complete the Faddeev equations, Eqs.\,(\ref{FEone}) \& (\ref{FEDelta}), one must specify the dressed-quark propagator, the diquark Bethe-Salpeter amplitudes and the diquark propagators that appear in the kernels.
\medskip

\hspace*{-\parindent}\underline{Dressed-quark propagator}. \hspace*{1em} This propagator has the general form given in Eq.\,(\ref{Sgeneral}) and herein we have already provided an overview of its properties.  In solving the Faddeev equation described above one is required to repeatedly evaluate some eight-dimensional integrals.  Thus, while it is straightforward to obtain a numerical solution of a truncation of QCD's gap equation, the utility of an algebraic form for $S(p)$ is self-evident.  An efficacious parametrisation, which exhibits the features described above and has been used extensively in studies of hadron properties, is expressed via
\begin{eqnarray} 
\bar\sigma_S(x) & =&  2\,\bar m \,{\check F}(2 (x+\bar m^2)) + {\check
F}(b_1 x) \,{\check F}(b_3 x) \,  
\left[b_0 + b_2 {\check F}(\epsilon x)\right]\,,\label{ssm} \\ 
\label{svm} \bar\sigma_V(x) & = & \frac{1}{x+\bar m^2}\, \left[ 1 - {\check F}(2 (x+\bar m^2))\right]\,, 
\end{eqnarray}
with $x=p^2/\lambda^2$, $\bar m$ = $m/\lambda$, ${\check F}(x)= (1-\mbox{\rm e}^{-x})/x$, $\bar\sigma_S(x) = \lambda\,\sigma_S(p^2)$ and $\bar\sigma_V(x) =
\lambda^2\,\sigma_V(p^2)$.  The mass-scale, $\lambda=0.566\,$GeV, and
parameter values\footnote{$\epsilon=10^{-4}$ in Eq.\ (\ref{ssm}) serves only to
decouple the large- and intermediate-$p^2$ domains.}
\begin{equation} 
\label{tableA} 
\begin{array}{ccccc} 
   \bar m& b_0 & b_1 & b_2 & b_3 \\\hline 
   0.00897 & 0.131 & 2.90 & 0.603 & 0.185 
\end{array}\;, 
\end{equation} 
were fixed in a least-squares fit to light-meson observables\,\cite{mark,valencedistn}.  The dimensionless $u=d$ current-quark mass in Eq.~(\ref{tableA}) corresponds to
\begin{equation} 
\label{mcq}
m=5.1\,{\rm MeV}\,,
\end{equation} 
and the parametrisation yields a Euclidean constituent-quark mass, Eq.\,(\ref{CQM}),
\begin{equation} 
\label{MEq} M_{u,d}^E = 0.33\,{\rm GeV},
\end{equation}
which agrees semiquantitatively with that produced by a renormalisation-group-improved rainbow-ladder truncation based on Eq.\,(\ref{gk2}) [see Eq.\,(\ref{rainbowCQmasses})].
\smallskip

\hspace*{-\parindent}\underline{Diquark Bethe-Salpeter amplitudes}. \hspace*{1em} The rainbow-ladder DSE truncation yields asymptotic diquark states in the strong interaction spectrum.  Such states are not observed and their appearance is an artefact of the truncation.  Higher order terms in the quark-quark scattering kernel, whose analogue in the quark-antiquark channel do not much affect the properties of vector and flavour non-singlet pseudoscalar mesons, ensure that QCD's quark-quark scattering matrix does not exhibit singularities which correspond to asymptotic diquark states.\cite{bhagwatvertex,truncscheme,detmold,njldiquark}  Nevertheless, studies with kernels that do not produce diquark bound states, do support a physical interpretation of the masses, $m_{(qq)_{J^P}}$, obtained using the rainbow-ladder truncation: the quantity $l_{(qq)_{J^P}}=1/m_{(qq)_{J^P}}$ may be interpreted as a range over which the diquark correlation can persist inside a baryon.  These observations motivate the {\it Ansatz} for the quark-quark scattering matrix that is employed in deriving the Faddeev equation: 
\begin{equation} 
[G_{qq}(k,q;K)]_{rs}^{tu} = \sum_{J^P=0^+,1^+,\ldots} \bar\Gamma^{J^P}\!(k;-K)\, \Delta^{J^P}\!(K) \, \Gamma^{J^P}\!(q;K)\,. \label{AnsatzMqq} 
\end{equation}  

One practical means of specifying the $\Gamma^{J^P}\!\!$ in Eq.\,(\ref{AnsatzMqq}) is to employ the solutions of a rainbow-ladder quark-quark BSE.  Using the properties of the Gell-Mann matrices one finds easily that $\Gamma^{J^P}_C:= \Gamma^{J^P}C^\dagger$ satisfies exactly the same equation as the $J^{-P}$ colour-singlet meson {\it but} for a halving of the coupling strength\,\cite{regdq}.  This makes clear that the interaction in the ${\bar 3_c}$ $(qq)$ channel is strong and attractive.\footnote{The same analysis shows the interaction to be strong and repulsive in the ${6_c}$ $(qq)$ channel.}  Moreover, it follows as a feature of the rainbow-ladder truncation that, independent of the specific form of a model's interaction, the calculated masses satisfy 
\begin{equation}
m_{(qq)_{J^P}} > m_{(\bar q q)_{J^{-P}}}\,.
\end{equation}
This is a useful guide for all but scalar diquark correlations because the partnered mesons in that case are pseudoscalars whose ground state masses, as we have seen, are constrained to be small by Goldstone's theorem and which therefore provide a weak lower bound.  For the correlations relevant herein, models typically give masses (in GeV, recall Eq.\,(\ref{dqmass}) and the associated discussion):
\begin{equation}
\label{diquarkmass}
m_{(ud)_{0^+}} = 0.74 - 0.82 \,,\; m_{(uu)_{1^+}}=m_{(ud)_{1^+}}=m_{(dd)_{1^+}}=0.95 - 1.02\,.
\end{equation}

A solution of the BSE equation requires a simultaneous solution of the quark-DSE.\cite{marisdq}  However, since in the model we're describing the calculations are simplified by parametrising $S(p)$, that expedient should also be employed with $\Gamma^{J^P}\!$:
\begin{eqnarray} 
\label{Gamma0p} \Gamma^{0^+}(k;K) &=& \frac{1}{{\mathbf N}^{0^+}} \, 
H^a\,C i\gamma_5\, i\tau_2\, {\check F}(k^2/\omega_{0^+}^2) \,, \\ 
\label{Gamma1p} {\tt t}^i \Gamma^{1^+}_\mu (k;K) &=& \frac{1}{{\mathbf N}^{1^+}}\, 
H^a\,i\gamma_\mu C\,{\tt t}^i\, {\check F}(k^2/\omega_{1^+}^2)\,, 
\end{eqnarray} 
with the normalisation, ${\mathbf N}^{J^P}\!$, fixed by
\begin{eqnarray}
\label{BSEnorm} 
2 \,K_\mu & = & 
\left[ \frac{\partial}{\partial Q_\mu} \Pi(K,Q) \right]_{Q=K}^{{K^2=-m_{J^P}^2}},
\end{eqnarray}
where
\begin{equation}
\Pi(K,Q) = tr\!\! \int\!\! 
\frac{d^4 q}{(2\pi)^4}\, \bar\Gamma(q;-K) \, S(q+Q/2) \, \Gamma(q;K) \, S^{\rm T}(-q+Q/2) .
\end{equation}
These {\it Ans\"atze} retain only that single Dirac-amplitude which would represent a point particle with the given quantum numbers in a local Lagrangian density: they are usually the dominant amplitudes in a solution of the rainbow-ladder BSE for the lowest mass $J^P$ mesons\,\cite{marisphotonvertex,mr97,maristandy1} and diquarks.\cite{cjbsep,marisdq}
\smallskip

\hspace*{-\parindent}\underline{Diquark propagators}. \hspace*{1em} Solving for the quark-quark scattering matrix using the rainbow-ladder truncation yields free particle propagators for $\Delta^{J^P}$ in Eq.\,(\ref{AnsatzMqq}).  Higher order contributions remedy that defect, eliminating asymptotic diquark states from the spectrum.  The attendant modification of $\Delta^{J^P}$ can be modelled efficiently by simple functions that are free-particle-like at spacelike momenta but pole-free on the timelike axis (see the discussion of confinement on page~\pageref{confpage}); namely, 
\begin{eqnarray} 
\Delta^{0^+}(K) & = & \frac{1}{m_{0^+}^2}\,{\check F}(K^2/\omega_{0^+}^2)\,,\\ 
\Delta^{1^+}_{\mu\nu}(K) & = & 
\left(\delta_{\mu\nu} + \frac{K_\mu K_\nu}{m_{1^+}^2}\right) \, \frac{1}{m_{1^+}^2}\, {\check F}(K^2/\omega_{1^+}^2) \,,
\end{eqnarray} 
where the two parameters $m_{J^P}$ are diquark pseudoparticle masses and 
$\omega_{J^P}$ are widths characterising $\Gamma^{J^P}\!$.  It is useful to require additionally that
\begin{equation}
\label{DQPropConstr}
\left. \frac{d}{d K^2}\,\left(\frac{1}{m_{J^P}^2}\,{\check F}(K^2/\omega_{J^P}^2)\right)^{-1} \right|_{K^2=0}\! = 1 \; \Rightarrow \; \omega_{J^P}^2 = \sfrac{1}{2}\,m_{J^P}^2\,,
\end{equation} 
which is a normalisation that accentuates the free-particle-like propagation characteristics of the diquarks {\it within} the hadron. 

\subsection{Nucleon and \mbox{\boldmath $\Delta$} Masses}
\label{NDmasses}
The Faddeev equations, Eqs.\,(\ref{FEone}) \& (\ref{FEDelta}), are now completely specified.  The three-body problem is intrinsically complex and thus it may appear to have been a complicated process.  However, we have merely combined existing information about the one- and two-body sectors of QCD. 

\subsubsection{Meson loops and baryon masses}
\label{loopmass}
Before reporting the solution of the Faddeev equation it is worthwhile to explicate the effect of pseudoscalar meson loops on baryon masses.  A thorough discussion is provided in Ref.\,[\refcite{hechtfe}], from which we draw some excerpts.  One can begin with the leading term in a pion-nucleon chiral Lagrangian:\footnote{That part which describes the pseudoscalar field alone has been neglected. NB.\ For this discussion we return to Minkowski metric for ease of comparison with textbook material.}
\begin{equation} 
\label{LpiNL} \bar N(x)\left[ i \not \!\partial - M  +\frac{g}{2 M} \gamma_5 
\gamma^\mu \,\vec{\tau} \cdot \partial_\mu\vec{\pi}(x) + \ldots \right] N(x)\,. 
\end{equation} 
The rainbow truncation DSE for the nucleon in this theory is
\begin{equation} 
\Sigma(P) = 3i \frac{g^2}{4 M^2} \int \frac{d^4 k}{(2\pi)^4} \, 
 \Delta(k^2,m_\pi^2) \! \not\! k\,\gamma_5\, G(P-k)  \! \not\! k\,\gamma_ 5\,,
\label{AVDSE} 
\end{equation} 
wherein the nucleon propagator 
\begin{eqnarray} 
\lefteqn{G(P) = \frac{1}{\not\! P - M -\Sigma(P)} =:  G^+(P) + G^-(P)}\\
\nonumber & & = \frac{M}{\omega_N(\vec{P})} \left[ \Lambda_+(\vec{P}) 
\frac{1}{P_0 - \omega_N(\vec{P}) + i \varepsilon} + \, \Lambda_-(\vec{P}) \frac{1}{P_0 + \omega_N(\vec{P}) - i 
\varepsilon} \right]\,, \\
\end{eqnarray} 
with $\omega_N^2(\vec{P}) = \vec{P}^2 + M^2$, and where $\Lambda_\pm(\vec{P}) = (\not\!\!\! \tilde{P} \pm M)/(2M)$, $\tilde P=(\omega(\vec{P}),\vec{P})$, are the Minkowski space positive and negative energy projection operators, respectively; and the pion propagator ($\omega_\pi^2(\vec{k}) = \vec{k}^2 + m_\pi^2$)
\begin{eqnarray} 
\lefteqn{ \Delta(k^2,m_\pi^2)  = \frac{1}{k^2 - m_\pi^2 + i \varepsilon}}\\ 
& = & \frac{1}{2\,\omega_\pi(\vec{k})}\left[ 
\frac{1}{k_0 - \omega_\pi(\vec{k}) + i\varepsilon} - 
\frac{1}{k_0 + \omega_\pi(\vec{k}) - i\varepsilon} \right] .\\ 
\end{eqnarray} 

As written, the integral in Eq.\,(\ref{AVDSE}) is divergent.  It must be regularised to give it meaning.  In this case the Poincar\'e invariant Paul-Villars procedure is useful.  It may be effected by modifying the $\pi$-propagator: 
\begin{equation} 
\label{DpiPV} \Delta(k^2,m_\pi^2) \to \bar\Delta_\pi(k^2) 
= \Delta(k^2,m_\pi^2) + \sum_{i=1,2} c_i\, \Delta(k^2,\lambda_i^2)\,, 
\end{equation} 
and then, with 
\begin{equation} 
c_1= - \,\frac{\lambda_2^2-m_\pi^2}{\lambda^2_2-\lambda^2_1}\,,\; 
c_2= \frac{\lambda_1^2-m_\pi^2}{\lambda^2_2-\lambda^2_1} \,,
\end{equation} 
Eq.~(\ref{DpiPV}) yields 
\begin{eqnarray} 
\nonumber \bar\Delta_\pi(k^2) & = & \Delta(k^2,m_\pi^2)\, 
\prod_{i=1,2} \, (\lambda_i^2-m_\pi^2) \, \Delta(k^2,\lambda_i^2)\,,\\ 
&& 
\end{eqnarray} 
in which case the integrals are convergent for any fixed $\lambda_{1,2}$.  Furthermore, for $m_\pi \ll \lambda_1\to \lambda_2 = \lambda$ 
\begin{eqnarray} 
\label{PVtovertex} 
\bar\Delta_\pi(k^2)& = & \Delta(k^2,m_\pi^2) \, \Delta^2(k^2/\lambda^2,1)
\end{eqnarray} 
i.e., the Pauli-Villars regularisation is equivalent to employing a monopole form factor at each $\pi N N$ vertex: $g \to g\, \Delta(k^2/\lambda^2,1)$, where $k$ is the pion's momentum.  Since this procedure modifies the pion propagator it may be interpreted as expressing compositeness of the pion and regularising its off-shell contribution (a related effect is identified in Refs.\,[\refcite{reglaws,rhopipipeter}]) but that interpretation is not unique. 

The contribution to the nucleon's mass from a positive-energy nucleon, $G^+(P)$, in the loop described by Eq.\,(\ref{AVDSE}) is
\begin{equation} 
\delta^A M_+^+  =  -\,\frac{3 g^2}{16 M^2} \int\frac{d^3 k}{(2\pi)^3} \frac{1}{\omega_N} \sum_{i=0,1,2} c_i\, \frac{\lambda_i^2 (\omega_N - M) + 2 \vec{k}^2 
(\omega_{\lambda_i} + \omega_N)} {\omega_{\lambda_i} [ \omega_{\lambda_i} + 
\omega_N - M ]}\, , 
\label{deltaMApp} 
\end{equation} 
with $\omega_N=\omega_N(\vec{k}^2)$, etc.  The connection between this and other mass-shift calculations can be made transparent by writing Eq.\,(\ref{deltaMApp}) in the form 
\begin{equation} 
\delta^A M_+^+ =  -\,6\pi\, \frac{f^2_{NN\pi}}{m_\pi^2} \int\frac{d^3 
k}{(2\pi)^3} \, \frac{\vec{k}^2\,u^2(\vec{k}^2)}{\omega_\pi(\vec{k}^2) [ 
\omega_\pi(\vec{k}^2) + \omega_N(\vec{k}^2) - M]} \,,
\label{deltaMppCBM} 
\end{equation} 
where $f_{NN\pi}^2 = g^2 m_\pi^2/(16 \pi M^2)$ and
\begin{equation} 
\vec{k}^2\, u^2(\vec{k}^2) :=
\frac{\omega_{\lambda_0}}{2\,\omega_N} \,  
[ \omega_{\lambda_0} + \omega_N - M ] 
\sum_{i=0,1,2} c_i\, \frac{\lambda_i^2 (\omega_N - M) + 2 \vec{k}^2 
(\omega_{\lambda_i} + \omega_N)} {\omega_{\lambda_i} [ \omega_{\lambda_i} + 
\omega_N - M ]}\,.  \label{ukdef} 
\end{equation} 
This is useful because for $m_\pi \ll  \lambda_1 \to \lambda_2= \lambda$; i.e., on the domain in which Eq.\,(\ref{PVtovertex}) is valid, one finds algebraically that 
\begin{equation} 
\label{uklimit} 
u(\vec{k}^2) = 1/(1+\vec{k}^2/\lambda^2)\,, 
\end{equation} 
which firmly establishes the qualitative equivalence between Eq.\,(\ref{deltaMApp}) and the calculation, e.g., in Refs.\,[\refcite{tonyCBM,tonyANU}]. 

It is instructive to consider Eq.\,(\ref{deltaMApp}) further.  Suppose that $M$ 
is very much greater than the other scales, then on the domain in which the 
integrand has significant support 
\begin{equation} 
\omega_N(\vec{k}^2) - M \approx \frac{\vec{k}^2}{2 M} 
\end{equation} 
and Eq.\,(\ref{deltaMApp}) yields
\begin{equation} 
\nonumber \delta_A M_+^+ \approx  - \frac{3 g^2}{8 M^2} \int \frac{d^3 k}{(2\pi)^3} \, 
\vec{k}^2  \sum_{i=0,1,2} 
\frac{c_i}{\omega_{\lambda_i}^2\!(\vec{k}^2)} \,. \label{deltaMAppNR} 
\end{equation} 
It follows that
\begin{equation} 
\frac{d^2 \,\delta_A M_+^+}{(d m_\pi^2)^2} \approx -\,\frac{3 g^2}{4 M^2} \, 
\int \frac{d^3 k}{(2\pi)^3} \frac{\vec{k}^2}{\omega_\pi^6(\vec{k}^2)}
=-\, \frac{9}{128 \pi} \frac{g^2}{M^2} \frac{1}{m_\pi}\,.
\end{equation} 
Hence on the domain considered, 
\begin{equation} 
\label{LNA} \delta_A M_+^+ = -\,\frac{3}{32\pi} \frac{g^2}{M^2} m_\pi^3 + 
f^+_{(1)}(\lambda_1,\lambda_2)\,m_\pi^2 + f^+_{(0)}(\lambda_1,\lambda_2)\,, 
\end{equation} 
where, as the derivation makes transparent, $f_{(0,1)}$ are scheme-dependent functions of (only) the regularisation parameters but the first term is regularisation-scheme-independent.  Given that $m_\pi^2 \propto \hat m$, Eq.\,(\ref{gmor}), this first term is nonanalytic in the current-quark mass.  It is the leading nonanalytic contribution, a much touted feature of effective field theory, and its coefficient is fixed by chiral symmetry and the pattern by which that symmetry is dynamically broken.  NB.\ The contribution from $G^-(P)$, which produces the so-called $Z$-diagram, is suppressed by $1/M$.\cite{hechtfe}

While the leading nonanalytic contribution is model-independent, and thus provides a constraint on models that purport to represent QCD, it is not of particular quantitative use in this or related studies.  The pion and nucleon are both of finite size and hence the regularisation parameters $\lambda_{1,2}$, which set a compositeness scale for the $\pi N N$ vertex, must assume soft values; e.g.,\cite{oettel2,jacquesmyriad,tonysoft} $\lsim 600\,$MeV.  Therefore the actual value of the pion-loop contribution to the nucleon's mass is completely determined by the regularisation-scheme-dependent terms.  

This is thoroughly explored in Ref.\,[\refcite{hechtfe}], wherein a self-consistent solution of the nucleon's gap equation shows that the one-loop result is accurate to within $95$\%.  A full consideration leads to the conclusion that the shift in the nucleon's mass owing to the $\pi N$-loop is (in GeV, for $g_A=1.26$): 
\begin{equation} 
\label{shiftF} - \delta M_+ \simeq  ( 0.039 - 0.063 ) \, g_A^2 = (0.061 - 
0.099)\,. 
\end{equation} 
Thus one arrives at a robust result: the $\pi N$-loop reduces the nucleon's mass by $100\,$-$200\,$MeV.  Extant calculations; e.g., Refs.\,[\refcite{tonyCBM,tonyANU}], show that the contribution from the analogous $\pi \Delta$-loop is of the same sign and no greater in magnitude so that the likely total reduction is $200$-$400\,$MeV.  Based on these same calculations one anticipates that the $\Delta$ mass is also reduced by $\pi$ loops but by a smaller amount ($\sim 50\,$-$\,100\,$MeV less).
 
How is that effect to be incorporated into the quark-diquark picture of baryons?  Reference [\refcite{hechtfe}] argued that it may be included by solving the Faddeev equation with target nucleon and $\Delta$ masses that are inflated to allow for the loop corrections.  Namely, that a physically sensible picture of the quark-diquark piece of the nucleon can be obtained if its parameters are chosen not so as to give the experimental masses, but higher values; e.g., $M_N=0.94 + 0.2=1.14\,$GeV, $M_\Delta= 1.232+0.1=1.332\,$GeV. 

\subsubsection{\mbox{\boldmath $N$} and \mbox{\boldmath $\Delta$} masses from the Faddeev equation}
\label{faddeevsolution}
The method described in Ref.\,[\refcite{oettelcomp}] is effective for solving Eqs.\,(\ref{FEone}) \& (\ref{FEDelta}).  Owing to Eq.\,(\ref{DQPropConstr}), the masses of the scalar and axial-vector diquarks are the only variable parameters in the kernels of the Faddeev equations.  It is natural to choose the axial-vector mass so as to obtain a desired mass for the $\Delta$, and set the scalar mass subsequently by requiring a particular nucleon mass.  Two primary parameter sets are presented in Table~\ref{ParaFix}.  Set~A is obtained by requiring a precise fit to the experimental nucleon and $\Delta$ masses, while Set~B was obtained by fitting to nucleon and $\Delta$ masses that are inflated so as to allow for the additional attractive contribution from the pion cloud, as described in Sec.\,\ref{loopmass}.

\begin{table}[t]
\tbl{Mass-scale parameters (in GeV) for the scalar and axial-vector diquark correlations, fixed by fitting nucleon and $\Delta$ masses: Set~A provides a fit to the actual masses; whereas Set~B provides masses that are offset to allow for ``pion cloud'' contributions, Sec.\,\protect\ref{loopmass}.  Also listed is $\omega_{J^{P}}=m_{J^{P}}/\surd 2$, which is the width-parameter in the $(qq)_{J^P}$ Bethe-Salpeter amplitude, Eqs.\,(\protect\ref{Gamma0p}) \& (\protect\ref{Gamma1p}):  its inverse is an indication of the diquark's matter radius.  Sets A$^\ast$ and B$^\ast$ illustrate effects of omitting the axial-vector diquark correlation: the $\Delta$ cannot be formed and $M_N$ is significantly increased.  It is thus plain that the axial-vector diquark provides significant attraction in the Faddeev equation's kernel. (Adapted from Ref.\,[\protect\refcite{sigmaterms}].)}
{\normalsize
\begin{tabular*}{1.0\textwidth}{
l@{\extracolsep{0ptplus1fil}}c@{\extracolsep{0ptplus1fil}}c@{\extracolsep{0ptplus1fil}}
c@{\extracolsep{0ptplus1fil}} c@{\extracolsep{0ptplus1fil}}c@{\extracolsep{0ptplus1fil}}c@{\extracolsep{0ptplus1fil}}}
\hline
set & $M_N$ & $M_{\Delta}$~ & $m_{0^{+}}$ & $m_{1^{+}}$~ &
$\omega_{0^{+}} $ & $\omega_{1^{+}}$ \\
\hline
A & 0.94 & 1.23~ & 0.63 & 0.84~ & 0.44=1/(0.45\,{\rm fm}) & 0.59=1/(0.33\,{\rm fm}) \\
B & 1.18 & 1.33~ & 0.79 & 0.89~ & 0.56=1/(0.35\,{\rm fm}) & 0.63=1/(0.31\,{\rm fm}) \\\hline
A$^\ast$ & 1.15 &  & 0.63 &  & 0.44=1/(0.45\,{\rm fm}) &  \\
B$^\ast$ & 1.46 &  & 0.79 &  & 0.56=1/(0.35\,{\rm fm}) &  \\
\hline
\end{tabular*}\label{ParaFix} }
\end{table}

It is apparent in Table~\ref{ParaFix} that a baryon's mass increases with increasing diquark mass, and the fitted diquark mass-scales are commensurate with the anticipated values, cf.\ Eq.\,(\ref{diquarkmass}), with Set~B in better accord.  If coupling to the axial-vector diquark channel is omitted from Eq.\,(\ref{FEone}), then $M_N^{\rm Set\,A} = 1.15\,$GeV and $M_N^{\rm Set\,B} = 1.46\,$GeV, rows labelled A$^\ast$, B$^\ast$, respectively.  It is thus clear that axial-vector diquark correlations provide significant attraction in the nucleon.  Of course, using the Faddeev equation approach, the $\Delta$ does not exist without axial-vector correlations.  In Set~B the amount of attraction provided by axial-vector correlations must be matched by that provided by the pion cloud.  This highlights the constructive interference between the contribution of these two effects to a baryons' mass.  It is related and noteworthy that $m_{1^+}-m_{0^+}$ is only a reasonable approximation to $M_\Delta - M_N=0.29\,$GeV when pion cloud effects are ignored: Set~A, $m_{1^+}-m_{0^+}=0.21\,$GeV cf.\ Set~B, $m_{1^+}-m_{0^+}=0.10\,$GeV.  Plainly, understanding the $N$-$\Delta$ mass splitting requires more than merely reckoning the mass-scales of constituent degrees of freedom.  It is curious that for Set~B the matter radius, $1/\omega_{J^P}$, of the diquarks is smaller than for Set~A.  One might view this as a contraction in the hadron's quark-core in response to the presence of a longer range pion cloud.  

\subsection{Nucleon Electromagnetic Form Factors}
\label{nucleonform}
\subsubsection{Nucleon-photon vertex}
\label{Ncurrent}
\begin{figure}[t]
\begin{minipage}[t]{\textwidth}
\begin{minipage}[t]{0.45\textwidth}
\leftline{\includegraphics[width=0.90\textwidth]{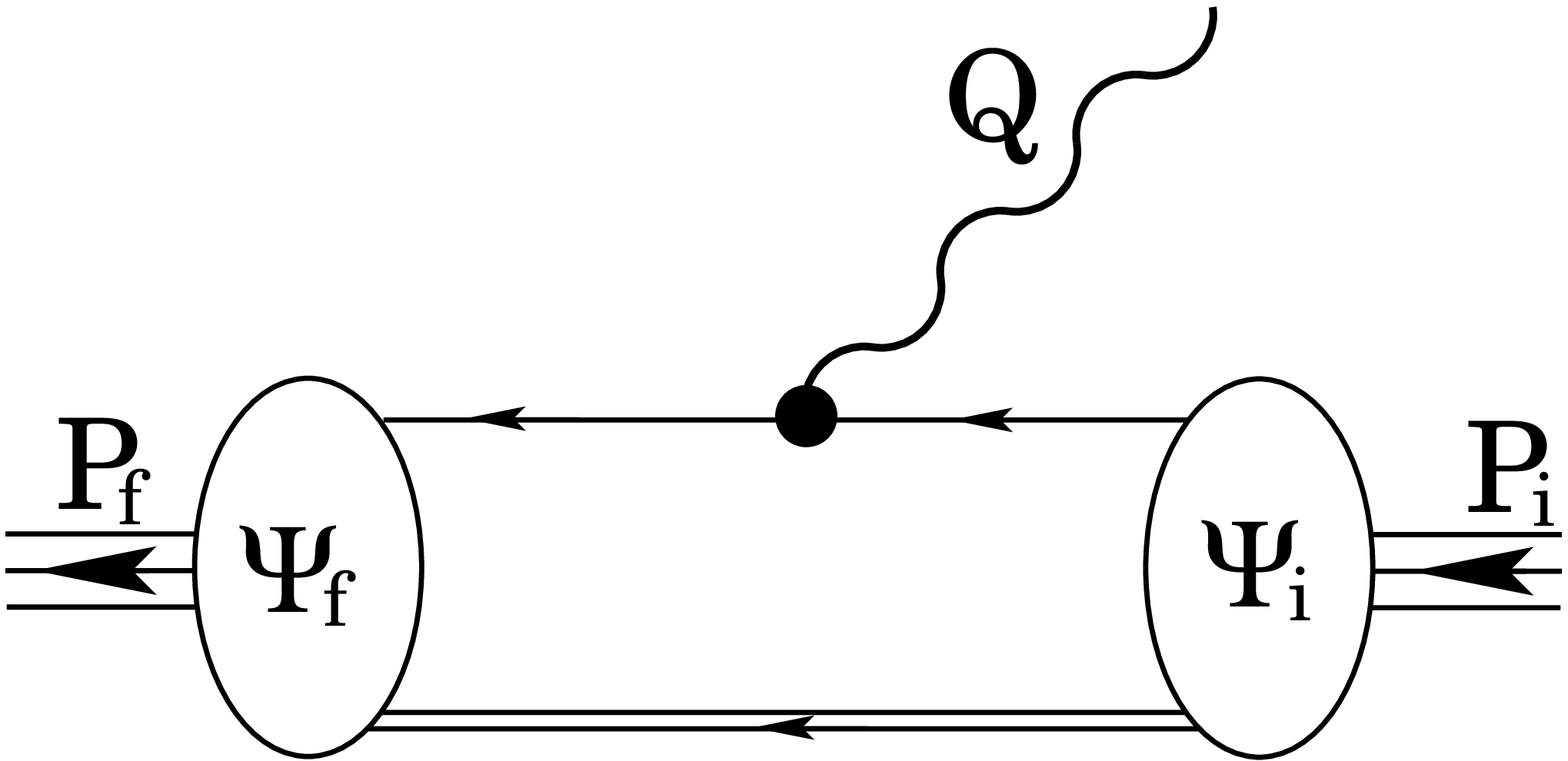}}
\end{minipage}
\begin{minipage}[t]{0.45\textwidth}
\rightline{\includegraphics[width=0.90\textwidth]{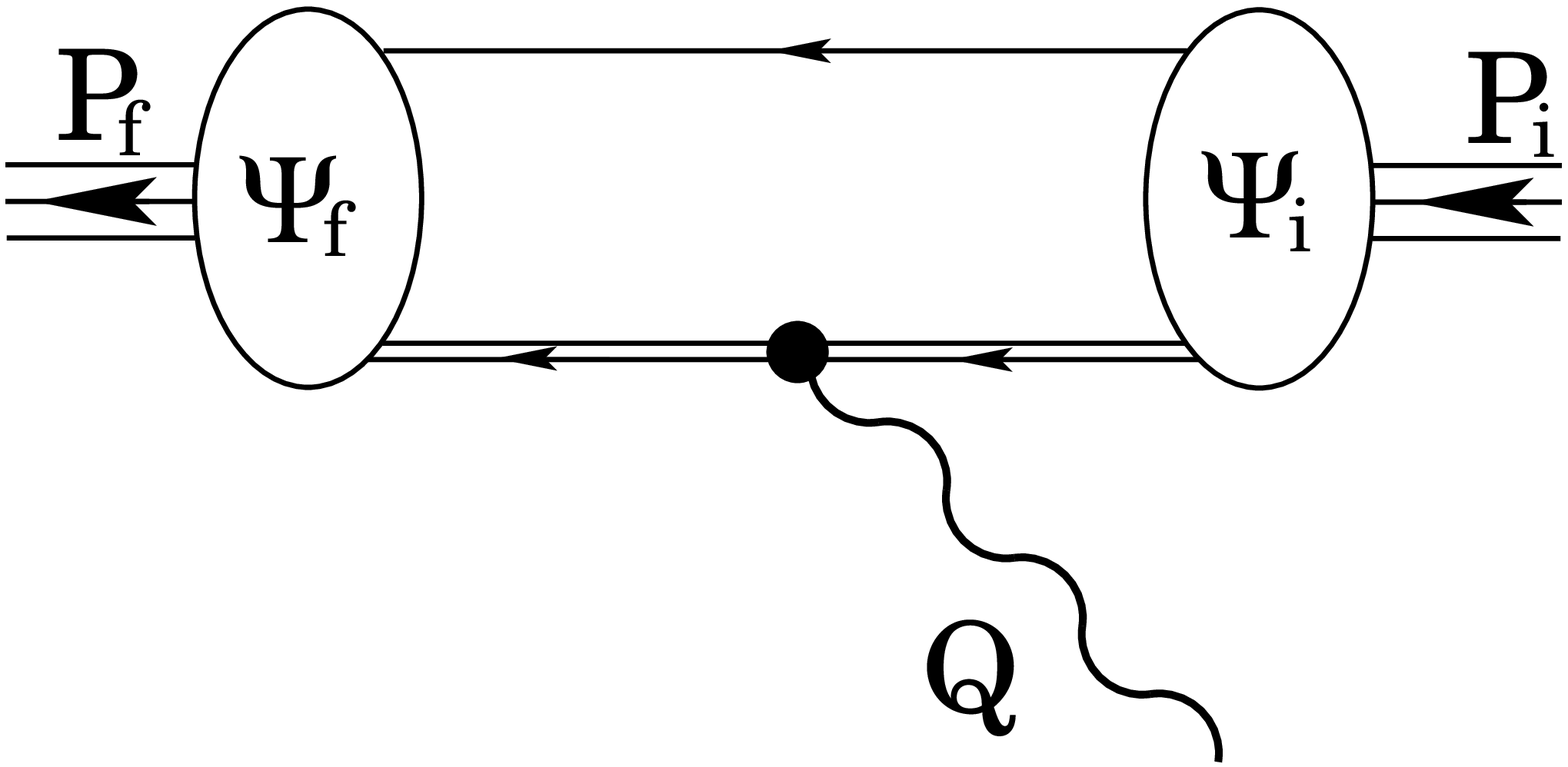}}
\end{minipage}\vspace*{3ex}

\begin{minipage}[t]{0.45\textwidth}
\leftline{\includegraphics[width=0.90\textwidth]{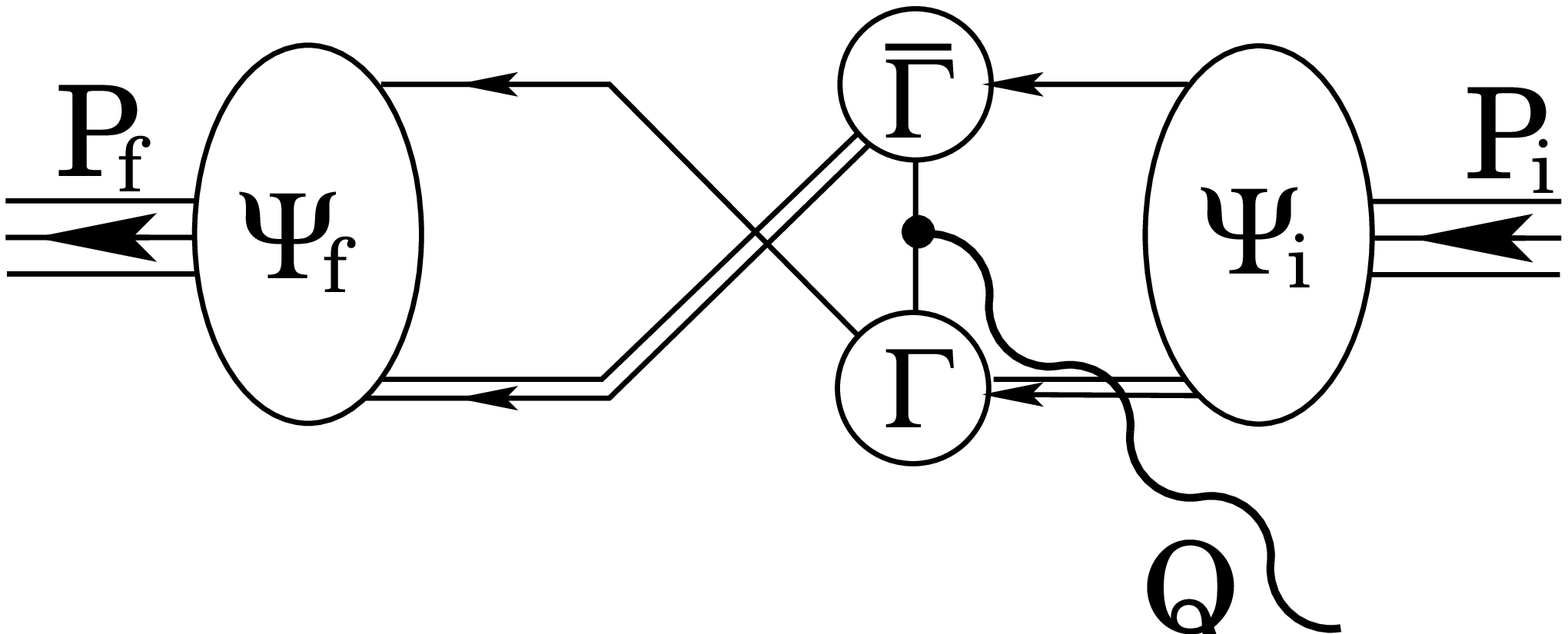}}
\end{minipage}
\begin{minipage}[t]{0.45\textwidth}
\rightline{\includegraphics[width=0.90\textwidth]{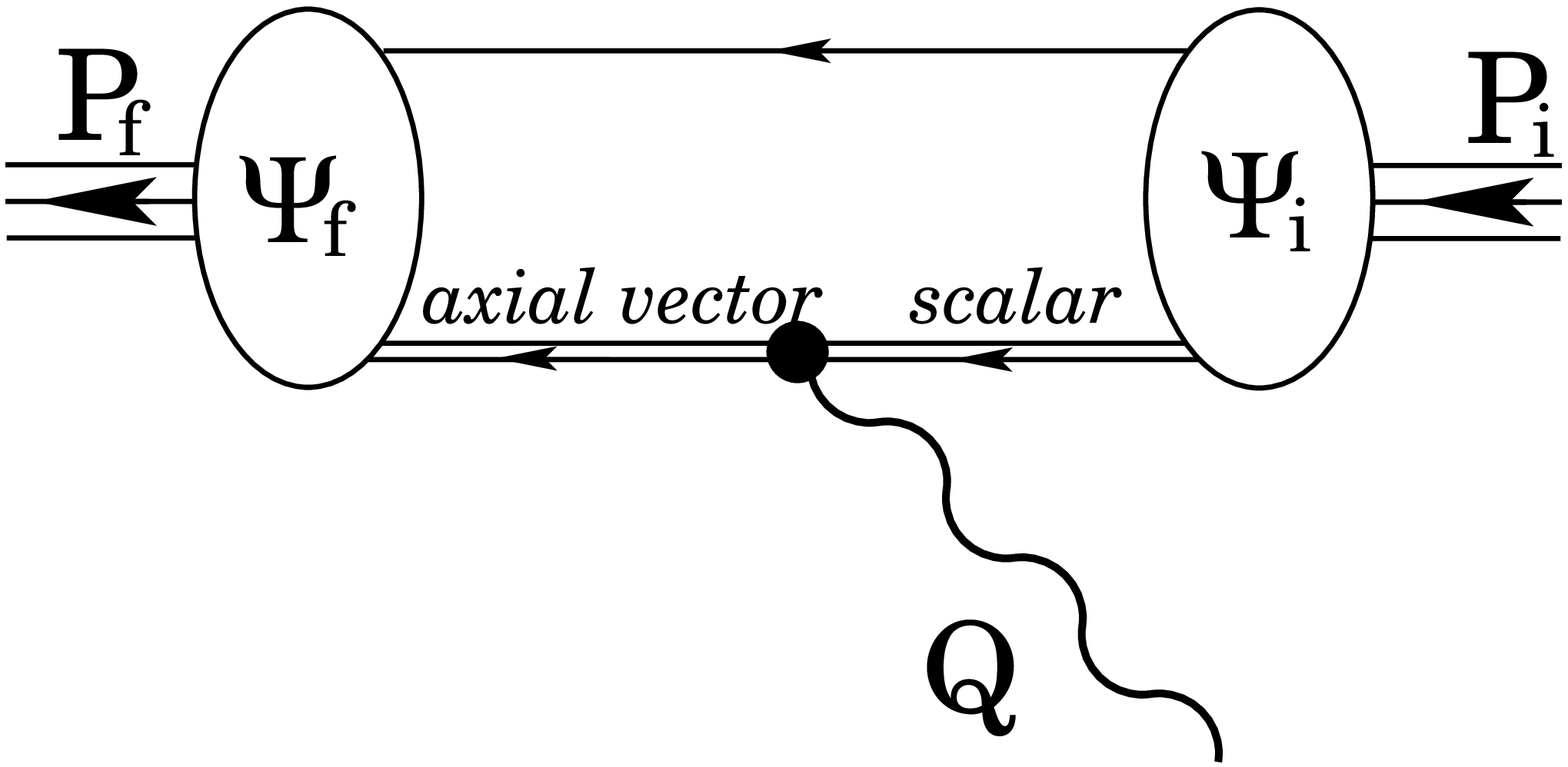}}
\end{minipage}\vspace*{3ex}

\begin{minipage}[t]{0.45\textwidth}
\leftline{\includegraphics[width=0.90\textwidth]{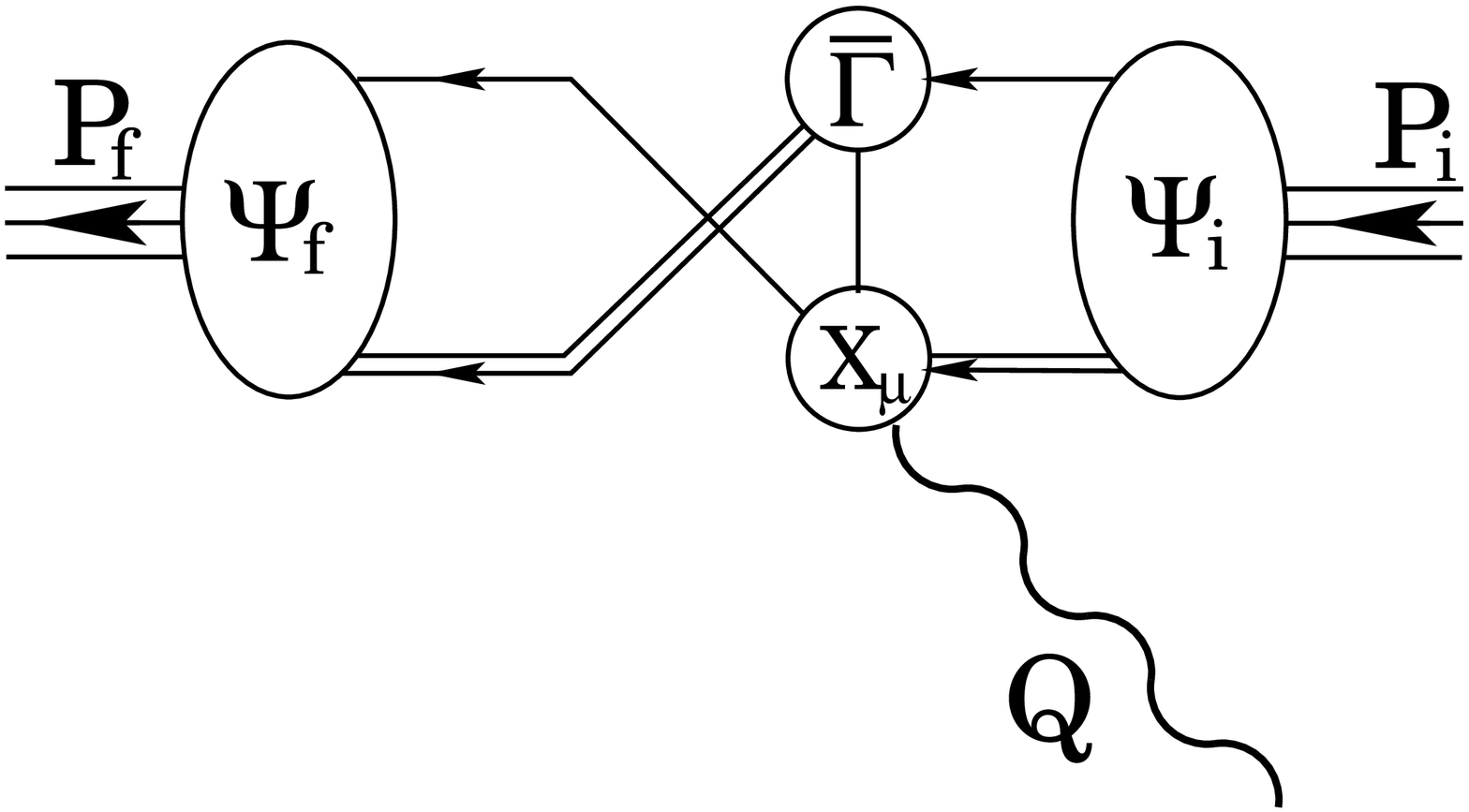}}
\end{minipage}
\begin{minipage}[t]{0.45\textwidth}
\rightline{\includegraphics[width=0.90\textwidth]{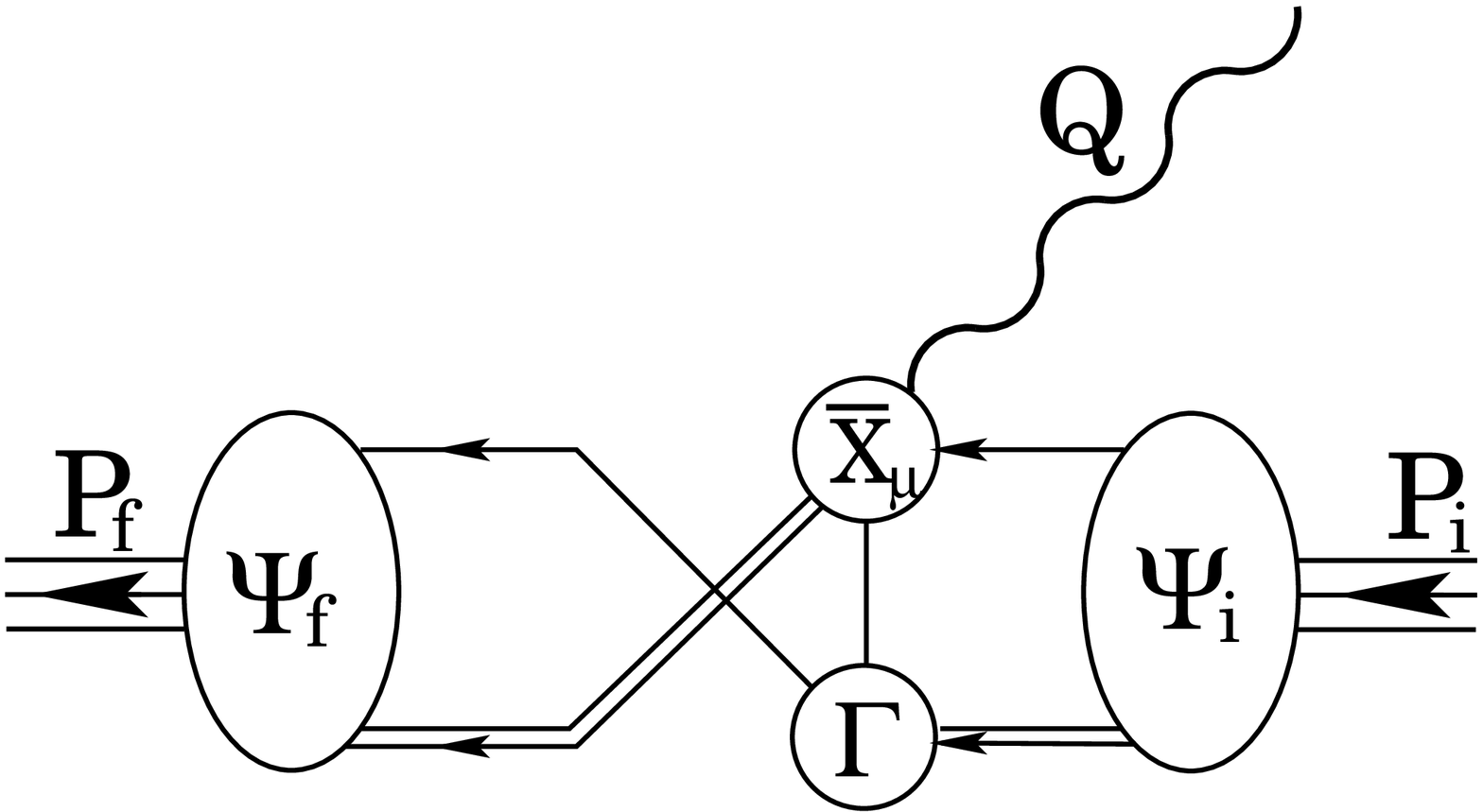}}
\end{minipage}
\end{minipage}
\caption{\label{vertex} Nucleon-photon vertex which ensures a conserved current for on-shell nucleons described by the Faddeev amplitudes, $\Psi_{i,f}$, calculated in Sec.\,\protect\ref{faddeevsolution}.  The single line represents $S(p)$, the dressed-quark propagator, the double line, the diquark propagator, and $\Gamma$ is the diquark Bethe-Salpeter amplitude, all of which are described in Sec.\,\protect\ref{completing}.  Aspects of the remaining vertices are described in Sec.\,\protect\ref{nucleonform}: the top-left image is diagram~1; the top-right, diagram~2; and so on,  with the bottom-right image, diagram~6.  (Adapted from Ref.\,[\protect\refcite{arneJ}].)}
\end{figure}

The nucleon's electromagnetic current is given in Eq.\,(\ref{jmuN}), which introduces the Dirac and Pauli form factors, and also the electric and magnetic form factors, which are related to the electric-charge-density distribution and the magnetic-current-density distribution.  In Eq.\,(\ref{jmuN}), $\Lambda_\mu$ is the nucleon-photon vertex.  It may be constructed following the systematic procedure of Ref.\,[\refcite{oettelpichowsky}].  That approach has the merit of automatically providing a conserved current for on-shell nucleons described by the Faddeev amplitudes which are obtained simultaneously with the mass.  Moreover, the canonical normalisation condition for the nucleons' Faddeev amplitude is equivalent to requiring $F_1(Q^2=0)=1$ for the proton.  The vertex has six terms, which are depicted in Fig.~\ref{vertex}.  Hereafter we describe them briefly.  A full explanation is provided in Ref.\,[\refcite{arneJ}].  
\smallskip

\hspace*{-\parindent}\underline{Diagram~1}. \hspace*{1em} This represents the photon coupling directly to the bystander quark.  It is a necessary condition for current conservation that the dressed-quark-photon vertex satisfy the Ward-Takahashi identity:
\begin{equation}
\label{vwti}
Q_\mu \, i\Gamma_\mu(\ell_1,\ell_2) = S^{-1}(\ell_1) - S^{-1}(\ell_2)\,,
\end{equation}
where $Q=\ell_1-\ell_2$ is the photon momentum flowing into the vertex.  Since the quark is dressed, Sec.\,\ref{completing}, the vertex cannot be bare; i.e., $\Gamma_\mu(\ell_1,\ell_2) \neq \gamma_\mu$.  It can be obtained by solving Eq.\,(\ref{photonvertex}), which was the procedure employed in Sec.\,\ref{modelindependent}.  However, since $S(p)$ is parametrised, Ref.\,[\refcite{blochff}] can be followed and the vertex written\,\cite{bc80}
\begin{equation}
\label{bcvtx}
i\Gamma_\mu(\ell_1,\ell_2)  =  
i\Sigma_A(\ell_1^2,\ell_2^2)\,\gamma_\mu +
2 k_\mu \left[i\gamma\cdot k_\mu \,
\Delta_A(\ell_1^2,\ell_2^2) + \Delta_B(\ell_1^2,\ell_2^2)\right] \!;
\end{equation}
with $k= (\ell_1+\ell_2)/2$ and
\begin{equation}
\Sigma_F(\ell_1^2,\ell_2^2) = \sfrac{1}{2}\,[F(\ell_1^2)+F(\ell_2^2)]\,,\;
\Delta_F(\ell_1^2,\ell_2^2) =
\frac{F(\ell_1^2)-F(\ell_2^2)}{\ell_1^2-\ell_2^2}\,,
\label{DeltaF}
\end{equation}
where $F= A, B$; viz., the scalar functions in Eq.\,(\ref{Sgeneral}).  It is
critical that $\Gamma_\mu$ in Eq.\ (\ref{bcvtx}) satisfies Eq.\ (\ref{vwti})
and very useful that it is completely determined by the dressed-quark
propagator.  
\smallskip

\hspace*{-\parindent}\underline{Diagram~2}. \hspace*{1em} This represents the photon coupling directly to a diquark correlation.  In the case of a scalar correlation, the general form of the diquark-photon vertex is
\begin{equation}
\Gamma_\mu^{0^+}(\ell_1,\ell_2) = 2\, k_\mu\, f_+(k^2,k\cdot Q,Q^2) + Q_\mu  \, f_-(k^2,k\cdot Q,Q^2)\,,
\end{equation}
and it must satisfy the Ward-Takahashi identity: 
\begin{equation}
\label{VWTI0}
Q_\mu \,\Gamma_\mu^{0^+}(\ell_1,\ell_2) = \Pi^{0^+}(\ell_1^2)  - \Pi^{0^+}(\ell_2^2)\,,\; \Pi^{J^P}(\ell^2) = \{\Delta^{J^P}(\ell^2)\}^{-1}.
\end{equation} 
The adaption of Eq.\,(\ref{bcvtx}) to this case is
\begin{equation}
\label{Gamma0plus}
\Gamma_\mu^{0^+}(\ell_1,\ell_2) =  k_\mu\,
\Delta_{\Pi^{0^+}}(\ell_1^2,\ell_2^2)\,,
%
\end{equation}  
with the definition of $\Delta_{\Pi^{0^+}}(\ell_1^2,\ell_2^2)$ apparent from Eq.\,(\ref{DeltaF}).  Equation~(\ref{Gamma0plus}) is the minimal \textit{Ansatz} that: satisfies Eq.\,(\ref{VWTI0}); is completely determined by quantities introduced already; and is free of kinematic singularities.  It also guarantees a valid normalisation of electric charge; viz., 
\begin{equation}
\lim_{\ell^\prime\to \ell} \Gamma_\mu^{0^+}(\ell^\prime,\ell) = 2 \, \ell_{\mu} \, \frac{d}{d\ell^2}\, \Pi^{0^+}(\ell^2) \stackrel{\ell^2\sim 0}{=} 2 \, \ell_{\mu}\,,
\end{equation}
owing to Eq.\,(\ref{DQPropConstr}).  NB.\ The fractional diquark charge has been factored.  It therefore appears subsequently as a simple multiplicative factor. 

For the case in which the struck diquark correlation is axial-vector, the vertex structure is more involved.  Nonetheless, there are many constraints that may be employed to build a realistic \textit{Ansatz}.  That composed in Ref.\,[\refcite{arneJ}] has two parameters: the magnetic dipole and electric quadrupole moments of the axial-vector diquark, $\mu_{1^+}$ and $\chi_{1^+}$, respectively.  
\smallskip

\hspace*{-\parindent}\underline{Diagram~3}. \hspace*{1em}  This image depicts a photon coupling to the quark that is exchanged as one diquark breaks up and another is formed.  While this is the first two-loop diagram in the current, no new elements appear in its specification: the quark-photon vertex was described above.  It is noteworthy that the process of quark exchange provides the attraction necessary in the Faddeev equation to bind the baryon.  It also guarantees that the Faddeev amplitude has the correct antisymmetry under the exchange of any two dressed-quarks.  This key feature is absent in models with elementary (noncomposite) diquarks.
\smallskip

\hspace*{-\parindent}\underline{Diagram~4}. \hspace*{1em}  This differs from Diagram~2 in expressing the contribution to a nucleon's form factors owing to an electromagnetically induced transition between scalar and axial-vector diquarks.  The transition vertex is a rank-2 pseudotensor, kindred to the matrix element describing the $\rho\, \gamma^\ast \pi^0$  transition \cite{maristandy4}, and can therefore be expressed 
\begin{equation}
\label{SAPhotVertex}
\Gamma_{SA}^{\gamma\alpha}(\ell_1,\ell_2) = -\Gamma_{AS}^{\gamma\alpha}(\ell_1,\ell_2) 
= \frac{i}{M_N} \, {\check T}(\ell_1,\ell_2) \, \varepsilon_{\gamma\alpha\rho\lambda}\ell_{1\rho} \ell_{2 \lambda}\,,
\end{equation}
where $\gamma$, $\alpha$ are, respectively, the vector indices of the photon and axial-vector diquark.  For simplicity, Ref.\,[\refcite{arneJ}] proceeded under the assumption
\begin{equation}
{\check T}(\ell_1,\ell_2) = \kappa_{\check T}\,;
\end{equation}
viz., a constant, for which a typical value is\,\cite{oettel2}: 
\begin{equation}
\label{kTbest}
\kappa_{\check T} \sim 2\,.
\end{equation}

In the nucleons' rest frame, a conspicuous piece of the Faddeev amplitude that describes an axial-vector diquark inside the bound state can be characterised as containing a bystander quark whose spin is antiparallel to that of the nucleon, with the axial-vector diquark's parallel.  The interaction pictured in this diagram does not affect the bystander quark but the transformation of an axial-vector diquark into a scalar effects a flip of the quark spin within the correlation.  After this transformation, the spin of the nucleon must be formed by summing the spin of the bystander quark, which is still aligned antiparallel to that of the nucleon, and the orbital angular momentum between that quark and the scalar diquark.\footnote{A less prominent component of the amplitude has the bystander quark's spin parallel to that of the nucleon while the axial-vector diquark's is antiparallel: this $q^\uparrow \oplus (qq)_{1^+}^{\downarrow} $ system has one unit of angular momentum.  That momentum is absent in the $q^\uparrow \oplus (qq)_{0^+}$ system.  Other combinations also contribute via Diagram~3 but all mediated processes inevitably require a modification of spin and/or angular momentum.}   Diagram~4 may therefore be expected to impact strongly on the nucleons' magnetic form factors.

\hspace*{-\parindent}\underline{Diagram~5 and 6}. \hspace*{1em}  These two-loop diagrams are the so-called ``seagull'' terms, which appear as partners to Diagram~3 and arise because binding in the nucleons' Faddeev equations is effected by the exchange of \textit{nonpointlike} diquark correlations\,\cite{oettelpichowsky}.  The new elements in these diagrams are the couplings of a photon to two dressed-quarks as they either separate from (Diagram~5) or combine to form (Diagram~6) a diquark correlation.  As such they are components of the five point Schwinger function which describes the coupling of a photon to the quark-quark scattering kernel.  This Schwinger function could be calculated, as is evident from the recent computation of analogous Schwinger functions relevant to meson observables.\cite{mariscotanch}  However, such a calculation provides valid input only when a uniform truncation of the DSEs has been employed to calculate each of the elements described hitherto.  In the present context, it is appropriate instead to employ the algebraic parametrisation of Ref.\,[\refcite{oettelpichowsky}], which is simple, completely determined by the elements introduced already, and guarantees current conservation for on-shell nucleons.

\subsubsection{Calculated results}
\label{numerical}
In order to place the calculation of baryon observables on the same footing as the study of mesons, the proficiency evident in Ref.\,[\refcite{pieterrev}] will need to be applied to every line and vertex that appears in Fig.\,\ref{vertex}.  While that is feasible, it remains to be done.  In the meantime, we relate a study whose merits include a capacity to: explore the potential of the Faddeev equation truncation of the baryon three-body problem; and elucidate the role of additional correlations, such as those associated with pseudoscalar mesons.

It is worthwhile to epitomise the input before presenting results.  One element is the dressed-quark propagator, Sec.\,\ref{completing}.  The form used\,\cite{mark} both anticipated and expresses the features that are now known to be true\cite{alkoferdetmold,bhagwat}.  It carries no free parameters, because its behaviour was fixed in analyses of meson observables, and is basic to a description of light- and heavy-quark mesons that is accurate to better than 10\%.\cite{mishaSVY}

The nucleon is supposed at heart to be composed of a dressed-quark and nonpointlike diquark with binding effected by an iterated exchange of roles between the bystander and diquark-participant quarks.  The picture is realised via a Poincar\'e covariant Faddeev equation, Sec.\,\ref{faddeev}, which incorporates scalar and axial-vector diquark correlations.  There are two parameters, Sec.\,\ref{faddeevsolution}: the mass-scales associated with these correlations.  They were fixed by fitting to specified nucleon and $\Delta$ masses.  There are no free parameters at this point.

With the constituents and the bound states' structure defined, only a specification of the nucleons' electromagnetic interaction remained.  Its formulation, sketched in Sec.\,\ref{Ncurrent}, is guided almost exclusively by a requirement that the nucleon-photon vertex satisfy a Ward-Takahashi identity.  Since the scalar diquark's electromagnetic properties are readily resolved, the result, Fig.\,\ref{vertex}, depends on three parameters that are all tied to properties of the axial-vector diquark correlation: $\mu_{1^+}$ \& $\chi_{1^+}$, respectively, the axial-vector diquarks' magnetic dipole and electric quadrupole moments; and $\kappa_{\check T}$, the strength of electromagnetic axial-vector $\leftrightarrow$ scalar diquark transitions.  Hence, the study of Ref.\,[\refcite{arneJ}] exhibits and interprets the dependence of the nucleons' form factors on these three parameters, and also on the nucleons' intrinsic quark structure as expressed in the Poincar\'e covariant Faddeev amplitudes. 

\begin{table}[t]
\tbl{%
Magnetic moments, in nuclear magnetons, calculated with the diquark mass-scales in Table~\protect\ref{ParaFix} for a range of axial-vector-diquark--photon vertex parameters, centred on the point-particle values of $\mu_{1^+}=2$ \& $\chi_{1^+}=1$, and $\kappa_{\check T} = 2$, Eq.\,(\protect\ref{kTbest}).  Columns labelled $\sigma$ give the percentage-difference from results obtained with the reference parameters.
Experimental values are: $\kappa_p:=\mu_p-1 = 1.79$ \& $\mu_n = -1.91$.  (Adapted from Ref.\,[\protect\refcite{arneJ}].)
}
{\normalsize
\begin{tabular*}{1.0\textwidth}{ccccccccccc}
\multicolumn{3}{c}{} & \multicolumn{4}{c}{Set~A} & \multicolumn{4}{c}{Set~B} \\
\multicolumn{3}{c}{} & \multicolumn{4}{c}{\rule{10em}{0.1ex}} & \multicolumn{4}{c}{\rule{10em}{0.1ex}} \\
$\mu_{1^+}$ & $\chi_{1^+}$ & $\kappa_{\check T}$ & $\kappa_p$ & \rule{0em}{2.5ex}$\sigma_{\kappa_p}^A$ & $|\mu_n|$ & $\sigma_{|\mu_n|}^A$ & $\kappa_p$ & $\sigma_{\kappa_p}^B$ & $|\mu_n|$ & $\sigma_{|\mu_n|}^B$ \\\hline
1 & 1 & 2 & 1.79 & -15.3  & 1.70 & -5.1 & 2.24 & -21.9 & 2.00 & -6.2 \\
2 & 1 & 2 & 2.06 & ~~~~~  & 1.79 & ~~~~ & 2.63 & ~~~~~ & 2.13 & ~~~~ \\
3 & 1 & 2 & 2.33 & ~15.4  & 1.88 & ~5.1 & 3.02 & ~21.9 & 2.26 & ~6.1 \\\hline
2 & 0 & 2 & 2.06 & ~~0.0  & 1.79 & ~0.0 & 2.63 & ~~0.0 & 2.13 & ~0.0 \\
2 & 2 & 2 & 2.06 & ~~0.0  & 1.79 & ~0.0 & 2.63 & ~~0.0 & 2.13 & ~0.0 \\\hline
2 & 1 & 1 & 1.91 & ~-8.4  & 1.64 & -8.4 & 2.45 & -10.1 & 1.95 & -8.5 \\
2 & 1 & 3 & 2.21 & ~~8.4  & 1.85 & ~8.3 & 2.82 & ~10.1 & 2.31 & ~8.5 \\\hline
\end{tabular*}\label{moments}}
\end{table}

Table~\ref{moments} lists results for the nucleons' magnetic moments, $G_M^N(0)$, where $N=n,p$.\cite{arneJ}  It indicates that the moments are insensitive to the axial-vector diquarks' quadrupole moment but react to the diquarks' magnetic moment, increasing quickly in magnitude as $\mu_{1^+}$ increases.  As anticipated in connection with Eq.\,(\ref{kTbest}), the nucleons' moments respond strongly to alterations in the strength of the scalar $\leftrightarrow$ axial-vector transition, increasing rapidly as $\kappa_{\check T}$ is increased.   Set~A, which is fitted to the experimental values of $M_N$ \& $M_\Delta$, describes the nucleons' moments quite well: $\kappa_p$ is 15\% too large; and $|\mu_n|$, 16\% too small.  On the other hand, Set~B, which is fitted to baryon masses that are inflated so as to make room for pion cloud effects, overestimates $\kappa_p$ by 47\% and $|\mu_n|$ by 18\%.  

The nucleons' charge and magnetic radii
\begin{equation}
r_{N}^2:= \left.- \, 6\,\frac{d}{ds} \ln G_E^{N}(s) \right|_{s=0},\;
(r_{N}^\mu)^2:= \left.- \, 6\,\frac{d}{ds} \ln G_M^{N}(s) \right|_{s=0},
\end{equation}
were also reported in Ref.\,[\refcite{arneJ}].  The charge radii, particularly that of the neutron, were most sensitive to changes in the axial-vector diquarks' electric quadrupole moment, $\chi_{1^+}$.  That is not surprising given that $\chi_{1^+}$ is the only model parameter which speaks directly of the axial-vector diquarks' electric charge distribution.  With the point-particle reference values of $\mu_{1^+}=2$ \& $\chi_{1^+}=1$, and $\kappa_{\check T} = 2$, Set~A underestimates the proton radius by 30\% and the magnitude of the neutron radius by 43\%, while for Set~B these differences are 32\% and 50\%, respectively.  The magnetic radii are insensitive to the axial-vector diquarks' quadrupole moment but react to the diquarks' magnetic moment as one would anticipate: increasing in magnitude as $\mu_{1^+}$ increases.  Moreover, again consistent with expectation, these radii respond to changes in $\kappa_{\check T}$, decreasing as this parameter is increased.  With the reference parameters values both Sets~A \& B underestimate $r_N^\mu$ by approximately 40\%.

\begin{figure}[t]
\begin{minipage}{0.45\textwidth}
\centerline{\hspace*{2.5em}%
\includegraphics[width=0.95\textwidth,angle=270]{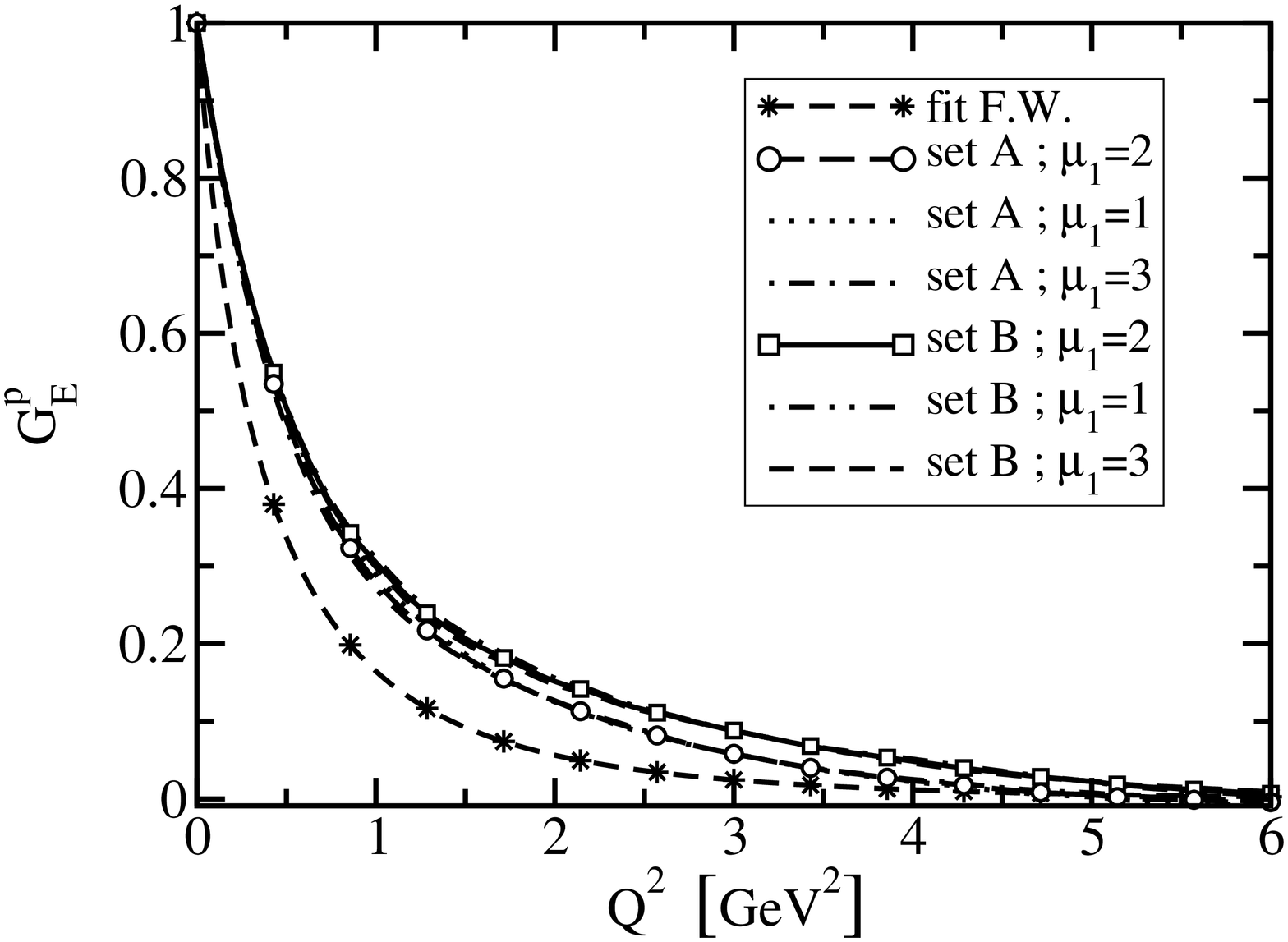}}
\end{minipage}
\hfill
\begin{minipage}{0.45\textwidth}
\centerline{\includegraphics[width=0.95\textwidth,angle=270]{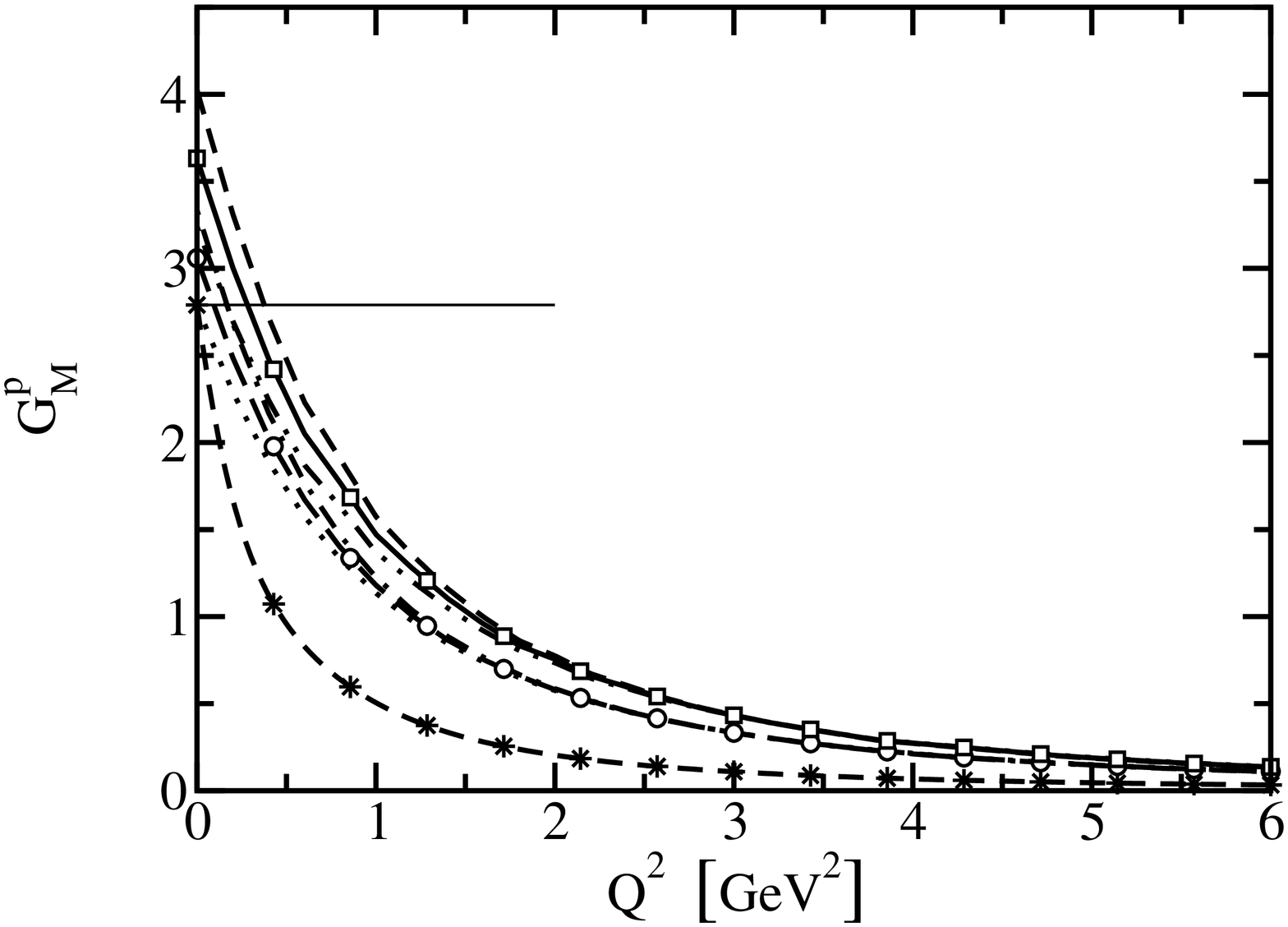}%
\hspace{1em}}
\end{minipage}
\\
\begin{minipage}{0.45\textwidth}
\centerline{\hspace*{2.5em}%
\includegraphics[width=0.95\textwidth,angle=270]{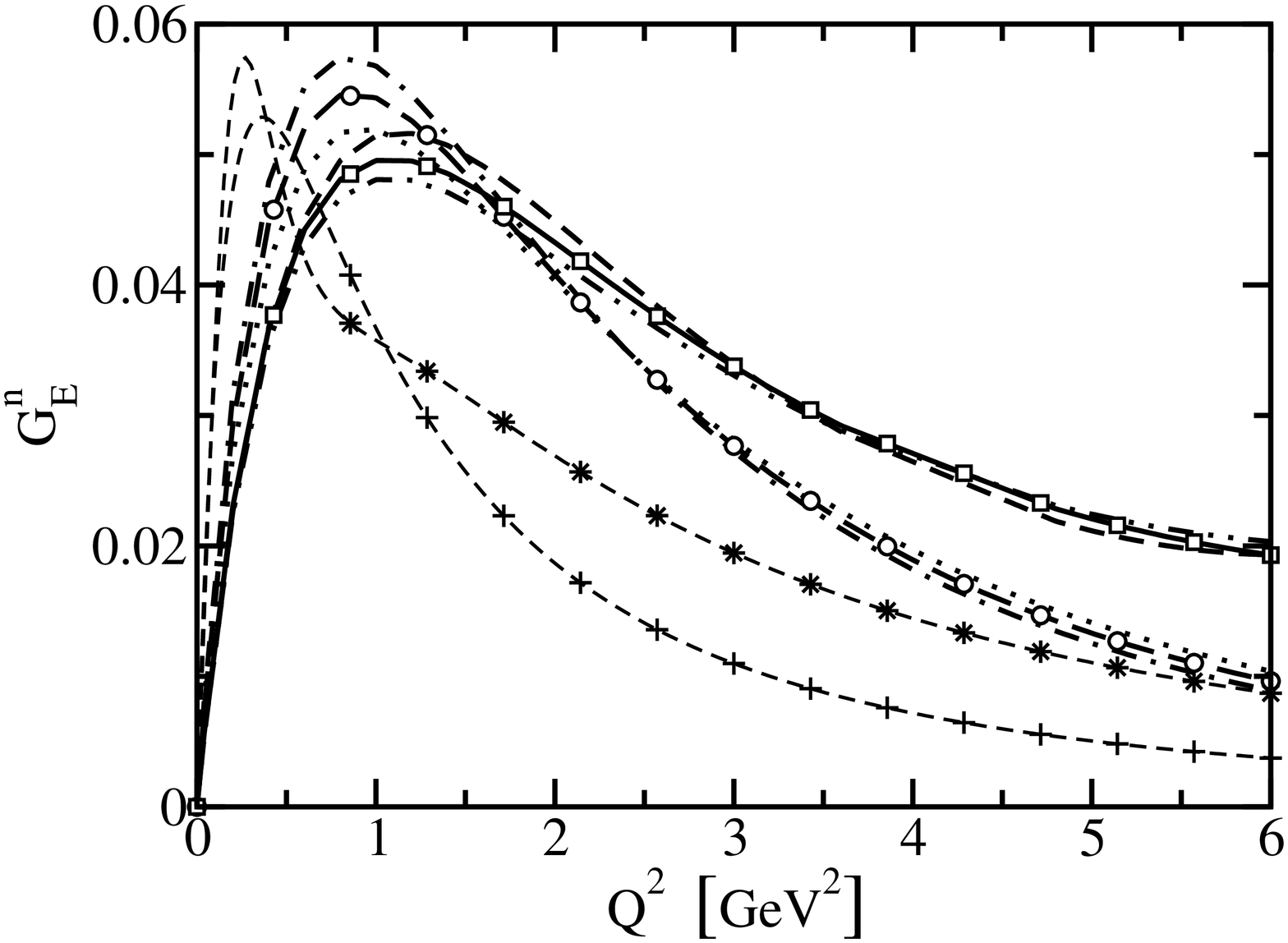}}
\end{minipage}
\hfill
\begin{minipage}{0.45\textwidth}
\centerline{\includegraphics[width=0.95\textwidth,angle=270]{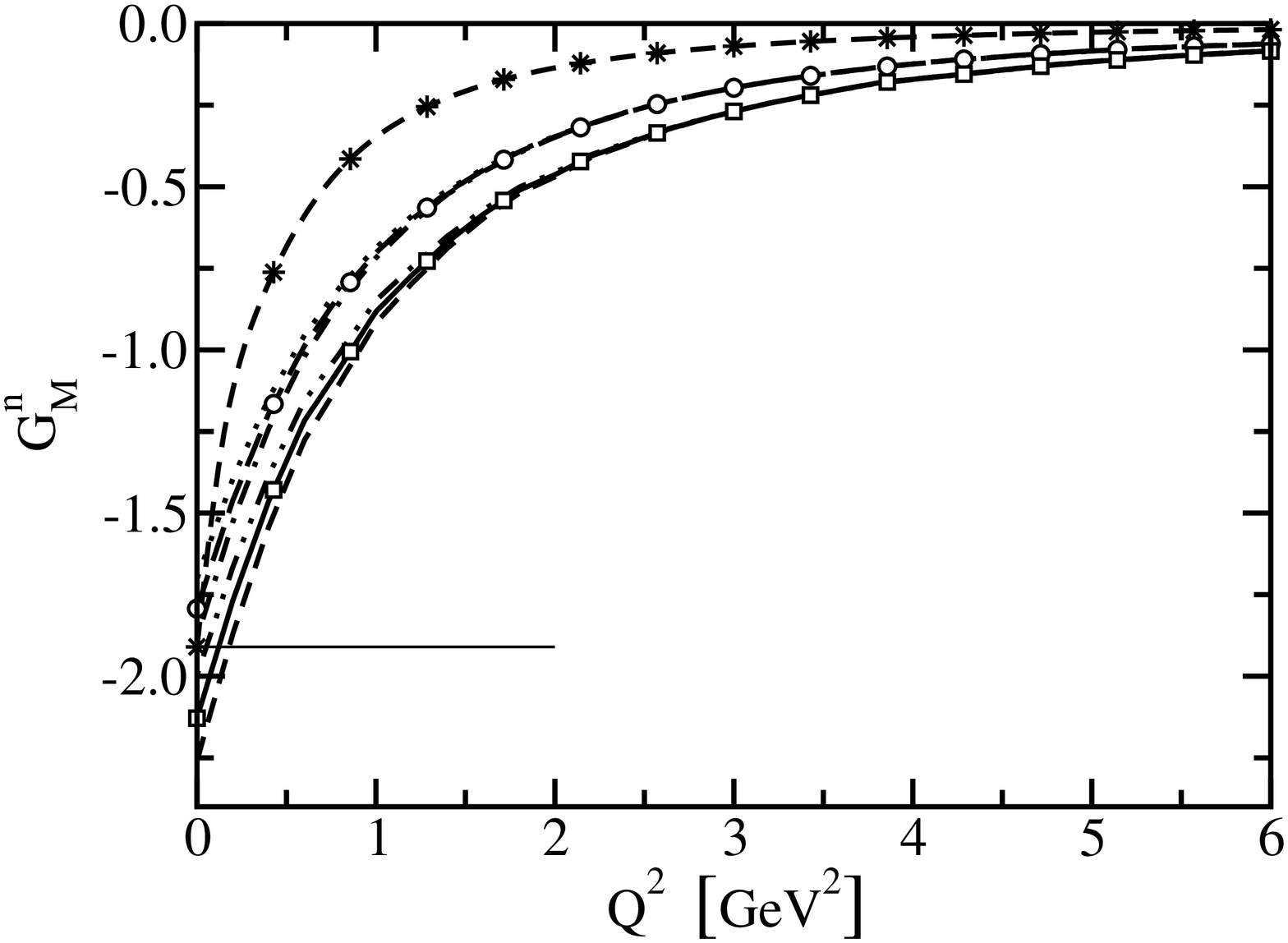}%
\hspace*{1em}}
\end{minipage}
\caption{\label{plot1}
Response of nucleon form factors to variations in the magnetic moment of the axial-vector diquark: $\mu_{1^+}=1,2,3$; with $\chi_{1^+}=1$, $\kappa_{\check T}=2$.  
The legend in the top-left panel applies to all; the dashed-line marked by ``$\ast$'' is a fit to experimental data\,\protect\cite{Walcher03} and the dashed-line marked by ``$+$'' in the lower-left panel is the fit to $G_E^n(Q^2)$ of Ref.\,[\protect\refcite{galster}]; and the horizontal lines in the right panels mark the experimental value of the nucleon's magnetic moment.  (Adapted from Ref.\,[\protect\refcite{arneJ}].)}
\end{figure}

Nucleon electromagnetic form factors associated with the tabulated values of static properties are presented in Fig.\,\ref{plot1}.  The figure confirms and augments the information in Table~\ref{moments}.  Consider, e.g., the electric form factors.  One observes that the differences between results obtained with Set~A and Set~B generally outweigh those delivered by variations in the parameters characterising the axial-vector diquark's electromagnetic properties.  The proton's electric form factor, in particular, is largely insensitive to these parameters.  The nucleons' magnetic form factors exhibit the greatest sensitivity to the axial-vector diquarks's electromagnetic properties but in this case, too, the differences between Set~A and Set~B are more significant.  For $Q^2 \gtrsim 4\,$GeV$^2$ there is little sensitivity to the diquarks' electromagnetic parameters in any curve because the model expresses the diquark current's perturbative limit.\cite{brodskyhiller92}  It is thus apparent from these plots that the behaviour of the nucleons' form factors is primarily determined by the information encoded in the Faddeev amplitudes.

The results show that the nucleons' electromagnetic properties are sensitive to the strength of axial-vector diquark correlations in the bound state and react to the electromagnetic properties of these correlations.  In all cases the dependence is readily understood intuitively.  However, taken together the results indicate that one cannot readily tune the model's parameters to provide a uniformly good account of nucleon properties: something more than dressed-quark and -diquark degrees of freedom is required.

\subsubsection{Chiral corrections}
\label{chiralem}
It is appropriate now to examine effects that arise through coupling to pseudoscalar mesons.  As with baryon masses, Sec.\,\ref{loopmass}, there are two types of contributions to electromagnetic form factors from meson loops: regularisation-scheme-dependent terms, which are analytic in the neighbourhood of $\hat m = 0$; and nonanalytic scheme-independent terms.  For nucleon static properties presented the leading-order scheme-independent contributions are\,\cite{kubis}
\begin{eqnarray}
\label{rpnpion}
\langle r_{p\atop n}^2\rangle^{1-loop}_{NA} &=& \mp\,\frac{1+5 g_A^2}{32 \pi^2 f_\pi^2} \,\ln (\frac{m_\pi^2}{M_N^2}) \,, \\
\label{rpnmpion}
\langle (r_{N}^\mu)^2\rangle^{1-loop}_{NA} &=& -\,\frac{1+5 g_A^2}{32 \pi^2 f_\pi^2} \,\ln (\frac{m_\pi^2}{M_N^2})+ \frac{g_A^2\, M_N}{16 \pi f_\pi^2 \mu_v} \frac{1}{m_\pi} \,,\\
\label{mpnpion}
(\mu_{p\atop n})_{NA}^{1-loop} & = & \mp \, \frac{g_A^2\, M_N}{4\pi^2 f_\pi^2}\, m_\pi\,,
\end{eqnarray}
where $g_A=1.26$, $f_\pi=0.0924$\,GeV\,$=1/(2.13 \,{\rm fm})$, $\mu_v=\mu_p-\mu_n$.  Clearly, the radii diverge in the chiral limit, a much touted aspect of chiral corrections.  While these scheme-independent terms are immutable, at physical values of the pseudoscalar meson masses they do not usually provide the dominant contribution to observables: that is provided by the regularisation-parameter-dependent terms, as we saw, e.g., with the masses in Sec.\,\ref{loopmass}.  

Since regularisation-parameter-dependent parts of the chiral loops are important it is sensible to follow Ref.\,[\refcite{ashley}] and estimate the corrections using modified formulae that incorporate a single parameter which mimics the effect of regularisation-dependent contributions from the integrals.  Thus Eqs.\,(\ref{rpnpion}) -- (\ref{mpnpion}) are rewritten
\begin{eqnarray}
\label{rpnpionR}
\langle r_{p\atop n}^2\rangle^{1-loop^R}_{NA} &=& \mp\,\frac{1+5 g_A^2}{32 \pi^2 f_\pi^2} \,\ln (\frac{m_\pi^2}{m_\pi^2+\lambda^2}) \,, \\
\nonumber \langle (r_{N}^\mu)^2\rangle^{1-loop^R}_{NA} &=& -\,\frac{1+5 g_A^2}{32 \pi^2 f_\pi^2} \,\ln (\frac{m_\pi^2}{m_\pi^2+\lambda^2}) + \frac{g_A^2\, M_N}{16 \pi f_\pi^2 \mu_v} \frac{1}{m_\pi} \,
\frac{2}{\pi}\arctan(\frac{\lambda}{m_\pi})\,,\\
\label{rpnmpionR}\\
\label{mpnpionR}
(\mu_{p\atop n})_{NA}^{1-loop^R}& = & \mp \, \frac{g_A^2\, M_N}{4\pi^2 f_\pi^2}\, m_\pi\, \frac{2}{\pi}\arctan(\frac{\lambda^3}{m_\pi^3})\,,
\end{eqnarray}
wherein $\lambda$ is a regularisation mass-scale, for which a typical value is\,\cite{ashley} $\sim 0.4\,$GeV.  NB.\ The loop contributions vanish when the pion mass is much larger than the regularisation scale, as required: very massive states must decouple from low-energy phenomena.  

\begin{table}[t]
\tbl{Row~1 -- static properties calculated with Set~B diquark masses, Table~\protect\ref{ParaFix}, and $\mu_{1^+}=2$, $\chi_{1^+}=1$, $\kappa_{\check T}=2$: charge radii in fm, with $r_n:= -\sqrt{-\langle r_n^2\rangle}$; and magnetic moments in nuclear magnetons.  Row~2 adds the corrections of Eqs.\,(\protect\ref{rpnpionR})--(\protect\ref{mpnpionR}) with $\lambda=0.3\,$GeV.  $\varsigma$ in row~$n$, is the rms relative-difference between the entries in rows $n$ and 3.  (Adapted from Ref.\,[\protect\refcite{arneJ}].)}
{\normalsize
\begin{tabular*}{1.0\textwidth}{l@{\extracolsep{0ptplus1fil}}
c@{\extracolsep{0ptplus1fil}}
c@{\extracolsep{0ptplus1fil}}c@{\extracolsep{0ptplus1fil}} 
c@{\extracolsep{0ptplus1fil}}c@{\extracolsep{0ptplus1fil}}
c@{\extracolsep{0ptplus1fil}}c@{\extracolsep{0ptplus1fil}}
c@{\extracolsep{0ptplus1fil}}}\\\hline
  & $r_p$ & $r_n$ & $r_p^\mu$ & $r_n^\mu$ & $\mu_p$ & $-\mu_n$ && $\varsigma$\\\hline
$q$-$(qq)$ core & 0.595 & 0.169 & 0.449 & 0.449 & 3.63 & 2.13 && 0.39\\
$+\pi$-loop correction & 0.762 & 0.506 & 0.761 & 0.761 & 3.05 & 1.55 && 0.23 \\\hline
experiment & 0.847 & 0.336 & 0.836 & 0.889 & 2.79& 1.91 && \\\hline
\end{tabular*}\label{picorrected}}
\end{table}

One may now return to the calculated values of the nucleons' static properties.  Consider the Set~B results obtained with $\mu_{1^+}=2$, $\chi_{1^+}=1$, $\kappa_{\check T}=2$.  Set~B was chosen to give inflated values of the nucleon and $\Delta$ masses in order to make room for chiral corrections, and therefore one may consistently apply the corrections in Eqs.\,(\ref{rpnpionR}) -- (\ref{mpnpionR}) to the static properties.  With $\lambda=0.3\,$GeV this yields the second row in Table~\ref{picorrected}: the regularised chiral corrections reduce the rms relative-difference significantly.  

This crude analysis, complementing Sec.\,\ref{loopmass}, suggests strongly that a veracious description of baryons can be obtained using dressed-quark and -diquark degrees of freedom augmented by a sensibly regulated pseudoscalar meson cloud.  The inverse of the regularisation parameter is a length-scale that may be viewed as a gauge of the distance from a nucleon's centre-of-mass to which the pseudoscalar meson cloud penetrates: $1/\lambda \approx \mbox{\small $\frac{2}{3}$}\,{\rm fm}$ is an intuitively reasonable value that indicates the cloud is expelled from the nucleon's core but materially affects its properties at distances $\gsim r_p$; viz., in the vicinity of the nucleon's surface and farther out.
 
\begin{figure}[t]
\centerline{\includegraphics[width=0.7\textwidth,angle=270]{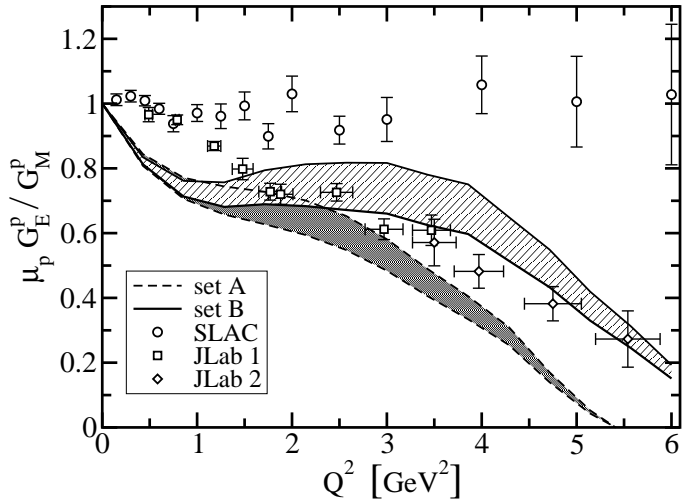}}
\caption{\label{plotGEpGMp} Form factor ratio: $\mu_p\, G_E^p(Q^2)/G_M^p(Q^2)$.  Calculated results: \textit{lower band} - Set~A in Table~\protect\ref{ParaFix}; and \textit{upper band} - Set~B.  For both, $G_E^p(Q^2)$ was calculated using the point-particle values: $\mu_{1^+}=2$ \& $\chi_{1^+}=1$, and $\kappa_{\check T} = 2$; i.e., the reference values in Table~\ref{moments}.  Changes in the axial-vector diquark parameters used to evaluate $G_E^p$ have little effect on the plotted results.  The width of the bands reflects the variation in $G_M^p$ with axial-vector diquark parameters.  In both cases, the upper border is obtained with $\mu_{1^+}=3$, $\chi_{1^+}=1$ and $\kappa_{\check T}= 2$, while the lower has $\mu_{1^+}= 1$.  The data are: \textit{squares} - Ref.\,[\protect\refcite{jones}]; \textit{diamonds} - Ref.\,[\protect\refcite{gayou}]; and \textit{circles} - Ref.\,[\protect\refcite{walker}].  (Adapted from Ref.\,[\protect\refcite{arneJ}].)}
\end{figure}

\subsubsection{Form factor ratios}
\label{FFratios}
We are now in a position to return to our discussion of Fig.\,\ref{gepgmpdata}, and in Fig.\,\ref{plotGEpGMp} plot the calculated ratio $\mu_p\, G_E^p(Q^2)/G_M^p(Q^2)$.  The behaviour of the experimental data at small $Q^2$ is now readily understood.  In the neighbourhood of $Q^2=0$, 
\begin{equation}
\mu_p\,\frac{ G_E^p(Q^2)}{G_M^p(Q^2)} \stackrel{Q^2\sim 0}{=} 1 - \frac{Q^2}{6} \,\left[ (r_p)^2 - (r_p^\mu)^2 \right]\,,
\end{equation}
and because experimentally $r_p\approx r_p^\mu$ the ratio varies by less than 10\% on $0<Q^2< 0.6\,$GeV$^2$, if the form factors are approximately dipole.  In the calculation described herein, $r_p> r_p^\mu$ without chiral corrections.  Hence the ratio must fall immediately with increasing $Q^2$.  Incorporating pion loops, one obtains the results in Row~2 of Table\,\ref{picorrected}, which have $r_p\approx r_p^\mu$.  The small $Q^2$ (long-range) behaviour of this ratio is thus materially affected by the proton's pion cloud.

We have emphasised that true pseudoscalar mesons are not pointlike and therefore pion cloud contributions to form factors diminish in magnitude with increasing $Q^2$.  To exemplify this further, it is notable that in a study of the $\gamma N \to \Delta$ transition,\cite{sato} pion cloud contributions to the $M1$ form factor fall from 50\% of the total at $Q^2=0$ to $\lsim 10$\% for $Q^2\gtrsim 2\,$GeV$^2$.  Hence, the evolution of $\mu_p\, G_E^p(Q^2)/G_M^p(Q^2)$ on $Q^2\gtrsim 2\,$GeV$^2$ is primarily determined by the quark core of the proton.  This is evident in Fig.\,\ref{plotGEpGMp}, which illustrates that, on $Q^2\in (1,5)\,$GeV$^2$, $\mu_p\, G_E^p(Q^2)/G_M^p(Q^2)$ is sensitive to the parameters defining the axial-vector-diquark--photon vertex.  The response diminishes with increasing $Q^2$ because the representation of the diquark current expresses the perturbative limit.  

The behaviour of $\mu_p\, G_E^p(Q^2)/G_M^p(Q^2)$ on $Q^2\gtrsim 2\,$GeV$^2$ is determined either by correlations expressed in the Faddeev amplitude, the electromagnetic properties of the constituent degrees of freedom, or both.  The issue is decided by the fact that the magnitude and trend of the results are not materially affected by the axial-vector-diquarks' electromagnetic parameters.  This observation suggests strongly that the ratio's evolution is due primarily to spin-isospin correlations in the nucleon's Faddeev amplitude.  It is notable that while Set~A is ruled out by the data, Set~B, which anticipates pion cloud effects, is in reasonable agreement with both the trend and magnitude of the polarisation transfer data \cite{jones,roygayou,gayou}.  NB.\ Neither this nor the Rosenbluth\,\cite{walker} data played any role in developing the Faddeev equation and nucleon current.  The agreement between calculation and experiment therefore yields clear understanding.\footnote{It is currently believed that the discrepancy between the two classes of experiment may be removed by the inclusion of two-photon exchange contributions in the analysis of the $e p$ scattering cross-section; i.e., by improving on the Born approximation.  Estimates indicate that such effects materially reduce the magnitude of the ratio inferred from Rosenbluth separation and slightly increase that inferred from the polarisation transfer measurements.\protect\cite{JLabRosenbluth}}

\begin{figure}[t]
\centerline{%
\includegraphics[clip,height=0.5\textwidth]{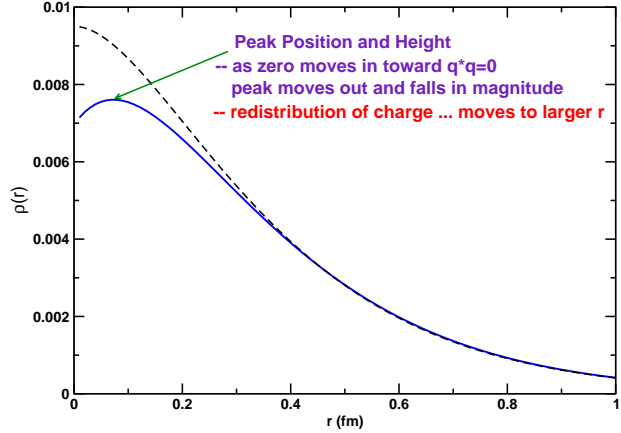}}
\caption{\label{GEzero} Three dimensional Fourier transform of a dipole charge distribution compared with that of a distribution which exhibits a zero; viz., $(1- a Q^2/m_D^2)/(1+Q^2/m_D^2)^3$, where $m_D$ is the dipole mass and $a=0.1$.  The zero indicates a depletion of charge at $r=0$, and its relocation to larger $r$.  This profile is familiar from nuclear physics: correlations in the proton wave functions of nuclei effect a similar redistribution of charge.\protect\cite{wiringa}}
\end{figure}

The calculation is extended to larger $Q^2$ in Ref.\,[\refcite{arneJ2}], with a prediction that the ratio will pass through zero at $Q^2\approx 6.5\,$GeV$^2$.  Experiments are planned at JLab that within three years will test this prediction.  If one adheres to a simple interpretation of $G_E^p(Q^2)$ as a Fourier-transform of the electric charge distribution within a proton, then the meaning of a zero in this form factor is depicted in Fig.\,\ref{GEzero}: it corresponds to a depletion of electric charge at the heart of the nucleon.  The distribution of magnetic current exhibits no such effect.  While this picture is not truly valid because the zero appears far into the relativistic domain, the parallel it draws between the effects of correlations in the wave functions of nuclei and those within the nucleon's Faddeev amplitude are useful.

\begin{figure}[t]
\centerline{
\includegraphics[width=0.63\textwidth,angle=270]{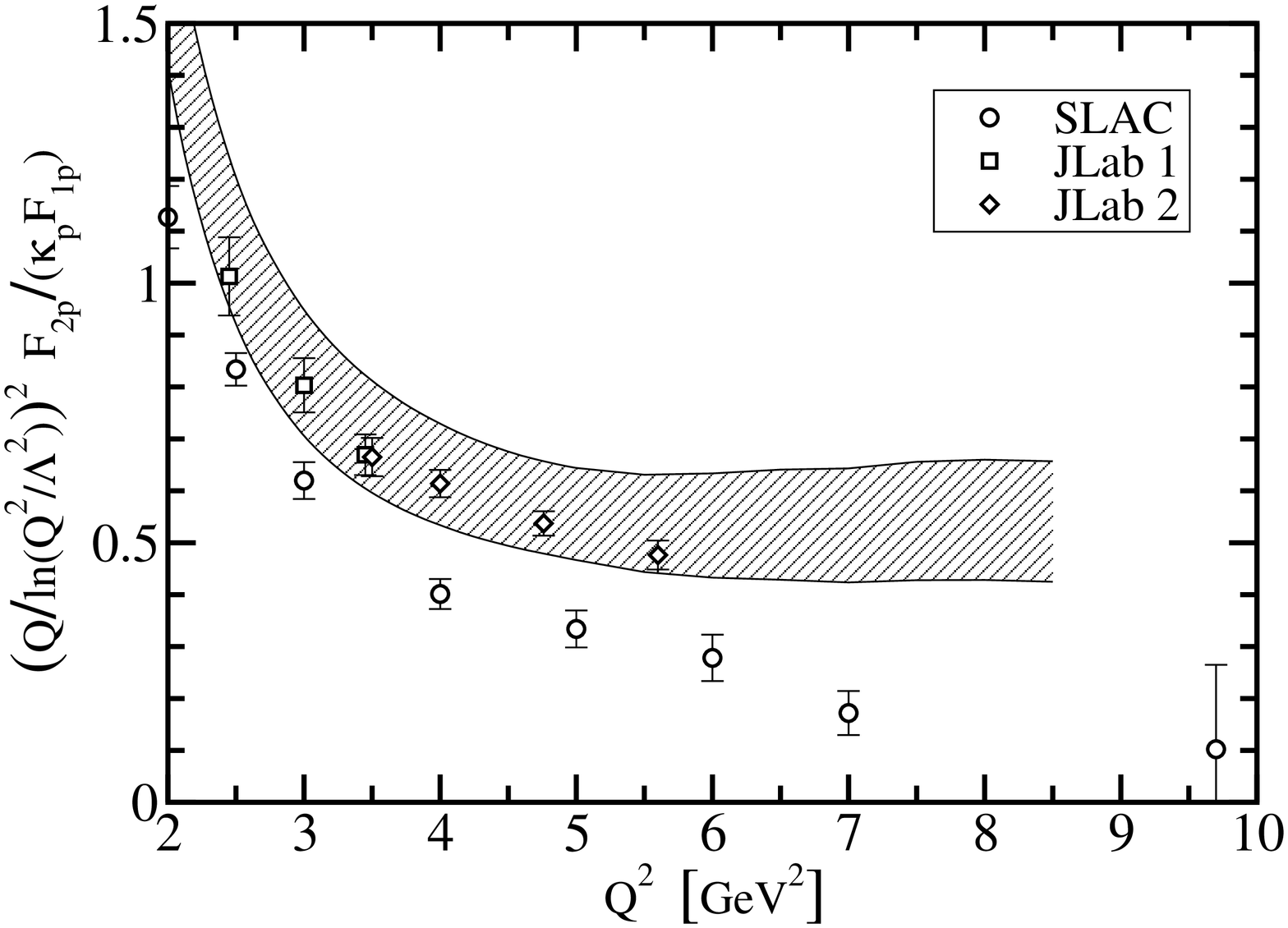}}
\caption{\label{plotF2F1log}  Weighted proton Pauli$/$Dirac form factor ratio, calculated with $\Lambda= 0.94\,$GeV.  The data are described in Fig.\,\protect\ref{plotGEpGMp}.  The band was calculated using the point-particle values: $\mu_{1^+}=2$ and $\chi_{1^+}=1$, and $\kappa_{\check T} = 2$.  Here the upper border is obtained with $\mu_{1^+}=1$, $\chi_{1^+}=1$ and $\kappa_{\check T}= 2$, and the lower with $\mu_{1^+}=3$.  (Adapted from Ref.\,[\protect\refcite{arneJ2}].)}
\end{figure}

In Fig.\,\ref{plotF2F1log} we plot a weighted ratio of Pauli and Dirac form factors.  A perturbative QCD analysis\,\cite{belitsky} that considers effects arising from both the proton's leading- and subleading-twist light-cone wave functions, the latter of which represents quarks with one unit of orbital angular momentum, suggests
\begin{equation}
\label{scaling}
\frac{Q^2}{[\ln Q^2/\Lambda^2]^2} \, \frac{F_2(Q^2)}{F_1(Q^2)} =\,{\rm constant,}\;\; Q^2\gg \Lambda^2\,,
\end{equation}  
where $\Lambda$ is a mass-scale that corresponds to an upper-bound on the domain of soft momenta.  An argument may be made that a judicious estimate of the least-upper-bound on this domain is\,\cite{arneJ} $\Lambda = M$.  The figure hints that Eq.\,(\ref{scaling}) may be valid for $Q^2 \gsim 6\,$GeV$^2$.  NB.\ The model for the nucleon and its current reviewed herein is consistent with quark counting rules, albeit neglecting the anomalous dimensions that arise via renormalisation.  However, they were also omitted in deriving Eq.\,(\ref{scaling}).

We reiterate that orbital angular momentum is not a Poincar\'e invariant.  However, if absent in a particular frame, it will in general appear in another frame related via a Poincar\'e transformation.  Nonzero quark orbital angular momentum is a necessary outcome of a Poincar\'e covariant description, and a nucleon's covariant Faddeev amplitude possesses structures that correspond in the rest frame to $s$-wave, $p$-wave and even $d$-wave components.  The result in Fig.\,\ref{plotF2F1log} is not significantly influenced by details of the diquarks' electromagnetic properties.  Instead, the behaviour is primarily governed by correlations expressed in the proton's Faddeev amplitude and, in particular, by the amount of intrinsic quark orbital angular momentum.\cite{blochff}  NB.\ This phenomenon is analogous to that observed in connection with the pion's electromagnetic form factor.  In that instance the so-called axial-vector components of the pion's Bethe-Salpeter amplitude, Eqs.\,(\ref{fwti}) \& (\ref{gwti}), are responsible for the large $Q^2$ behaviour of the form factor: they alone ensure $Q^2 F_\pi(Q^2) \approx \,$constant for truly ultraviolet momenta,\cite{mrpion} which is the result anticipated from perturbative QCD.\cite{pQCDpionFF}  These components are required by covariance\,\cite{mrt98} and signal the presence of quark orbital angular momentum in the pseudoscalar pion.

\section{Epilogue}
\label{epilogue}
\setcounter{equation}{0}
Protons and neutrons are the seeds of all the universe's observable matter.  The standard model of particle physics is supposed to explain their properties.  However, this theory's perturbative formulation fails spectacularly to account for even the simplest bulk properties.  Two fundamental, emergent phenomena are responsible: confinement and dynamical chiral symmetry breaking.  Their importance is difficult to overestimate.  They determine which chemical elements are stable and hence influence even the existence of life.  

Dynamical chiral symmetry breaking (DCSB) is a singularly effective mass generating mechanism.  It can take the almost massless light-quarks of perturbative QCD and turn them into the massive constituent-quarks whose mass sets the scale which characterises the spectrum of the strong interaction accessible at JLab.  The phenomenon is understood via QCD's gap equation, the solution of which delivers a quark mass function with a momentum-dependence that connects the perturbative and nonperturbative, constituent-quark domains.  

Despite the fact that light-quarks are made heavy, the mass of the pseudoscalar mesons remains unnaturally small.  That, too, owes to DCSB, expressed this time in a remarkable relationship between QCD's gap equation and those colour singlet Bethe-Salpeter equations which have a pseudoscalar projection.  Goldstone's theorem is a natural consequence of this connection.

These features may only be veraciously understood in relativistic quantum field theory.  They can be viewed as an essential consequence of the presence and role of particle-antiparticle pairs in an asymptotically free theory.  This is apparent via a self-consistent solution of the appropriate Dyson-Schwinger equations (DSEs), which wrap each of QCD's elementary excitations in a cloud of virtual particles that is exceedingly dense at low momentum.

Indeed, the DSEs provide a natural framework for the exploration of QCD's emergent phenomena.  They are a generating tool for perturbation theory and thus give a clean connection with processes that are well understood.  Moreover, they admit a systematic, symmetry preserving and nonperturbative truncation scheme, and thereby give access to 
strong QCD in the continuum.  On top of this, a quantitative comparison and feedback between DSE and lattice-QCD studies is today proving fruitful.

The existence of a sensible truncation scheme enables the proof of exact results using the DSEs.  That the truncation scheme is also tractable provides a means by which the results may be illustrated, and furthermore a practical tool for the prediction of observables that are accessible at contemporary experimental facilities.  The consequent opportunities for rapid feedback between experiment and theory brings within reach an intuitive understanding of nonperturbative strong interaction phenomena.

Modern, high-luminosity experimental facilities employ large momentum transfer reactions to probe the structure of hadrons.  They are providing remarkable and intriguing new information.  For an example one need only look so far as the discrepancy between the ratio of electromagnetic proton form factors extracted via Rosenbluth separation and that inferred from polarisation transfer.  This discrepancy is marked for $Q^2\gsim 2\,$GeV$^2$ and grows with increasing $Q^2$.  At such values of momentum transfer, $Q^2 > M^2$, where $M$ is the nucleon's mass, a true understanding of these and other contemporary data require a Poincar\'e covariant description of the nucleon.  This can be obtained with a Faddeev equation that describes a baryon as composed primarily of a quark core, constituted of confined quark and confined diquark correlations, but augmented by pseudoscalar meson cloud contributions that are sensed by long wavelength probes.  Short wavelength probes pierce the cloud, and expose spin-isospin correlations and quark orbital angular momentum within the baryon.  The veracity of the elements in this description makes plain that a picture of baryons as a bag of three constituent-quarks is profoundly misleading.

While there are indications that confinement may be expressed in the analyticity properties of the dressed propagators, one cannot say that it is understood.  Consequently, one pressing task, to which the methods described herein can be applied, is the drawing of an accurate map of the confinement force between light-quarks within mesons.  This will enable a clear connection to be established between this force and the realisation of dynamical chiral symmetry breaking, and an accounting of the distribution of mass within mesons.  In addition one may then begin to determine whether the confinement force-field can be excited to produce exotic systems of light-quarks and glue.  The same must be done for baryons, but here the difficulties are greater because one is confronted with a Poincar\'e covariant three-body problem.  New and improved tools must be developed, which may extend beyond a rigorous grounding of the Faddeev equation.  

It is also crucial to develop the tools necessary for charting the pointwise distribution of quarks and gluons within hadrons.  With such in hand, one might lay out and apportion the pointwise distribution of mass within the hadron, and its evolution with the resolving scale of the probe.  That knowledge could be used to elucidate the impact of confinement and DCSB on these distributions; e.g., by exhibiting the effects of quenching these emergent phenomena.  

It should now be plain that contemporary nuclear physics poses numerous challenges and may reasonably be expected to provide many new surprises.

\section*{Acknowledgments}
In preparing this article we benefited from conversations with A.~Bashir, L.~Chang, P.~Jaikumar, A.~Krassnigg, Y.-X.~Liu, P.~Maris, A.~Raya, S.\,M.~Schmidt and P.\,C.~Tandy; and from the solitude to be found in Nutteln.
This work is supported by: 
Department of Energy, Office of Nuclear Physics, contract no.\ W-31-109-ENG-38; 
\textit{Helmholtz-Gemeinschaft} Virtual Theory Institute VH-VI-041; 
the \textit{A.\,v.\ Humboldt-Stiftung} via a \textit{F.\,W.\ Bessel Forschungspreis}; 
and benefited from the facilities of ANL's Computing Resource Center.

\appendix

\section{Euclidean Space}
\label{Appendix1}
It is possible to view the Euclidean formulation of a quantum field theory as  definitive.\cite{glimm,symanzik,streater,seiler}  That decision is crucial when a consideration of nonperturbative effects becomes important.  In addition, the discrete lattice formulation in Euclidean space has allowed some progress to be made in attempting to answer existence questions for interacting gauge field theories.  NB.\ A lattice formulation is impossible in Minkowski space because the integrand is not non-negative and hence does not provide a probability measure.  An heuristic exposition of probability measures in quantum field theory can be found in Ref.\,[\refcite{rivers}], Chap.\,6, while Ref.\,[\refcite{glimm}], Chaps.\,3 and 6, provides a more rigorous discussion in the context of quantum mechanics and quantum field theory.

Our Euclidean conventions are easily made plain.  For $4$-vectors $a$, $b$:
\begin{equation}
a\cdot b := a_\mu\,b_\nu\,\delta_{\mu\nu} := \sum_{i=1}^4\,a_i\,b_i\,,
\end{equation}
where $\delta_{\mu\nu}$ is the Kronecker delta and the metric tensor.  Hence, a spacelike vector, $Q_\mu$, has $Q^2>0$.  The Dirac matrices are Hermitian and defined by the algebra 
\begin{equation}
\{\gamma_\mu,\gamma_\nu\} = 2\,\delta_{\mu\nu}\,.
\end{equation}
We use 
\begin{equation}
\gamma_5 := -\,\gamma_1\gamma_2\gamma_3\gamma_4\,,
\end{equation}
so that 
\begin{equation}
{\rm tr}\left[ \gamma_5 \gamma_\mu\gamma_\nu\gamma_\rho\gamma_\sigma \right] = 
- 4 \,\varepsilon_{\mu\nu\rho\sigma}\,,\; \varepsilon_{1234}= 1\,.
\end{equation}

A Dirac-like representation of these matrices is: 
\begin{equation} 
\vec{\gamma}=\left( 
\begin{array}{cc} 
0 & -i\vec{\tau}  \\ 
i\vec{\tau} & 0 
\end{array} 
\right),\; 
\gamma_4=\left( 
\begin{array}{cc} 
\tau^0 & 0 \\ 
0 & -\tau^0 
\end{array} 
\right), 
\end{equation} 
where the $2\times 2$ Pauli matrices are: 
\begin{equation} 
\label{PauliMs} 
\rule{-4ex}{0ex}
\tau^0 = \left( 
\begin{array}{cc} 
1 & 0 \\ 
0 & 1 
\end{array}\right),\; 
\tau^1 = \left( 
\begin{array}{cc} 
0 & 1 \\ 
1 & 0 
\end{array}\right),\; 
\tau^2 = \left( 
\begin{array}{cc} 
0 & -i \\ 
i & 0 
\end{array}\right),\; 
\tau^3 = \left( 
\begin{array}{cc} 
1 & 0 \\ 
0 & -1 
\end{array}\right). 
\end{equation} 

It is possible to derive the Euclidean version of every equation introduced above assuming certain analytic properties of the integrands.  However, the derivations can be sidestepped using the following \textit{transcription rules}:
\begin{center} 
\parbox{30em}{ 
\parbox{16em}{Configuration Space 
\begin{enumerate} 
\item $\displaystyle \int^M \!d^4x^M \, \rightarrow \,-i \int^E \!d^4x^E$ 
\item $\slash\!\!\! \partial \,\rightarrow \, i\gamma^E\cdot \partial^E $ 
\item $\slash \!\!\!\! A \, \rightarrow\, -i\gamma^E\cdot A^E$ 
\item $A_\mu B^\mu\,\rightarrow\,-A^E\cdot B^E$ 
\item $x^\mu\partial_\mu \to x^E\cdot \partial^E$ 
\end{enumerate}}\hspace*{0.5em} 
\parbox{16em}{Momentum Space 
\begin{enumerate} 
\item $\displaystyle \int^M\! d^4k^M \, \rightarrow \,i \int^E\! d^4k^E$ 
\item $\slash\!\!\! k \,\rightarrow \, -i\gamma^E\cdot k^E $ 
\item $\slash \!\!\!\! A \, \rightarrow\, -i\gamma^E\cdot A^E$ 
\item $k_\mu q^\mu \, \rightarrow\, - k^E\cdot q^E$ 
\item $k_\mu x^\mu\,\rightarrow\,-k^E\cdot x^E$ 
\end{enumerate}}} 
\end{center} 
These rules are valid in perturbation theory; i.e., the correct Euclidean space integral for a given diagram will be obtained by applying these rules to the Minkowski integral.  The rules take account of the change of variables and Wick rotation of the contour.  When one begins with Euclidean space, as we do, the reverse is also true.  However, for diagrams that represent DSEs which involve dressed $n$-point functions, whose analytic structure is not known \textit{a priori}, the Minkowski space equation obtained using this prescription will have the right appearance but it's solutions may bear no relation to the analytic continuation of the solution of the Euclidean equation.  Any such differences will be nonperturbative in origin.  It is this fact that makes a choice of metric crucial at the outset.

To return to practical matters, a positive energy spinor satisfies 
\begin{equation} 
\label{DiracN}
\bar u(P,s)\, (i \gamma\cdot P + M) = 0 = (i\gamma\cdot P + M)\, u(P,s)\,, 
\end{equation} 
where $M$ is the mass obtained by solving the Faddeev equation and $s=\pm$ is the spin label.  The spinor is normalised: 
\begin{equation} 
\bar u(P,s) \, u(P,s) = 2 M \,,
\end{equation} 
and may be expressed explicitly: 
\begin{equation} 
u(P,s) = \sqrt{M- i {\check{E}}}\left( 
\begin{array}{l} 
\chi_s\\ 
\displaystyle \frac{\vec{\sigma}\cdot \vec{P}}{M - i \check{E}} \chi_s 
\end{array} 
\right)\,, 
\end{equation} 
with $\check{E} = i \sqrt{\vec{P}^2 + M^2}$, 
\begin{equation} 
\chi_+ = \left( \begin{array}{c} 1 \\ 0  \end{array}\right)\,,\; 
\chi_- = \left( \begin{array}{c} 0\\ 1  \end{array}\right)\,. 
\end{equation} 
For the free-particle spinor, $\bar u(P,s)= u(P,s)^\dagger \gamma_4$. 
 
The spinor can be used to construct a positive energy projection operator: 
\begin{equation} 
\label{Lplus} \Lambda_+(P):= \frac{1}{2 M}\,\sum_{s=\pm} \, u(P,s) \, \bar 
u(P,s) = \frac{1}{2M} \left( -i \gamma\cdot P + M\right). 
\end{equation} 
 
A negative energy spinor satisfies 
\begin{equation} 
\bar v(P,s)\,(i\gamma\cdot P - M) = 0 = (i\gamma\cdot P - M) \, v(P,s)\,, 
\end{equation} 
and possesses properties and satisfies constraints obtained via obvious analogy 
with $u(P,s)$. 
 
A charge-conjugated Bethe-Salpeter amplitude is obtained via 
\begin{equation} 
\label{chargec}
\bar\Gamma(k;P) = C^\dagger \, \Gamma(-k;P)^{\rm T}\,C\,, 
\end{equation} 
where ``T'' denotes a transposing of all matrix indices and 
$C=\gamma_2\gamma_4$ is the charge conjugation matrix, $C^\dagger=-C$. 
 
In describing the $\Delta$ resonance we employ a Rarita-Schwinger spinor to 
unambiguously represent a covariant spin-$3/2$ field.  The positive energy 
spinor is defined by the following equations: 
\begin{equation} 
\label{rarita}
(i \gamma\cdot P + M)\, u_\mu(P;r) = 0\,,\;
\gamma_\mu u_\mu(P;r) = 0\,,\;
P_\mu u_\mu(P;r) = 0\,, 
\end{equation} 
where $r=-3/2,-1/2,1/2,3/2$.  It is normalised: 
\begin{equation} 
\bar u_{\mu}(P;r^\prime) \, u_\mu(P;r) = 2 M\,, 
\end{equation} 
and satisfies a completeness relation 
\begin{equation} 
\frac{1}{2 M}\sum_{r=-3/2}^{3/2} u_\mu(P;r)\,\bar u_\nu(P;r) = 
\Lambda_+(P)\,R_{\mu\nu}\,, 
\end{equation} 
where 
\begin{equation} 
R_{\mu\nu} = \delta_{\mu\nu} \mbox{\boldmath $I$}_{\rm D} -\frac{1}{3} \gamma_\mu \gamma_\nu + 
\frac{2}{3} \hat P_\mu \hat P_\nu \mbox{\boldmath $I$}_{\rm D} - i\frac{1}{3} [ \hat P_\mu 
\gamma_\nu - \hat P_\nu \gamma_\mu]\,, 
\end{equation} 
with $\hat P^2 = -1$, which is very useful in simplifying the positive energy 
$\Delta$'s Faddeev equation. 



\section*{References}
\begin{enumerate}

\bibitem{bd1} J.\,D.~Bjorken and S.\,D.~Drell,
\textit{Relativistic Quantum Mechanics} (McGraw-Hill, New York, 1964). \label{ibd1}

\bibitem{kp91} B.\,D.~Keister and W.\,N.~Polyzou, Adv.\ Nucl.\ Phys.\ \textbf{20}, 225.

\bibitem{co92} F.~Coester,
  Prog.\ Part.\ Nucl.\ Phys.\  {\bf 29}, 1 (1992).

\bibitem{nobelQCD} See, e.g., the official web page for the 2004 Nobel Prize: \href{http://nobelprize.org/physics/laureates/2004/}%
{http://nobelprize.org/physics/laureates/2004/}

\bibitem{cdrvalence} M.\,B.~Hecht, C.\,D.~Roberts and S.\,M.~Schmidt,
  Phys.\ Rev.\ \textbf{C\,63}, 025213 (2001);
  C.\,D.~Roberts,
  Nucl.\ Phys.\ Proc.\ Suppl.\  {\bf 108}, 227 (2002).

\bibitem{reimervalence} K.~Wijesooriya, P.\,E.~Reimer and R.\,J.~Holt,
  Phys.\ Rev.\ \textbf{C\,72}, 065203 (2005).

\bibitem{iz80} C.~Itzykson and J.-B.~Zuber, \textit{Quantum Field Theory} (McGraw-Hill, New York, 1980). \label{iiz80}

\bibitem{pt84} P.~Pascual and R.~Tarrach, Lecture Notes in Physics, Vol.\ {\bf 194}, \textit{QCD: Renormalization for the Practitioner} (Springer-Verlag, Berlin, 1984).\label{ipt84}

\bibitem{fh65} R.\,P.~Feynman and A.\,R.~Hibbs, \emph{Quantum Mechanics and Path Integrals} (McGraw-Hill, New York, 1965).

\bibitem{fb66} F.\,A.~Berezin, \emph{The Method of Second Quantization} (Academic Press, New York, 1966). \label{ifb66}

\bibitem{agwGribov} A.\,G.~Williams,
  Prog.\ Theor.\ Phys.\ Suppl.\  {\bf 151}, 154 (2003). \label{iagwGribov}
  
\bibitem{dyson49} F.\,J.~Dyson,
  Phys.\ Rev.\  {\bf 75} (1949) 1736.

\bibitem{schwinger51} J.\,S.~Schwinger,
  Proc.\ Nat.\ Acad.\ Sci.\  {\bf 37}, 452 (1951);
\textit{ibid}.\ 455.
  
\bibitem{cdragw} C.\,D.~Roberts and A.\,G.~Williams,
  Prog.\ Part.\ Nucl.\ Phys.\  {\bf 33}, 477 (1994). \label{icdragw}
  
\bibitem{bpr91} C.\,J.~Burden, J.~Praschifka and C.\,D.~Roberts,
  Phys.\ Rev.\ D {\bf 46}, 2695 (1992). \label{ibpr91}

\bibitem{br93} C.\,J.~Burden and C.\,D.~Roberts,
  Phys.\ Rev.\ D {\bf 47}, 5581 (1993). \label{ibr93}
  
\bibitem{dong} Z.-h.~Dong, H.\,J.~Munczek and C.\,D.~Roberts,
  Phys.\ Lett.\ B {\bf 333}, 536 (1994). \label{idong}
  
\bibitem{raya} A.~Bashir and A.~Raya, ``Gauge symmetry and its implications for the Schwinger-Dyson equations,'' in \textit{Trends in Boson Research}, ed.\ A.\,V.~Ling  (Nova Science, New York, 2006);
\href{http://www.arXiv.org/abs/hep-ph/0411310}{{\tt hep-ph/0411310}}. 

\bibitem{raya1} A.~Bashir and A.~Raya,
Phys.\ Rev.\ D {\bf 66}, 105005 (2002);
Nucl.\ Phys.\ B {\bf 709}, 307 (2005).

\bibitem{raya2} A.~Bashir, M.\,R.~Pennington and A.~Raya, ``On gauge independent dynamical chiral symmetry breaking,'' 
\href{http://www.arXiv.org/abs/hep-ph/0511291}{{\tt hep-ph/0511291}}. \label{iraya2}

\bibitem{bd2} J.\,D.~Bjorken and S.\,D.~Drell,
\textit{Relativistic Quantum Fields} (McGraw-Hill, New York, 1965). \label{ibd2}

\bibitem{nobelStern} See, e.g., the official web page for the 1943 Nobel Prize: \href{http://nobelprize.org/physics/laureates/1943/}%
{http://nobelprize.org/physics/laureates/1943/}.

\bibitem{gellmannNobel} See, e.g., the official web page for the 1969 Nobel Prize:
\href{http://nobelprize.org/physics/laureates/1969/index.html}%
{http://nobelprize.org/physics/laureates/1969/index.html}.

\bibitem{gao} H.\,y.~Gao,
Int.\ J.\ Mod.\ Phys.\ {\bf E\,12}, 1 (2003)
[Erratum-ibid.\ {\bf E\,12}, 567 (2003)].

\bibitem{leeburkert} V.\,D.~Burkert and T.-S.\,H.~Lee, 
Int.\ J.\ Mod.\ Phys.\ E {\bf 13}, 1035 (2004).

\bibitem{walker} R.\,C.~Walker {\it et al.},
Phys.\ Rev.\ {\bf D\,49}, 5671 (1994). \label{iwalker}

\bibitem{jones} M.\,K.\ Jones {\it et al.}  [JLab Hall A Collaboration],
Phys.\ Rev.\ Lett.\ {\bf 84}, 1398 (2000). \label{ijones} 

\bibitem{roygayou} O.\ Gayou, {\it et al.},
Phys.\ Rev.\ {\bf C\,64}, 038202 (2001).

\bibitem{gayou} O.\ Gayou, {\it et al.}  [JLab Hall A Collaboration],
Phys.\ Rev.\ Lett.\ {\bf 88}, 092301 (2002).  \label{igayou}

\bibitem{arrington} J.~Arrington,
Phys.\ Rev.\ {\bf C\,69}, 022201 (2004).

\bibitem{qattan} I.\,A.~Qattan {\it et al.}, 
Phys.\ Rev.\ Lett.\  {\bf 94}, 142301 (2005).

\bibitem{cdrANU} C.\,D.~Roberts, ``Nonperturbative QCD with modern tools,'' in \textit{Frontiers in Nuclear Physics: From Quark - Gluon Plasma to Supernova}, ed.\ S.~Kuyucak (World Scientific, Singapore, 1999) pp.~212-261. \label{icdrANU}

\bibitem{pdg} S.~Eidelman \textit{et al.}, 
Phys.\ Lett.\ \textbf{B\,592}, 1 (2004). \label{ipdg}

\bibitem{kurtcondensate} K.~Langfeld, H.~Markum, R.~Pullirsch, C.\,D.~Roberts and S.\,M.~Schmidt,
  Phys.\ Rev.\ C {\bf 67}, 065206 (2003).

\bibitem{bastirev} C.\,D.\ Roberts and S.\,M.\ Schmidt,
Prog.\ Part.\ Nucl.\ Phys.\  \textbf{45}, S1 (2000). \label{ibastirev}

\bibitem{cdrwien} C.\,D.~Roberts, ``Continuum strong QCD: Confinement and dynamical chiral symmetry  breaking,'' Contribution to Confinement Research Program at the Erwin Schodinger Institute, Vienna, Austria, 5 May - 17 Jul 2000,  
\href{http://www.arXiv.org/abs/nucl-th/0007054}{{\tt nucl-th/0007054}}. \label{icdrwien}

\bibitem{reinhardrev} R.~Alkofer and L.~von~Smekal, 
Phys.\ Rept.\ \textbf{353}, 281 (2001). \label{ireinhardrev}

\bibitem{pieterrev} P.\ Maris and C.\,D.\ Roberts,
Int.\ J.\ Mod.\ Phys.\ \textbf{E\,12}, 297 (2003). \label{ipieterrev}

\bibitem{weisenjl} U.~Vogl and W.~Weise,
Prog.\ Part.\ Nucl.\ Phys.\  \textbf{27}, 195 (1991). \label{iweisenjl}

\bibitem{klevanskynjl} S.\,P.~Klevansky, 
Rev.\ Mod.\ Phys.\  \textbf{64}, 649 (1992). \label{iklevanskynjl}
  
\bibitem{ebertnjl}
  D.~Ebert, H.~Reinhardt and M.~K.~Volkov,
  Prog.\ Part.\ Nucl.\ Phys.\  \textbf{33}, 1 (1994). \label{iebertnjl}
  
\bibitem{mn83} H.\,J.~Munczek and A.\,M.~Nemirovsky, 
Phys.\ Rev.\ \textbf{D\,28}, 181 (1983).

\bibitem{latticegluon} D.\,B.\ Leinweber, J.\,I.\ Skullerud, A.\,G.\ Williams and C.\ Parrinello  [UKQCD Collaboration],
Phys.\ Rev.\ D {\bf 60}, 094507 (1999)
[Erratum-ibid.\ D {\bf 61}, 079901 (2000)].

\bibitem{alkoferdetmold} R.~Alkofer, W.~Detmold, C.\,S.~Fischer and P.~Maris,
  Nucl.\ Phys.\ Proc.\ Suppl.\  {\bf 141}, 122 (2005). \label{ialkoferdetmold}
  
\bibitem{lane} K.\,D.\ Lane,
Phys.\ Rev.\ D {\bf 10}, 2605 (1974). \label{ilane}

\bibitem{politzer} H.\,D.\ Politzer,
Nucl.\ Phys.\ B {\bf 117}, 397 (1976). \label{ipolitzer}

\bibitem{bowman2} P.\,O.~Bowman, U.\,M.~Heller, D.\,B~Leinweber and A\,G.~Williams, 
  Nucl.\ Phys.\ Proc.\ Suppl.\ \textbf{119}, 323 (2003). \label{ibowman2}

\bibitem{bowman} J.\,B.~Zhang, P.\,O.~Bowman, R.\,J.~Coad, U.\,M.~Heller, D.\,B.~Leinweber and A.\,G.~Williams,
Phys.\ Rev.\ D {\bf 71}, 014501 (2005).

\bibitem{bhagwat} M.S.~Bhagwat, M.A.~Pichowsky, C.D.~Roberts and P.C.~Tandy,
Phys.\ Rev.\ {\bf C\,68}, 015203 (2003). \label{ibhagwat}

\bibitem{bhagwat2} M.\,S.~Bhagwat and P.\,C.~Tandy,
Phys.\ Rev.\ {\bf D\,70}, 094039 (2004). \label{ibhagwat2}

\bibitem{entire1} H.\,J.~Munczek, Phys.\ Lett.\ {\bf B\,175}, 215 (1986). \label{ientire1} 

\bibitem{entire2} C.\,J.~Burden, C.\,D.~Roberts and A.\,G.~Williams, Phys.\ Lett.\ {\bf B\,285}, 347 (1992). \label{ientire2} 

\bibitem{stingl} M.~Stingl,
Phys.\ Rev.\ {\bf D\,34}, 3863 (1986)
[Erratum-ibid.\ {\bf D\,36}, 651 (1987)]. \label{istingl}

\bibitem{krein} G.~Krein, C.\,D.~Roberts and A.\,G.~Williams,
Int.\ J.\ Mod.\ Phys.\ {\bf A\,7}, 5607 (1992). \label{ikrein}

\bibitem{glimm} J.~Glimm and A.~Jaffee, \textit{Quantum Physics.\ A Functional Point of View} (Springer-Verlag, New York, 1981). \label{iglimm}

\bibitem{dsenonzeroT} A.~Bender, D.~Blaschke, Yu.\,L.~Kalinovsky and C.\,D.~Roberts,
  Phys.\ Rev.\ Lett.\  {\bf 77}, 3724 (1996).
  
\bibitem{dsenonzeromu} A.~Bender, G.\,I.~Poulis, C.\,D.~Roberts, S.\,M.~Schmidt and A.\,W.~Thomas,
  Phys.\ Lett.\ B {\bf 431}, 263 (1998).
  
\bibitem{hawes} F.\,T.~Hawes, P.~Maris and C.\,D.~Roberts,
  Phys.\ Lett.\ B {\bf 440}, 353 (1998). \label{ihawes}
  
\bibitem{jisvertex} J.\,I.~Skullerud, P.\,O.~Bowman, A.~Kizilersu, D.\,B.~Leinweber and A.\,G.~Williams,
  JHEP {\bf 0304}, 047 (2003). \label{ijisvertex}

\bibitem{bhagwatvertex} M.\,S.~Bhagwat, A.~H\"oll, A.~Krassnigg, C.\,D.~Roberts and P.\,C.~Tandy,
Phys.\ Rev.\ {\bf C\,70}, 035205 (2004). \label{ibhagwatvertex}

\bibitem{marisphotonvertex} P.~Maris and P.\,C.~Tandy,
  Phys.\ Rev.\ C {\bf 61}, 045202 (2000). \label{imarisphotonvertex}

\bibitem{munczek} H.\,J.~Munczek,
  Phys.\ Rev.\ \textbf{D\,52} 4736 (1995).
  
\bibitem{truncscheme} A.~Bender, C.\,D.~Roberts and L.~von Smekal, 
Phys.\ Lett.\ {\bf B 380}, 7 (1996). \label{itruncscheme}

\bibitem{mrt98} P.~Maris, C.\,D.~Roberts and P.\,C.~Tandy,
Phys.\ Lett.\ {\bf B\,420}, 267 (1998).

\bibitem{mrpion} P.~Maris and C.\,D.~Roberts,
Phys.\ Rev.\ C {\bf 58}, 3659 (1998). \label{imrpion}

\bibitem{marismisha} M.\,A.~Ivanov, Yu.\,L.~Kalinovsky, P.~Maris and C.\,D.~Roberts,
  Phys.\ Lett.\ B {\bf 416}, 29 (1998).

\bibitem{neubert93} M.~Neubert,
  Phys.\ Rept.\  {\bf 245}, 259 (1994).

\bibitem{marisAdelaide} P.~Maris and C.\,D.~Roberts, ``QCD bound states and their response to extremes of temperature and density,'' in \emph{Proc.\ of Wkshp.\ on Nonperturbative Methods in Quantum Field Theory},  eds.\ A.\,W.~Schreiber, A.\,G.~Williams and A.\,W.~Thomas (World Scientific, Singapore, 1998) pp.~132-151.

\bibitem{mishaSVY} M.\,A.~Ivanov, Yu.\,L.~Kalinovsky and C.\,D.~Roberts,
Phys.\ Rev.\ {\bf D\,60}, 034018 (1999). \label{imishaSVY}

\bibitem{experiment} G.\ Bellini {\it et al.},
Phys.\ Rev.\ Lett.\ {\bf 48}, 1697 (1982).

\bibitem{a1b1} J.\,C.\,R.~Bloch, Yu.\,L.~Kalinovsky, C.\,D.~Roberts and S.\,M.~Schmidt,
Phys.\ Rev.\ \textbf{D\,60}, 111502 (1999).

\bibitem{krassnigg1} A.~H\"oll, A.~Krassnigg and C.\,D.~Roberts,
  Phys.\ Rev.\ {\bf C\,70}, 042203(R) (2004). \label{ikrassnigg1}

\bibitem{krassnigg2} A.~H\"oll, A.~Krassnigg, P.~Maris, C.\,D.~Roberts and S.\,V.~Wright,
  Phys.\ Rev.\ C {\bf 71}, 065204 (2005). \label{ikrassnigg2}

\bibitem{dominguez} C.\,A.~Dominguez,
Phys.\ Rev.\ {\bf D\,15}, 1350 (1977);
{\it ibid.} {\bf D\,16}, 2313 (1977). \label{idominguez}

\bibitem{volkov} M.\,K.\ Volkov and V.\,L.\ Yudichev,
Phys.\ Part.\ Nucl.\  {\bf 31}, 282 (2000)
[Fiz.\ Elem.\ Chast.\ Atom.\ Yadra {\bf 31}, 576 (2000)]. \label{ivolkov}

\bibitem{bicudo} P.~Bicudo,
Phys.\ Rev.\ \textbf{C\,67}, 035201 (2003).

\bibitem{lc03} A.~H\"oll, A.~Krassnigg and C.\,D.~Roberts, ``DSEs and pseudoscalar mesons: an aper\c{c}u,'' in \textit{Light-Cone Physics: Hadrons and Beyond}, ed.\ S.\ Dalley, pp.~158-165,  
\href{http://www.arXiv.org/abs/nucl-th/0311033}{{\tt nucl-th/0311033}}. 

\bibitem{maristandy4} P.~Maris and P.\,C.~Tandy,
  Phys.\ Rev.\ {\bf C\,65}, 045211 (2002). \label{imaristandy4}
  
\bibitem{kekez} D.~Kekez and D.~Klabu\v{c}ar,
Phys.\ Lett.\ {\bf B\,457}, 359 (1999).

\bibitem{jainmunczek} P.~Jain and H.\,J.~Munczek,
  Phys.\ Rev.\ {\bf D\,48}, 5403 (1993). \label{ijainmunczek}

\bibitem{mr97} P.~Maris and C.\,D.~Roberts,
  Phys.\ Rev.\ {\bf C\,56}, 3369 (1997). \label{imr97}

\bibitem{klabucar} D.~Klabu\v{c}ar and D.~Kekez,
  Phys.\ Rev.\ {\bf D\,58}, 096003 (1998). \label{iklabucar}
  
\bibitem{maristandy1} P.~Maris and P.\,C.~Tandy,
  Phys.\ Rev.\ {\bf C\,60}, 055214 (1999). \label{imaristandy1}

\bibitem{maristandy3} P.~Maris and P.\,C.~Tandy,
  Phys.\ Rev.\ C {\bf 62}, 055204 (2000). \label{imaristandy3}
  
\bibitem{marisji} C.-R.~Ji and P.~Maris,
  Phys.\ Rev.\ D {\bf 64}, 014032 (2001). \label{imarisji}
  
\bibitem{marisbicudo} P.~Bicudo, S.~Cotanch, F.~Llanes-Estrada, P.~Maris, E.~Ribeiro and A.~Szczepaniak,
  Phys.\ Rev.\ D {\bf 65}, 076008 (2002). \label{imarisbicudo}

\bibitem{maristandy5} D.~Jarecke, P.~Maris and P.\,C.~Tandy,
  Phys.\ Rev.\ C {\bf 67}, 035202 (2003). \label{imaristandy5}
  
\bibitem{mariscotanch} S.~R.~Cotanch and P.~Maris,
  Phys.\ Rev.\ {\bf D\,66}, 116010 (2002). \label{imariscotanch}
\textit{ibid} \textbf{D\,68}, 036006 (2003). 

\bibitem{wright05}  A.~H\"oll, P.~Maris, C.\,D.~Roberts and S.\,V.~Wright, ``Schwinger functions and light-quark bound states, and sigma terms,'' 
\href{http://www.arXiv.org/abs/nucl-th/0512048}{{\tt nucl-th/0512048}}. \label{iwright05}
  
\bibitem{raya3} P.~Maris, A.~Raya, C.\,D.~Roberts and S.\,M.~Schmidt,
  Eur.\ Phys.\ J.\ {\bf A\,18}, 231 (2003).

\bibitem{Volmer2000} J.~Volmer, \textit{et al.}  [JLab $F_\pi$
Collaboration],
Phys.\ Rev.\ Lett.\  \textbf{86}, 1713 (2001). \label{iVolmer2000} 

\bibitem{benderpion} R.~Alkofer, A.~Bender and C.\,D.~Roberts, 
Int.\ J.\ Mod.\ Phys.\  \textbf{A\,10}, 3319 (1995). 

\bibitem{detmold} A.~Bender, W.~Detmold, C.\,D.~Roberts and A.\,W.~Thomas, 
Phys.\ Rev.\ \textbf{C\,65}, 065203 (2002). \label{idetmold} 

\bibitem{mariscairns} P.~Maris and P.\,C.~Tandy, ``QCD modeling of hadron physics,'' \href{http://www.arXiv.org/abs/nucl-th/0511017}{{\tt nucl-th/0511017}}. \label{imariscairns}

\bibitem{millerfrank} G.\,A.~Miller and M.\,R.~Frank,
Phys.\ Rev.\ {\bf C\,65}, 065205 (2002). \label{imillerfrank}

\bibitem{boffi} S.~Boffi, L.\,Y.~Glozman, W.\,H.~Klink, W.~Plessas, M.~Radici and R.\,F.~Wagenbrunn,  
Eur.\ Phys.\ J.\ {\bf A\,14}, 17 (2002).\label{iboffi}

\bibitem{fuda} M.\,G.~Fuda and H.~Alharbi,
Phys.\ Rev.\ {\bf C\,68}, 064002 (2003).\label{ifuda}

\bibitem{stan} S.\,J.~Brodsky, J.\,R.~Hiller, D.\,S.~Hwang and V.\,A.~Karmanov,
Phys.\ Rev.\ {\bf D\,69}, 076001 (2004).\label{istan}

\bibitem{bruno} B.~Julia-Diaz, D.\,O.~Riska and F.~Coester,
Phys.\ Rev.\ {\bf C\,69}, 035212 (2004).\label{ibruno}

\bibitem{regfe} R.\,T.~Cahill, C.\,D.~Roberts and J.~Praschifka,
Austral.\ J.\ Phys.\  {\bf 42}, 129 (1989).

\bibitem{hugofe} H.~Reinhardt,
Phys.\ Lett.\ {\bf B\,244}, 316 (1990).

\bibitem{regdq} R.\,T.~Cahill, C.\,D.~Roberts and J.~Praschifka,
Phys.\ Rev.\ {\bf D\,36}, 2804 (1987).

\bibitem{njldiquark} G.~Hellstern, R.~Alkofer and H.~Reinhardt,
  Nucl.\ Phys.\ {\bf A\,625}, 697 (1997).

\bibitem{cjbfe} C.\,J.~Burden, R.\,T.~Cahill and J.~Praschifka,
Austral.\ J.\ Phys.\  {\bf 42}, 147 (1989). \label{icjbfe}

\bibitem{bentzfe} H.~Asami, N.~Ishii, W.~Bentz and K.~Yazaki,
Phys.\ Rev.\ {\bf C\,51}, 3388 (1995). \label{ibentzfe}

\bibitem{oettelfe} M.~Oettel, G.~Hellstern, R.~Alkofer and H.~Reinhardt,
Phys.\ Rev.\ {\bf C\,58}, 2459 (1998).\label{ioettelfe}

\bibitem{hechtfe} M.\,B.~Hecht, M.~Oettel, C.\,D.~Roberts, S.\,M.~Schmidt, P.\,C.~Tandy and A.\,W.~Thomas,
Phys.\ Rev.\ {\bf C\,65}, 055204 (2002).\label{ihechtfe}

\bibitem{birsefe} A.\,H.~Rezaeian, N.\,R.~Walet and M.\,C.~Birse, 
Phys.\ Rev.\ {\bf C\,70}, 065203 (2004).

\bibitem{cjbsep} C.\,J.~Burden, L.~Qian, C.\,D.~Roberts, P.\,C.~Tandy and M.\,J.~Thomson,
Phys.\ Rev.\ {\bf C\,55}, 2649 (1997).

\bibitem{marisdq} P.~Maris,
Few Body Syst.\  {\bf 32}, 41 (2002).

\bibitem{oettel2} M.~Oettel, R.~Alkofer and L.~von~Smekal, 
Eur.\ Phys.\ J.\ A {\bf 8}, 553 (2000). \label{ioettel2}

\bibitem{jacquesA} J.\,C.\,R~Bloch, C.\,D.~Roberts, S.\,M.~Schmidt, A.~Bender and 
M.\,R.~Frank, 
Phys.\ Rev.\ {\bf C\,60}, 062201 (1999). \label{ijacquesA}
%
\bibitem{jacquesmyriad} J.\,C.\,R.~Bloch, C.\,D.~Roberts and S.\,M.~Schmidt, 
Phys. Rev. {\bf C\,61}, 065207 (2000). \label{ijacquesmyriad}
 
\bibitem{cdrqciv} M.\,B.~Hecht, C.\,D.~Roberts and S.\,M.~Schmidt, ``Contemporary applications of Dyson-Schwinger equations,'' in \textit{Wien 2000, Quark Confinement and the Hadron Spectrum --- Proc.\ of the 4th Int.\ Conf.}, eds.\ W.~Lucha, K.~Maung-Maung (World Scientific, Singapore, 2002) pp.~27-39. \label{icdrqciv}

\bibitem{nedm} M.\,B.~Hecht, C.\,D.~Roberts and S.\,M.~Schmidt, 
Phys.\ Rev.\ {\bf C\,64}, 025204 (2001). \label{inedm}

\bibitem{tonyCBM} A.\,W.~Thomas, S.~Theberge and G.\,A.~Miller, 
Phys.\ Rev.\ D {\bf 24}, 216 (1981); 
A.\,W.~Thomas, 
Adv.\ Nucl.\ Phys.\  {\bf 13}, 1 (1984);
G.\,A.~Miller,
Int.\ Rev.\ Nucl.\ Phys.\  {\bf 1}, 189 (1985). \label{itonyCBM}

\bibitem{tonyANU} A.\,W.~Thomas and S.\,V.~Wright, ``Classical Quark Models: An
Introduction,'' in \textit{Frontiers in Nuclear Physics: From Quark - Gluon Plasma to Supernova}, ed.\ S.~Kuyucak (World Scientific, Singapore, 1999) pp.~171-211. \label{itonyANU}

\bibitem{bruceCBM} B.C.~Pearce and I.R.~Afnan, 
Phys.\ Rev.\ {\bf C\,34}, 991 (1986). 

\bibitem{ishii} N.~Ishii, 
Phys.\ Lett.\ {\bf B\,431}, 1 (1998). 

\bibitem{radiiCh} E.\,J.~Hackett-Jones, D.\,B.~Leinweber and A.\,W.~Tho\-mas, 
Phys.\ Lett.\ {\bf B\,489}, 143 (2000); 
{\it ibid}. {\bf 494}, 89 (2000).

\bibitem{young} D.\,B.~Leinweber, A.\,W.~Thomas and R.\,D.~Young, 
Phys.\ Rev.\ Lett.\  {\bf 86}, 5011 (2001). 

\bibitem{gunner} R.T.~Cahill and S.M.~Gunner, 
Phys.\ Lett.\ {\bf B\,359}, 281 (1995). 
 
\bibitem{hess} M.~Hess, F.~Karsch, E.~Laermann and I.~Wetzorke, 
Phys.\ Rev.\ {\bf D\,58}, 111502 (1998). 

\bibitem{Close:br} F.\,E.~Close and A.\,W.~Thomas, 
Phys.\ Lett.\ {\bf B\,212}, 227 (1988); 
A.\,W.~Schreiber, A.\,I.~Signal and A.\,W.~Thomas, 
Phys.\ Rev.\ {\bf D\,44}, 2653 (1991); 
M.~Alberg, E.\,M.~Henley, X.\,D.~Ji and A.\,W.~Thomas, 
Phys.\ Lett.\ {\bf B\,389}, 367 (1996). 

\bibitem{oettelthesis} M.~Oettel, ``Baryons as relativistic bound states of quark and diquark,'' PhD Thesis, University of T\"ubingen, 
\href{http://www.arXiv.org/abs/nucl-th/0012067}{{\tt nucl-th/0012067}}. \label{ioettelthesis}

\bibitem{mark} C.\,J.\ Burden, C.\,D.\ Roberts and M.\,J.\ Thomson, Phys.\
Lett.\ {\bf B\,371}, 163 (1996).

\bibitem{valencedistn} M.~B.~Hecht, C.~D.~Roberts and S.~M.~Schmidt,
  Phys.\ Rev.\ {\bf C\,63}, 025213 (2001).

\bibitem{reglaws} R.\,T.~Cahill, 
Nucl.\ Phys.\ {\bf A\,543}, 63C (1992). \label{ireglaws}
 
\bibitem{rhopipipeter} K.\,L.~Mitchell and P.\,C.~Tandy, Phys.\ Rev.\ {\bf C\,55}, 
1477 (1997). \label{irhopipipeter}

\bibitem{tonysoft} A.\,W.~Thomas and K.~Holinde, 
Phys.\ Rev.\ Lett.\  {\bf 63}, 2025 (1989). 

\bibitem{oettelcomp} M.~Oettel, L.~Von Smekal and R.~Alkofer,
Comput.\ Phys.\ Commun.\ {\bf 144}, 63 (2002). \label{ioettelcomp}

\bibitem{sigmaterms} V.\,V.~Flambaum, A.~H\"oll, P.~Jaikumar, C.\,D.~Roberts, and S.\,V.~Wright, ``Sigma Terms of Light-Quark Hadrons,'' to appear in \emph{Few Body Systems}, 
\href{http://www.arXiv.org/abs/nucl-th/0510075}{{\tt nucl-th/0510075}}. \label{isigmaterms}

\bibitem{changlei} L.~Chang, Y.-X.~Liu and H.~Guo,
  Phys.\ Rev.\ D {\bf 72}, 094023 (2005).

\bibitem{oettelpichowsky} M.~Oettel, M.\,A.~Pichowsky and L.~von Smekal,
Eur.\ Phys.\ J.\ {\bf A\,8}, 251 (2000). \label{ioettelpichowsky}

\bibitem{arneJ} R.~Alkofer, A.~H\"oll, M.~Kloker, A.~Krassnigg and C.~D.~Roberts,
Few Body Syst.\  {\bf 37}, 1 (2005). \label{iarneJ}

\bibitem{blochff} J.\,C.\,R.~Bloch, A.~Krassnigg and C.\,D.~Roberts,
Few Body Syst.\ {\bf 33}, 219 (2003). \label{iblochff}

\bibitem{bc80} J.\,S.\ Ball and T.-W.\ Chiu, Phys.\ Rev.\ {\bf D\,22}, 2542
(1980).

\bibitem{Walcher03} J.~Friedrich and T.~Walcher,
Eur.\ Phys.\ J.\ {\bf A\,17}, 607 (2003). \label{iWalcher03}

\bibitem{galster} S.~Galster, H.~Klein, J.~Moritz, K.~H.~Schmidt, D.~Wegener and J.~Bleckwenn,
Nucl.\ Phys.\ {\bf B\,32}, 221 (1971). \label{igalster}

\bibitem{brodskyhiller92} S.\,J.~Brodsky and J.\,R.~Hiller,
Phys.\ Rev.\ {\bf D\,46}, 2141 (1992).

\bibitem{kubis} B.~Kubis and U.~G.~Meissner,
Nucl.\ Phys.\ {\bf A\,679}, 698 (2001).

\bibitem{ashley} J.\,D.~Ashley, D.\,B.~Leinweber, A.\,W.~Thomas and R.\,D.~Young,
Eur.\ Phys.\ J.\ {\bf A\,19}, 9 (2004). \label{iashley}

\bibitem{sato} T.~Sato and T.-S.\,H.~Lee,
Phys.\ Rev.\ {\bf C\,63}, 055201 (2001).

\bibitem{JLabRosenbluth} V.~Tvaskis, J.~Arrington, M.\,E.~Christy, R.~Ent, C.\,E.~Keppel, Y.~Liang and G.~Vittorini, ``Experimental constraints on non-linearities induced by two-photon effects in elastic and inelastic Rosenbluth separations,'' \href{http://www.arXiv.org/abs/nucl-ex/0511021}{{\tt nucl-ex/0511021}};
and references therein.

\bibitem{arneJ2} A.~H\"oll, R.~Alkofer, M.~Kloker, A.~Krassnigg, C.\,D.~Roberts and S.\,V.~Wright,
  Nucl.\ Phys.\ {\bf A\,755}, 298 (2005). \label{iarneJ2}
  
\bibitem{wiringa} S.\,C.~Pieper, K.~Varga and R.\,B.~Wiringa,
  Phys.\ Rev.\ {\bf C\,66}, 044310 (2002).

\bibitem{belitsky} A.\,V.~Belitsky, X.-d.~Ji and F.~Yuan,
Phys.\ Rev.\ Lett.\  {91} (2003) 092003.

\bibitem{pQCDpionFF} A.\,V.~Efremov and A.\,V.~Radyushkin,
  Phys.\ Lett.\ {\bf B\,94}, 245 (1980);
G.\,P.~Lepage and S.\,J.~Brodsky,
  Phys.\ Rev.\ D {\bf 22}, 2157 (1980).

\bibitem{symanzik} K.~Symanzik, in \textit{Local Quantum Theory}, ed.\ R.~Jost (Academic, New York, 1963). \label{isymanzik}

\bibitem{streater} R.\,F.~Streater and A.\,S.~Wightman, \textit{PCT, Spin and Statistics, and All That}, 3rd edition (Addison-Wesley, Reading, Mass, 1980).

\bibitem{seiler} E.~Seiler, \textit{Gauge Theories as a Problem
of Constructive Quantum Theory and Statistical Mechanics} (Springer-Verlag,
New York, 1982).

\bibitem{rivers} R.\,J.~Rivers, \textit{Path integral methods in quantum field theory} (Cambridge University Press, Cambridge, 1987). \label{irivers}

\end{enumerate}
\end{document}